\documentclass[a4paper,10pt]{article}
\usepackage[utf8]{inputenc}
\usepackage[T1]{fontenc}
\usepackage[x11names,dvipsnames]{xcolor}
\usepackage{amssymb}
\usepackage{amsmath}
\usepackage{verbatim}
\usepackage{booktabs}
\usepackage{multicol}
\usepackage{xspace}
\usepackage{underscore}
\usepackage{upgreek}
\usepackage[textsize=scriptsize]{todonotes}
\usepackage{lineno}
\usepackage{pxfonts}
\usepackage{algorithm}
\usepackage{algpseudocode}
\usepackage{bm}
\usepackage{mathtools}
\usepackage[normalem]{ulem}
\usepackage[draft]{fixme}
\usepackage{comment}
\usepackage{url}

\usepackage[font=small,skip=0pt]{caption}

\usepackage{hyperref}

\hypersetup{
  pdfstartview      = {FitH},
  pdfpagelayout     = {OneColumn},
  pdfpagelabels     = true,             % for the reader to display the page number as 'ii (4 of 40)' rather than simply '4 of 40'
  plainpages        = false,            % page anchors with formatted form of the page number (i.e., different anchors for pages 'ii' and '2')
  bookmarksnumbered = {true},
  naturalnames      = {true},
  colorlinks        = {true},           % ligas con color; si es falso pone un recuadro que no aparece en la impresion
  linkcolor         = blue,             % color de las ligas dentro del documento (tabla de contenido, secciones, etc.)
  anchorcolor       = red,              % ?
  urlcolor          = NavyBlue,         % color de las ligas externas
  citecolor         = BrickRed          % color de las referencias
}

\usepackage{newmacros}

\renewcommand{\arraystretch}{1.3}
\renewcommand{\emptyset}{\varnothing}

\includecomment{proof:defining.A.in.ULTS} % Proof of Proposition 3

\includecomment{proof:KhE.KhA.derivables.short} % Short proof of Proposition 10
\excludecomment{proof:KhE.KhA.derivables.detailed} % Detailed proof of Proposition 10

\excludecomment{proof:simple-lts-ltsu-corresp} % Proof of Proposition 12

\includecomment{proof:kh-sound-complete-cults} % Proof of Theorem 5

\begin{document}

\title{Uncertainty-Based Knowing How Logic}
%\title{Knowing How Logic over Indistinguishable Plans}

\author{
	{\renewcommand{\arraystretch}{1.5}
\begin{tabular}{c@{\qquad}c}
  Carlos Areces$^{1,2}$ & Raul Fervari$^{1,2,3}$\\
 Andr\'es R. Saravia$^{1,2}$ & Fernando R. Vel\'azquez-Quesada$^{4}$
\end{tabular} }
}

\date{
\begin{small}
$^1$FAMAF, Universidad Nacional de C\'ordoba, Argentina \\
$^2$Consejo Nacional de Investigaciones Cient\'ificas y T\'ecnicas (CONICET), Argentina\\
$^3$Guangdong Technion - Israel Institute of Technology (GTIIT), China\\
$^4$Universitetet i Bergen, Norway
\end{small}}

\maketitle

\begin{abstract}
We introduce a novel semantics for a multi-agent epistemic operator of
\emph{knowing how}, based on an indistinguishability relation between
plans. Our proposal is, arguably, closer to the standard presentation
of \emph{knowing that} modalities in classical epistemic logic.  We
study the relationship between this new semantics and
previous approaches, showing that our setting is general enough to
capture them. We also study the logical properties of the new semantics.
First, we define a sound and complete axiomatization. Second, we define a
suitable notion of bisimulation and prove correspondence theorems.
Finally, we investigate the computational complexity
of the model checking and satisfiability problems for the new logic. 

%%% Local Variables:
%%% mode: latex
%%% TeX-master: "main"
%%% End:

\end{abstract}

%------------------------------------------------------------------------------------------------
\section{Introduction}\label{sec:intro}
% main = main.tex
Epistemic logic~(EL; \cite{Hintikka:kab,RAK}) is a logical formalism tailored
for reasoning about the knowledge of abstract autonomous entities commonly
called agents (e.g., a human being, a robot, a vehicle). It has contributed to the formal study of complex multi-agent epistemic notions not only in philosophy \cite{Hendricks2006} but also in computer science \cite{RAK,MeyervanDerHoek1995elaics} and economics \cite{Perea2012}.

Standard epistemic logics deal with an agent's knowledge about the truth-value
of propositions (the notion of \emph{knowing that}). Thus, they focus
on the study of sentences like \emph{``the agent knows that it is sunny in
  Paris''} or \emph{``the robot knows that it is standing next to a
  wall''}. For doing so, at the semantic level, EL formulas are typically interpreted over
relational models~\cite{mlbook,blackburn06}: essentially, labeled
directed graphs. The elements of the domain (called \emph{states} or
\emph{worlds}) represent different possible situations, and they fix the facts an agent might or might not know. Then, the knowledge of each agent is given by her \emph{epistemic
  indistinguishability} relation, used to represent her uncertainty about the truth: related states are considered indistinguishable for the
agent. Finally, an agent is said to know that a proposition
$\varphi$ is true at a given state $s$ if and only if $\varphi$ holds
in all states she cannot distinguish from $s$ (i.e., in all states accessible from $s$). In order to capture properly the properties of knowledge, it is typically assumed
that the indistinguishability relation is an equivalence relation.

In spite of its simplicity, this indistinguishability-based
representation of knowledge has several advantages. First, it captures the agent's \emph{high-order} knowledge (knowledge about
her own knowledge and that of other agents). Moreover, due to its generality, it opens the way to study other epistemic notions, such as the notion of \emph{belief}~\cite{Hintikka:kab}. Finally,
it allows a very natural
representation of actions through which knowledge changes~\cite{DELbook,vanBenthem2011ldii}.

%\medskip

In recent years, other forms of knowledge have been studied (see the discussion in~\cite{Wang16}). Some authors have studied knowledge of propositions using rather the notion of \emph{knowing whether}~\cite{Hart:1996,FWvD15}; some others have focused in the reasons/justifications for propositional knowledge, exploring the notion of \emph{knowing why}~\cite{artemov2008logic,XuW16}; some more have looked at more general scenarios, proposing logics for \emph{knowing the value}~\cite{GW16,Baltag16,EijckGW17}. A further and particularly interesting form of knowledge, motivated by different scenarios in philosophy and AI, is one that focuses rather on the agent's abilities: the notion of \emph{knowing how}~\cite{fantl2012introduction}. Intuitively, an agent knows how to achieve $\varphi$ given $\psi$ if she has the \emph{ability} to guarantee that $\varphi$ will be the case whenever she is in a situation in which~$\psi$ holds. Arguably, this notion is particularly important as it provides the formal foundations of automated planning and strategic reasoning within AI.

%\smallskip

%\begin{textcolor}{blue}{
		Historically, the concept of knowing how has been  considered different from knowing that, as posed e.g.\ in~\cite{Ryle1949}. Knowing how is often seen as a reflection of actions or abilities that agents may take, in an intelligent manner, in order to achieve a certain goal. In turn, 
%\end{textcolor} 
there is a large literature connecting \emph{knowing how} with logics of knowledge and action (see, e.g.,~\cite{Mccarthy69,Moore85,Les00,Hoek00,HerzigT06}). However, the way in which these proposals represent \emph{knowing how} has been the target of criticisms. The main issue is that a simple combination of standard operators expressing \emph{knowing that} and \emph{ability} (see, e.g.,~\cite{wiebeetal:2003}) does not seem to lead to a natural notion of \emph{knowing how} (see~\cite{JamrogaA07,Herzig15} for a discussion).
%One requires a \emph{de re} reading: there should be a `course 
%of action' that, when executed in $\psi$-situations, leads only to 
%$\varphi$-states. 

%\medskip

Taking these considerations into account, \cite{Wang15lori,Wang16,Wang2016} introduced a novel framework based on a binary \emph{knowing how} modality that is not defined in terms of \emph{knowing that}. At the semantic level, this language is also interpreted over relational models --- called in this context labeled transition systems (LTSs). Yet, relations do not represent
indistinguishability anymore; they rather describe the actions the agent has at her disposal (similar to what is done in, e.g., propositional dynamic logic~\cite{HKT00}). Indeed, an edge labeled $a$ from state $w$ to state $u$ indicates now that the agent can execute action $a$ to transform state $w$ into $u$. In the proposed semantics, the new modality $\kh(\psi,\varphi)$ holds if and only if there is a ``plan'' --- a sequence of actions satisfying a constraint called strong executability (SE) --- leading from $\psi$-states to $\varphi$-states. Intuitively, SE implies that the plan is ``fail-proof'' in the LTS; it unerringly leads from every $\psi$-state only to $\varphi$-states. Other variants of this \emph{knowing how} operator follow a similar approach (see~\cite{Li17,LiWang17,FervariHLW17,Wang19a}). Further motivation for these semantics can be found in the referred papers.

%\smallskip

It is interesting to notice how LTSs have no epistemic component: their relations are interpreted as actions, and then the abilities of an agent are defined only in terms what these actions can achieve. This is in sharp contrast with standard EL, where relational models provide two kinds of information: ontic facts about the given situation (the model's evaluation point) and the particular way an agent `sees' this situation (both the possible states available in the model and the agent's indistinguishability relation among them). In particular, in a multi-agent scenario, all agents share the same ontic information, and differ on their \emph{epistemic interpretation} of it. %We will come back to this later.}
If one wants to mirror the situation in EL, it seems natural that \emph{knowing how} should be defined in terms of some kind of indistinguishability over the actual situation. Such an extended model would then be able to capture both the abilities of an agent as given by her available actions (the ontic information) as well as the knowledge (or lack of it) that arises from her uncertainty (the epistemic information). 
%when considering two different actions/plans/executions indistinguishable .

%\medskip

This paper investigates a new semantics for $\kh_i(\psi,\varphi)$, a multi-agent version of the \emph{knowing how} modality, first presented in~\cite{AFSVQ21}.    
%\begin{textonuevo}
These semantics introduces two ideas. The first, the crucial, is the use of a notion of \emph{epistemic indistinguishability over plans}, in the spirit of the \emph{strategy indistinguishability} of, e.g.,~\cite{JamrogaH04,Belardinelli14}. The intuition behind it is that, under the original \lts semantics, the only reason why an agent might not know how to achieve a goal %\emph{lack} certain ability 
is because there is no adequate action available. However, one can think of scenarios in which the lack of knowledge arises for a different reason: the agent might have an adequate plan available (she has the ability to do something), and yet she might not be able to distinguish it from a non-adequate one, in the sense of not being able to tell that, in general, these plans produce different outcomes. \autoref{sec:kh-ults} provides a deeper discussion on this. In this way, these \emph{uncertainty-based} $\ltss$ reintroduce the notion of epistemic indistinguishability.

Now, although indistinguishability over plans is the main idea behind the new semantics, this proposal incorporates a second insight. One can also think of scenarios in which some of the actions, despite being \emph{ontically} available, are not \emph{epistemically} accessible to the agent. There might be several reasons for this, but an appealing one is that the agent might not be \emph{aware of} all available actions. In such cases, the epistemically inaccessible actions are then not even under consideration when the agent looks for a plan to reach a goal. The idea of awareness is not new in the EL literature: it has been used for dealing with the EL problem of logical omniscience~\cite{Vardi1986,Stalnaker1991,HP11:LogOm} by allowing the agent not to be aware of all involved atoms/formulas, thus bringing it closer to what a `real' resource-bounded agent is (see~\cite{FaginHalpern1988}).
%\fer{This part tries to motivate the ``agent might not have an action available'' thing by using the notion of awareness. What do you think? If it looks reasonable, we still have to adapt the last part of the introduction (``our contributions''). For the sake of uniformity, I also added some lines going slightly deeper into this on \autoref{sec:kh-ults}. Any changes here should be adapted there.}  

%\smallskip

Notice that, the ideas discussed above are in line with a reading that has a consensus among the literature (see, e.g.,~\cite{Hawley2003}): knowing how of an agent entails her ability (i.e., the capacity of actually doing it); but ability does not necessarily entail knowing how.
It is equally important to notice that, in the new semantics, the agent does not \emph{need} to be incapable of distinguishing certain actions, and she does not \emph{need} to be unaware of some of them. As it will be proved, the new semantics are a generalisation of the original ones in \cite{Wang15lori,Wang2016}. Thus, an agent in the new semantics who does not have uncertainty among plans and has full awareness of all of them is, knowledge-wise, exactly as an agent in the original semantics.
%We interpret formulas on \emph{uncertainty-based LTSs ($\ults$)}, which are LTSs equipped with an indistinguishability relation \emph{over plans}. The intuition is that an agent, when trying to achieve a goal, an agent may have different plans at her disposal, all of them ``as good as any other'' (and in that sense indistinguishable) \emph{as far as she can tell}.
%\end{textonuevo} 

%\bigskip

%Consider
%the following example. An agent knows how to get drunk: by drinking
%two bottles of wine, or by drinking three glasses of whiskey. The two
%plans are different, but lead to the same goal: drunkenness. An agent
%might consider them indistinguishable, and together they constitute a
%\emph{strategy} (from suitable preconditions of availability of wine
%or whiskey to the goal state of being drunk).  

% Moreover, the use of $\ults{s}$ leads to a
% natural definition of operators that represent dynamic aspects of
% knowing how (e.g., the concept of \emph{learning how}
% can be modeled by eliminating uncertainty between plans).\fer{This last sentence makes sense only if we say something about such operators. Otherwise, we might save it for the conclusions and further work.}

\paragraph{Our contributions.} Our work aims to shed new light on knowing how logics. In particular, we investigate a new multi-agent semantics for capturing the notion of knowing how, generalizing previous proposals~\cite{Wang15lori,Wang16,Wang2016,AFSVQ21}. 
Herein, we establish a distinction between ontic information shared by the agents (or abilities), and epistemic information for each individual agent (or awareness), at the level of models. In our semantics, knowing how is given by the latter, instead of by the former, as in existing approaches~\cite{Wang15lori,Wang16,Wang2016}.
Moreover, we present a thorough study of the metalogical properties of the new logic, and compare it with previous approaches.  
Our contributions can be summarized as follows: 
\begin{enumerate} 
  \item We introduce a new semantics for $\kh_i(\psi,\varphi)$ (for $i$ an agent) that reintroduces the notion of epistemic indistinguishability from classical EL. This dimension captures the awareness for each particular agent over the available abilities in the real world.
  \item We introduce a suitable notion of bisimulation for the new semantics, based on ideas from~\cite{FervariVQW17,FervariVQW21}. We prove an invariance result, and a Hennessy-Milner style theorem over finite models.
  \item We show that the logic obtained is strictly weaker (and this is an advantage, as we will discuss) than the logic from~\cite{Wang15lori,Wang16,Wang2016}. Still, the new semantics is general enough to capture the original proposal by imposing adequate conditions on the class of models.  Apart from the direct correspondence between models of each framework established already in~\cite{AFSVQ21}, we introduce a new general class of models that also does the job.
  \item We present a sound and complete axiomatization for the logic over the class of all models.
  % \item We introduce a notion of filtrations that, given a satisfiable formula, enables us to obtain a finite model satisfying such formula. 
  \item We study the computational properties of our logic. First, we provide a finite model property via filtrations. I.e., we show how, given an arbitrary model, it is possible to obtain a finite model satisfying the same set of formulas. A more careful selection argument can be used to prove that the satisfiability problem for the new logic is \NP-complete, whereas model checking is in \Poly. 
  %To our knowledge, this is the first time that exact complexity bounds for knowing how logics are given.
\end{enumerate}

This paper extends the work presented in~\cite{AFSVQ21}. Herein, we provide detailed discussions and motivations, and full proofs. Moreover, the results about bisimulations, expressive power and finite models via filtrations, are novel with respect to~\cite{AFSVQ21}. 

  \paragraph{Outline of the article.} \autoref{sec:kh-lts} recalls the framework
  of~\cite{Wang15lori,Wang16,Wang2016}, including its axiom
  system. \autoref{sec:kh-ults} introduces \emph{uncertainty-based LTSs},
  indicating how they can be used for interpreting a multi-agent version
  of the \emph{knowing how} language. In \autoref{sec:bisim} we introduce a suitable notion of bisimulation, together with correspondence theorems. We provide a sound and complete axiom
  system in \autoref{sec:axiom}. \autoref{sec:comparing} studies the correspondence between
  our semantics and the one in the original proposals. In particular, we present two different classes of models that capture the original semantics. In \autoref{sec:complexity} we investigate a finite model property via filtrations (\autoref{subsec:filtrations}), and the
  computational complexity of model checking and the satisfiability problem for our
  logic (\autoref{subsec:complexity}). We conclude in \autoref{sec:final} with some conclusions and future lines of research.

%%% Local Variables:
%%% mode: latex
%%% TeX-master: "main"
%%% End:

\section{A short review of the literature}

The ideas discussed in the previous section concerning the notion of \emph{knowing how} introduced in~\cite{Wang15lori,Wang16,Wang2016} have been successful, and have lead to different works in the literature. An earlier one is \cite{LiWang17}, which considers a ternary modality $\kh(\psi,\chi,\varphi)$ asking for a plan whose intermediate states satisfy $\chi$. Then, \cite{Li17} introduces a weaker binary modality $\kh^{\sf w}(\psi,\varphi)$ that allows plans that abort, and in which the states reached by these aborted executions should also satisfy the goal $\varphi$. Finally, \cite{Wang19a} uses a semantics under which intermediate actions in a given plan may be skipped.

The respective works introducing these variants also provide an axiom system (interestingly, the logic for the modality with skippable plans is the same as the logic for the original modality). Regarding computational behaviour, the satisfiability problem has been proved to be decidable (in their respective papers) for the basic system, the one allowing aborted executions and the one with skippable plans. However, no complexity bounds for any of the systems have been given. Finally, suitable notions of bisimulation for these systems can be found in \cite{FervariVQW17,FervariVQW21} (for all but the one with skippable plans) and \cite{Wang19a} (for the one with skippable plans). These bisimilarity tools have been useful to investigate the systems' relative expressive power. It has been shown that the original binary modality $\kh(\psi,\varphi)$ is strictly less expressive than the one with intermediate steps ($\kh(\psi,\chi,\varphi)$), and that they are both incomparable with the modality with aborted executions $\kh^{\sf w}(\psi,\varphi)$. Moreover, in~\cite{DF23} the computational complexity of the model-checking problem for different knowing how logics is characterized. In particular, it is established that model-checking for the basic knowing how logic from~\cite{Wang15lori,Wang16,Wang2016} is \PSPACE-complete, whereas for a variant with budget constraints is \EXPSPACE-hard. Other constraints over plans are also studied therein, concretely the variant of~\cite{AFSVQ21} (the one studied in this paper) with regularity constraints and budgets, for which model-checking is in \Poly. More recently, in~\cite{ACCFS23}, the framework of knowing how is extended to a deontic setting, formalizing the notion of \emph{knowingly complying}.
 
Further proposals explore new features. For instance, a natural extension is considering the interaction between \emph{knowing how} and standard \emph{knowing that} modalities. In~\cite{FervariHLW17}, a single-agent logic with the two modalities is introduced. The knowing how operator is, unlike previous approaches, a unary local modality $\kh(\varphi)$, and its interpretation allows branching plans. The interaction between both kinds of knowledge is studied via an axiom system, and it is proved that its satisfiability problem is decidable. The decidability result has been recently refined in~\cite{Li21}, where \PSPACE-completeness is proved for the satisfiability problem, via a tableau-based procedure. In~\cite{LiW21b} a neighbourhood semantics is provided for the \emph{knowing how} modality, as an alternative to the standard relational semantics. 

Other papers incorporate multi-agent behaviour for \emph{knowing how} and \emph{knowing that} modalities. For instance, in~\cite{NaumovT17tark,Naumov2018a} this is explored in the context of coalitions, i.e., the logic is used to describe different notions of collective knowledge. It is known that a fragment of this logic is incomparable in expressive power with the logic from~\cite{FervariHLW17} (the proof uses bisimulation, and it is presented in~\cite{FervariVQW21}). Other variants of this logic have been explored, including those relying on \emph{second-order knowing how} strategies~\cite{NaumovT18}, and \emph{knowing how} with degrees of uncertainty~\cite{NaumovT19}. Axiom systems are presented for each logic.
%An epistemic semantics based on indistinguishability relations over plans, was first introduced in~\cite{AFSVQ21}, exploring some preliminary results.

Finally, a multi-agent knowing how logic describing the behaviour of epistemic planning is investigated in~\cite{LiW19}. The main peculiarity is that the execution of an action is represented by an update in the model via epistemic action models~\cite{BMS}. The logic obtained is strictly weaker than the one in~\cite{FervariHLW17}. Again, its satisfiability problem is decidable. This work is extended in~\cite{LiW21}, which provides a unified approach for planning-based knowing how. More remarkably, the work in~\cite{LiW21c} establishes a connection between planning and knowing how, not just from the perspective of \emph{planning-based} know how, but also the other way around: a planning problem based on know how goals. To do so, the authors introduce a model checking algorithm running in $\Poly$ time.

% \begin{itemize}
% % \item Basic approach: \cite{Wang15lori,Wang16,Wang2016}, single-agent modality $\kh(\psi,\varphi)$ under 
% % \lts-based semantics.
% % \item Intermediate steps~\cite{LiWang17}: a ternary modality $\kh(\psi,\theta,\varphi)$ over \ltss, but each intermediate step must satisfy $\theta$.
% % %% intermediate is more expressive than standard knowing how.
% % \item So-called ``weakly knowing how''~\cite{Li17}: binary $\kh(\psi,\varphi)$ but the witness plan may abort.
% % % incomparable with kh and intermediate kw.
% %\item Knowing how with skippable plans~\cite{Wang19a}.
% % same as kh.
% \item Local knowing how and knowing that~\cite{FervariHLW17} (unary modality $\kh\varphi$ with branching plans).
% \item Multi-agent knowing how via Dynamic Epistemic Planning~\cite{LiW19}.
% % similar to IJCAI 2017, but multi-agent and where actions are epistemic action models, leads to a weaker logic. Claim to illustrate the behaviour of epistemic planning.
% \item One-step multi-aget knowing how with coalitions~\cite{NaumovT17tark,Naumov2018a}. 
% % incomparable with IJCAI 2017. the kh behaviour is simpler.
% Second-order know how strategies~\cite{NaumovT18}. With degrees of uncertainty~\cite{NaumovT19}.
% %\item Bisimulations and expressive power are studied in~\cite{FervariVQW17,FervariVQW21}.
% \end{itemize}

%------------------------------------------------------------------------------------------------
\section{A logic of knowing how}\label{sec:kh-lts}
%!TEX root = main.tex
This section recalls the basics of the \emph{knowing how} framework from \cite{Wang15lori,Wang16,Wang2016}.

\msparagraph{Syntax and semantics.} Throughout the text, let $\PROP$ be a countable non-empty set of propositional symbols.

\begin{definicion}\label{def:khsyntax}
  Formulas of the language \KHlogic are given by the grammar
  \begin{nscenter}
    $\varphi ::= p \mid \neg\varphi \mid \varphi\vee\varphi \mid \kh(\varphi,\varphi)$,
  \end{nscenter}
  with $p \in \PROP$. Boolean constants and other Boolean connectives are defined as usual. Formulas of the form $\kh(\psi,\varphi)$ are read as \emph{``when $\psi$ holds, the agent knows how to make $\varphi$ true''}.
\end{definicion}

In \cite{Wang15lori,Wang16,Wang2016} (and variations like \cite{LiWang17,Li17}), formulas of $\KHlogic$ are interpreted over \emph{labeled transition systems}: relational models in which the relations describe the state-transitions available to the agent. Throughout the text, let $\ACT$ be a denumerable set of (basic) action names. %First, some syntactic notions and then the structure.

\begin{definicion}[Actions and plans]
  Let $\ACT^*$ be the set of finite sequences over $\ACT$. Elements of
  $\ACT^*$ are called \emph{plans}, with $\epsilon$ being the
  \emph{empty plan}. %Let $\ACT^+ := \ACT^* \setminus  \cset{\epsilon}$. 
  Given $\sigma \in \ACT^*$, let $\card{\sigma}$ be
  the length of $\sigma$ (note: $\card{\epsilon} := 0$). For a plan $\sigma$ and 
  $0 \le k \le \card{\sigma}$, the \emph{plan} $\sigma_k$ is $\sigma$'s initial
  segment up to (and including) the $k$th position (with
  $\sigma_0 := \epsilon$). For $0 < k \le \card{\sigma}$, the \emph{action}
  $\sigma[k]$ is the one in $\sigma$'s $k$th position.
\end{definicion}

\begin{definicion}[Labeled transition systems]\label{def:abmap}
  A \emph{labeled transition system (\lts)} for $\PROP$ and $\ACT$ is
  a tuple $\modlts = \tup{\W,\R,\V}$ where $\W$ is a non-empty set of
  states (also denoted by $\D{\modlts}$),
  $\R = \csetsc{\R_a \subseteq \W \times \W}{a \in A, \text{ for some }A\subseteq\ACT}$ is a
  collection of binary relations on $\W$,\footnote{Thus, $\R_a$ might not be defined for some $a \in \ACT$.} and $\V:\W \to 2^\PROP$ is a
  labelling function. Given an \lts $\modlts$ and $w \in \D{\modlts}$,
  the pair $(\modlts,w)$ is a \emph{pointed \lts} (parentheses are
  usually dropped).
\end{definicion}

An \lts describes the \emph{abilities} of the agent; thus, sometimes (e.g., \cite{Wang15lori,Wang16,Wang2016}) it is also called an \emph{ability map}. Here we introduce some useful definitions. It is worth noticing that, although the signature is infinite (since $\ACT$ is a \emph{denumerable} set), the relations in the model might be defined only for a (possibly finite) subset of actions.

\begin{definicion}
  Let $\csetsc{\R_a \subseteq \W \times \W}{a \in A, \text{ for some } A\subseteq\ACT}$ be a collection of binary relations. Define $\R_\epsilon := \csetsc{(w,w)}{w \in \W}$ and, for $\sigma \in \ACT^*$ and $a \in \ACT$, $\R_{\sigma{a}} := \csetc{(w,u)}{\W \times \W}{ \exists v \in \W \text{ s.t. } (w,v) \in \R_{\sigma} \text{ and } (v,u) \in \R_{a} }$. Take a plan $\sigma \in \ACT^*$: for $u \in \W$ define $\R_{\sigma}(u) := \csetsc{v \in \W}{(u,v) \in \R_{\sigma}}$, and for $U \subseteq \W$ define $\R_{\sigma}(U) := \bigcup_{u \in U} \R_{\sigma}(u)$.
\end{definicion}

The idea in \cite{Wang15lori,Wang16,Wang2016} is that an agent
knows how to achieve $\varphi$ given $\psi$ when she has an
appropriate plan that allows her to go from any state in which
$\psi$ holds only to states in which $\varphi$
holds. A crucial part is, then, what ``appropriate'' is taken to be.

%Still, one can impose different restrictions on what qualifies as an adequate plan.

\begin{definicion}[Strong executability]\label{def:plans-exec}
  Let $\csetsc{\R_a \subseteq \W \times \W}{a\in A, \text{ for some } A\subseteq \ACT}$ be a collection of binary relations. A plan $\sigma \in \ACT^*$ is 
%\emph{weakly executable} at a given $u \in \W$ if and only if $\R_\sigma(u) \neq \emptyset$. The set $\wkexec(\sigma)$ contains the states in $\W$ where $\sigma$ is weakly executable.
  \emph{strongly executable} (SE) at $u \in \W$ if and only if $\R_\sigma$ is defined %\raul{agregue la condicion de 'is defined', ya que ahora podria no estar definido}
   and, additionally, $v \in \R_{\sigma_k}(u)$ implies $\R_{\sigma[k+1]}(v) \neq \emptyset$ for every $k \in \intint{0}{\card{\sigma}-1}$. We define the set $\stexec(\sigma):= \cset{w\in\W \mid \sigma \mbox{ is SE at }w}$. % contains the states in $\W$ where $\sigma$ is strongly executable.
\end{definicion}

Thus, %while weak executability asks only for \emph{some} partial execution of the plan to be completed,
strong executability asks for \emph{every} partial execution of the plan (including $\epsilon$) to be completed. %Note: the notion of weak executability is not used in the original \emph{knowing how} proposals. It is introduced her for the purposes of comparisons and as a basis for other executability notions. It is not hard to see that $\stexec(a) = \wkexec(a)$ for every $a \in \ACT$, and that $\stexec(\sigma) \subseteq \wkexec(\sigma)$ for every $\sigma \in \ACT^*$.
With this notion, formulas in $\KHlogic$ are interpreted over an LTS as follows. Notice that the semantic clause for the $\kh$ modality shown here is equivalent to the one found in the original papers.

\begin{definicion}[\KHlogic over \ltss]\label{def:sem-abmap}
  The relation $\models$ between a pointed \lts $\modlts, w$ (with $\modlts=\tup{\W,\R,\V}$ an \lts over \ACT and \PROP) and formulas in \KHlogic (over \PROP) is defined inductively as follows:
  % The cases for atoms and Boolean operators are as usual. For \emph{knowing how} formulas,
 \begin{nscenter}
   \begin{tabular}{@{}l@{\;\;\;}c@{\;\;\;}l@{}}
     $\modlts,w \models p$ & \iffdef & $p\in\V(w)$, \\
     $\modlts,w \models \neg\varphi$ & \iffdef & $\modlts,w \not\models\varphi$, \\ 
     $\modlts,w \models \varphi\vee\psi$ & \iffdef & $\modlts,w \models \varphi \,\mbox{ or }\, \modlts,w \models\psi$, \\
     $\modlts,w \models \kh(\psi,\varphi)$ & \iffdef & \begin{minipage}[t]{0.7\textwidth}
                                                         there exists $\sigma \in \ACT^*$ such that \\
                                                         {\centering
                                                           \begin{inline-cond-kh}\item $\truthset{\modlts}{\psi} \subseteq \stexec(\sigma)$ and \item $\R_\sigma(\truthset{\modlts}{\psi}) \subseteq \truthset{\modlts}{\varphi}$,\end{inline-cond-kh}
                                                          }
                                                       \end{minipage}
   \end{tabular}
 \end{nscenter}
  % \begin{nscenter}
  %     $\modlts,w \models p ~\iffdef~ p\in\V(w)$, \quad\quad
  %     $\modlts,w \models \neg\varphi  ~\iffdef~  \modlts,w \not\models\varphi$, \quad\quad 
  %     $\modlts,w \models \varphi\vee\psi  ~\iffdef~  \modlts,w \models \varphi \mbox{ or } \modlts,w \models\psi$, \\
  %     $\modlts,w \models \kh(\psi,\varphi)$  ~\iffdef~  \begin{minipage}[t]{0.725\textwidth}
  %                                                         $\exists \sigma \in \ACT^*$ such that \begin{inline-cond-kh} \item $\truthset{\modlts}{\psi} \subseteq \stexec(\sigma)$ and \item $\R_\sigma(\truthset{\modlts}{\psi}) \subseteq \truthset{\modlts}{\varphi}$,
  %                                                         \end{inline-cond-kh}
  %                                                       \end{minipage}
  % \end{nscenter}
  with $\truthset{\modlts}{\varphi} := \csetc{w}{\W}{\modlts,w \models \varphi}$ (the elements of $\truthset{\modlts}{\varphi}$ are sometimes called $\varphi$-states). %A formula $\varphi$ is satisfiable iff $\lts,w\models\varphi$ for some $\lts$ and $w\in \D{\lts}$, and it is valid (notation $\models\varphi$) iff $\lts,w\models\varphi$ for all $\lts$ and all $w \in \D{\lts}$.  A formula $\varphi$ holds (globally) in an LTS $\lts$ ($\lts \models \varphi$) iff $\lts,w \models \varphi$ for all $w \in \D{\lts}$.  For $\Psi \subseteq \KHlogic$, $\lts,w\models\Psi$ iff $\lts,w\models\psi$, for all $\psi\in\Psi$;   and $\Psi \models\varphi$ iff for all $\lts$, $w \in \D{\lts}$, $\lts,w\models\Psi$ implies $\lts,w\models\varphi$.
  The plan $\sigma$ in the semantic case for $\kh(\psi,\varphi)$  is often called the witness for
  $\kh(\psi,\varphi)$ in $\modlts$.
\end{definicion}

Thus, $\kh(\psi,\varphi)$ holds at a given $w$ when there is a plan $\sigma$ such that,
when it is executed at any $\psi$-state, it will always complete every partial
execution (condition \ITMKHi), ending unerringly in states satisfying
$\varphi$ (condition \ITMKHii). Since $w$ does not play any role in $\kh$'s semantic clause, the \emph{knowing how} operator acts \emph{globally}. Hence, $\truthset{\modlts}{\kh(\psi,\varphi)}$ is either $\D{\modlts}$ or $\emptyset$.

\msparagraph{Axiomatization.} For axiomatization purposes, note that the global universal modality \cite{GorankoP92}, interpreted as truth in every state of the model, is definable in $\KHlogic$ as $\A\varphi := \kh(\neg\varphi,\bot)$. This is justified by the proposition below, whose proof relies on the fact that $\ACT^*$ is never empty (it always contains $\epsilon$).

\begin{proposicion}[\cite{Wang15lori}]\label{prop:lts:universal}
  Let $\modlts, w$ be a pointed \lts. Then,
   \begin{center}
     $\modlts, w \models \kh(\neg\varphi,\bot)$ \qquad iff \qquad $\truthset{\modlts}{\varphi} = \D{\modlts}$. 
   \end{center} 
\end{proposicion}

\begin{table}[t]
$$
\begin{array}{l@{\ \ \ \ }ll}
\toprule
   \mbox{\underline{Block $\axset$:}}        & \axm{TAUT}   & \vdash \varphi \mbox{ for $\varphi$ a propositional tautology} \\
                             & \axm{DISTA}  & \vdash \A(\varphi\ra\psi) \ra (\A\varphi \ra \A\psi) \\
                              
                              & \axm{TA}     & \vdash \A\varphi \ra \varphi \\
                              & \axm{4KhA}   & \vdash \kh(\psi,\varphi) \ra \A\kh(\psi,\varphi) \\
                              &  \axm{5KhA}   & \vdash \neg\kh(\psi,\varphi) \ra \A\neg\kh(\psi,\varphi) \\
                         &  \axm{MP}     & \mbox{From $\vdash \varphi$ and $\vdash \varphi \limp \psi$ infer $\vdash \psi$ }\\
                         &  \axm{NECA}   & \mbox{From $\vdash \varphi$ infer $\vdash \A\varphi$} \\
     \midrule
    \mbox{\underline{Block $\axset_{\lts}$:}} & \axm{EMP}    & \vdash \A(\psi \ra \varphi) \ra \kh(\psi,\varphi)  \\
                               &           \axm{COMPKh} & \vdash (\kh(\psi,\varphi)\wedge\kh(\varphi,\chi))\ra\kh(\psi,\chi) \\
     \bottomrule
  \end{array}
  $$
  \caption{Axiom system \KHaxiom, for \KHlogic w.r.t. \ltss.}\label{tab:khaxiom}
\end{table}

The axiom system \KHaxiom (\autoref{tab:khaxiom}) shows the relationship between the global universal modality $\A$ and the knowing-how operator $\kh$. The first block is essentially a standard modal system for $\A$, additionally establishing that $\kh$ is global (see the discussion in \cite{Wang15lori}). The axioms in the second block deserve a further comment. Axiom \axm{EMP} states that, if $\psi \to \varphi$ is globally true, then given $\psi$ the agent knows how to make $\varphi$ true. In simpler words, global ontic information turns into knowledge. This is because the empty plan $\epsilon$ is always available. Axiom \axm{COMP} establishes that $\kh$ is compositional: if given $\psi$ the agent knows how to make $\varphi$ true, and given $\varphi$ she knows how to make $\chi$ true, then given $\psi$ she knows how to make $\chi$ true.

\begin{teorema}[\cite{Wang15lori}]\label{teo:kh-sound-complete-lts}
   The axiom system \KHaxiom (\autoref{tab:khaxiom}) is sound and strongly complete for \KHlogic w.r.t. the class of all \ltss.
\end{teorema}
 
Axioms in the second block might be questionable. First, one could argue that, contrary to what \axm{EMP} states, not all global truths about what is achievable in the model need to be considered as knowledge (how) of the agent. 
 %Technically, this requires situations in which the empty plan (i.e., the \emph{skip} action) is not available.
Second, notice that axiom \axm{COMPKh} implies also a certain level of  omniscience: it might as well be that an agent knows how to make $\varphi$ true given~$\psi$, and how to make $\chi$ true given $\varphi$, but still has not worked out how to put together the two witness plans to ensure $\chi$ given~$\psi$. These are the two properties that will be lost in the more general semantics introduced in the next section. In \autoref{sec:comparing} we will show how these formulas become valid when, in the new semantics, one make strong idealizations.

%%% Local Variables:
%%% mode: latex
%%% TeX-master: "tark21"
%%% End:

%------------------------------------------------------------------------------------------------
\section{Uncertainty-based semantics}\label{sec:kh-ults}
%!TEX root = tark21.tex
The \lts-based semantics provides a reasonable representation of an
agent's abilities: the agent knows how to achieve $\varphi$ given
$\psi$ if and only if there is a plan that, when executed at any
$\psi$-state, will always complete every partial execution, ending
unerringly in states satisfying $\varphi$.  Still, one could argue that
this representation involves a certain level of idealization. 

Take an agent that \emph{lacks} a certain ability. In the \lts-based
semantics, this can only happen when the environment does not provide the required
(sequence of) action(s). Still, there are situations in which an adequate plan
exists, and yet the agent lacks the ability for a different reason. Indeed, she might
\emph{fail to distinguish} an adequate plan from a non-adequate one, in the
sense of not being able to tell that, in general, those plans produce
different outcomes. Consider, for example, an agent baking a
cake. She might have the ability to do the
{nine different mixing methods}\footnote{\url{https://www.perfectlypastry.com/the-importance-of-the-mixing-method/}} (beating, blending, creaming, cutting,
folding, kneading, sifting, stirring, whipping), and she might even
recognize them as different actions. However, she might not be able to
perfectly distinguish one from the others: she might not recognize that,
sometimes, they produce different results. In such cases, one would say that the
agent does not know how to bake a cake: sometimes she gets good
outcomes (when she uses the adequate mixing method) and sometimes she
does not.

Indistinguishability among \emph{basic} actions can account for the
example above (with each mixing method a basic action). Still, one can also
think of situations in which a more general form of indistinguishability, one \emph{among plans}, is involved.  Consider the baking agent
again. It is reasonable to assume that she can tell the difference
between ``adding milk'' and ``adding flour'', but perhaps she does not
realize the effect that \emph{the order} of these actions might
have in the final result. Here, the issue is not that she cannot
distinguish between basic actions; rather, two plans are
indistinguishable because the order of their actions is being
  considered irrelevant. For a last possibility, the agent might
not know that, while opening the oven once to check whether the baking
goods are done is reasonable, this must not be done in excess. In this
case, the problem consists in not being able to tell the difference between
the effect of executing an action once and executing it multiple times.
Thus, plans of \emph{different lengths}
might be considered equivalent for the task at hand, for such an agent.
  
The previous examples suggest that one can devise a more general
representation of an agent's abilities. This involves
taking into account not only
the plans she has available (the \lts structure), but also her skills
for telling two different plans apart (a form of
\emph{indistinguishability among plans}).
%Moreover: although
%indistinguishability among plans aims to deal with additional reasons
%why an agent might lack certain ability, it also allows a more
%flexible representation of the abilities the agent does have. The
%baking agent might sometimes use a red bowl and sometimes use a green
%one, not being able to tell the difference between the two
%actions. Thus, she has actually two plans that she considers
%indistinguishable, and yet this still works because both plans produce
%always the same outcome.  The semantic structure below puts these
%ideas to work.
As we will see, this (in)ability for distinguishing plans will also let us define a natural model for a multi-agent
scenario. In this setting, agents share the same set of \emph{affordances} (provided
by the actual environment), but still have different \emph{abilities} depending
on and how well they can tell these affordances apart, or even which of these affordances they have available. 
To drive this last point home notice that, in principle, an agent does not need to have `epistemic access' to every available plan. Some might be so foreign to the agent, or  so complex, that she might not be \emph{aware of} them. Such plans are, then,  out of the agent's reach, not in the sense that she cannot distinguish them from others, but in that she does not even take them into consideration. This is similar to what~\cite{FaginHalpern1988} proposed for the epistemic notion of \emph{knowing that}: the agent might not be aware of (i.e., she might not entertain) every formula of the language, and thus she does not need to know that these formulas are indeed the case. 
%As mentioned before, this allows the representation of less idealised agents, thus making the system more suitable for modelling `real' resource-bounded human/computational agents.

\begin{definicion}[Uncertainty-based \lts]\label{def:ults}
  Let \AGT be a finite non-empty set of agents.  A \emph{multi-agent
    uncertainty-based \lts (\ults)} for $\PROP$, $\ACT$ and $\AGT$ is
  a tuple $\modults = \tup{\W,\R,\sim,\V}$ where $\tup{\W, \R, \V}$ is
  an \lts and $\sim$ assigns, to each agent $i \in \AGT$, an
  equivalence \emph{indistinguishability} relation over a non-empty
  set of plans $\DS{i} \subseteq \ACT^*$. Given an \ults $\modults$
  and $w \in \D{\modults}$, the pair $(\modults,w)$ (parenthesis
  usually dropped) is called a \emph{pointed \ults}.
\end{definicion}

Intuitively, $\DS{i}$ is the set of plans that agent $i$ has at her
disposal; it contains the plans the agent has access to. Then, similarly as in classical epistemic logic,
${\sim_i} \subseteq \DS{i} \times \DS{i}$ describes agent $i$'s
indistinguishability over her available plans.

\begin{nota}\label{rem:corresp}
  The following change in notation will simplify some definitions
  later on, and will make the comparison with the \lts-based semantics
  clearer.

  Let $\tup{\W, \R, \sim, \V}$ be an \ults and take
  $i \in \AGT$; for a plan $\sigma \in \DS{i}$, let $[\sigma]_i$ be
  its equivalence class in $\sim_i$ (i.e.,
  $[\sigma]_i := \csetc{\sigma'}{\DS{i}}{ \sigma \sim_i
    \sigma'}$). There is a one-to-one correspondence between each $\sim_i$ and its induced
  set of equivalence classes
  $\S_i := \csetsc{[\sigma]_i}{\sigma \in \DS{i}}$. Hence, from now on,
  an \ults will be presented as a tuple $\tup{\W, \R, \cset{\S_i}_{i\in\AGT}, \V}$. Notice the following
  properties of each $\S_i$: \begin{inlineenum} \item $\S_i \neq \emptyset$ (as $\DS{i} \neq \emptyset$), \item
    if $\strategy_1, \strategy_2 \in \S_i$ and $\strategy_1\neq\strategy_2$, then
    $\strategy_1 \cap \strategy_2 = \emptyset$ (equivalence classes are pairwise disjoint), \item
    $\DS{i} = \bigcup_{\strategy \in \S_i} \strategy$ (their union is exactly $\DS{1}$), and \item
    $\emptyset \notin \S_i$ (the empty set is not an equivalence class)\end{inlineenum}.
\end{nota}

Given her uncertainty over $\ACT^*$ (or, more precisely, over \emph{her} `domain of plans' $\DS{i} \subseteq \ACT^*$), the abilities of an agent $i$
depend not on what a single plan can achieve, but rather on what a set
of them can guarantee.

\begin{definicion}
  For $\strategy \subseteq \ACT^*$, $u \in \W$ and $U \subseteq \W$, 
  define
  \[
    \R_\strategy := \bigcup_{\sigma \in \strategy} \R_{\sigma},
    \qquad
    \R_{\strategy}(u) := \bigcup_{\sigma \in \strategy} \R_\sigma(u),
    \qquad
    \R_{\strategy}(U) := \bigcup_{u \in U} \R_{\strategy}(u).
  \]
\end{definicion}

We can now generalize the notion of strong executability for sets of plans.

\begin{definicion}[Strong executability]\label{def:strat-exec}
  A \emph{set of plans} $\strategy \subseteq \ACT^*$ is \emph{strongly
    executable} at $u \in \W$ if and only if \emph{every} plan
  $\sigma \in \strategy$ is \emph{strongly executable} at $u$.
  Thus, $\stexec(\strategy) := \bigcap_{\sigma \in \strategy}
  \stexec(\sigma)$ is the set of the states in $\W$ where $\strategy$ is strongly
  executable.
\end{definicion}

%There are other possibilities for defining adequate sets of plans. One might require for the \emph{existence} of a plan in $\strategy$ that is \emph{strongly} executable (`$\bigcup \stexec$'), or for \emph{all} plans in $\strategy$ to be \emph{weakly} executable (`$\bigcap \wkexec$'). This paper will focus on the two extreme cases defined above.

\begin{definicion}[$\khi$ over \ultss]\label{def:sem-esm}
  Let $\KHilogic$ be the multi-agent version of the language $\KHlogic$, obtained by replacing $\kh$ with $\khi$ (with $i\in\AGT$ for $\AGT \neq \emptyset$). The satisfiability relation $\models$ between a pointed \ults $\modults, w$ (with $\modults = \tup{\W, \R, \cset{\S_i}_{i\in\AGT}, \V}$ an \ults over \ACT, \PROP and \AGT) and formulas in $\KHilogic$ is defined inductively. The atomic and Boolean cases are as before. For
  \emph{knowing how} formulas,
  \begin{nscenter}
    \begin{tabular}{@{}l@{\;\;\;}c@{\;\;\;}l@{}}
      $\modults,w \models \khi(\psi,\varphi)$ & \iffdef & \begin{minipage}[t]{0.68\textwidth}
                                                         there exists $\strategy \in \S_i$ such that \\
                                                         {\centering
                                                           \begin{inline-cond-kh}\item $\truthset{\modults}{\psi} \subseteq \stexec(\strategy)$ and \item $\R_\strategy(\truthset{\modults}{\psi}) \subseteq \truthset{\modults}{\varphi}$,\end{inline-cond-kh}
                                                          }
                                                       \end{minipage}
    \end{tabular}
  \end{nscenter}
  with $\truthset{\modults}{\varphi} := \csetc{w}{\W}{\modults,w \models \varphi}$. The set of plans $\pi$ in the semantic case for $\khi(\psi,\varphi)$  is often called the witness for
  $\khi(\psi,\varphi)$ in $\modults$.
\end{definicion}

It is worth comparing~\autoref{def:sem-abmap} and~\autoref{def:sem-esm}.
As before, $\khi(\psi,\varphi)$ acts
\emph{globally}. But now, we require \emph{for
  agent $i$} to have a \emph{set of plans} satisfying strong
executability in every $\psi$-state (condition \ITMKHi). Still, the
set of plans should work as the single plan did before: when executed at
$\psi$-states, it should end unerringly in states satisfying $\varphi$
(condition \ITMKHii).

\smallskip

It is also important to notice that the global universal modality is also definable within $\KHilogic$ over \ults. (For this, it is crucial that $\S_i \neq \emptyset$ and $\emptyset \not\in \S_i$, as stated in \autoref{rem:corresp}.)  

\begin{proposicion}\label{pro:ults:universal}
  Let $\modults, w$ be a pointed \ults. Then,
  \begin{center}
    there is $i \in \AGT$ with $\modults, w \models \khi(\neg\varphi,\bot)$ \qquad iff \qquad $\truthset{\modults}{\varphi} = \D{\modults}$.
  \end{center}
  %\begin{center}
   % $\modults, w \models \khi(\neg\varphi,\bot)$ \qquad iff \qquad $\truthset{\modults}{\varphi} = \D{\modults}$. 
  %\end{center}  
\begin{proof:defining.A.in.ULTS}
\begin{demostracion}
  {\prooflr} Suppose there is $i \in \AGT$ with $\modults, w \models \khi(\neg\varphi,\bot)$. Then, there is $\strategy \in \S_i$ such that \begin{inline-cond-kh} \item\label{pro:esmiv-stexec-def-glo:i} $\truthset{\modults}{\lnot\varphi} \subseteq \stexec(\strategy)$ and \item\label{pro:esmiv-stexec-def-glo:ii} $\R_\strategy(\truthset{\modults}{\lnot\varphi}) \subseteq \truthset{\modults}{\bot}$\end{inline-cond-kh}. For a contradiction, suppose $\truthset{\modults}{\varphi} \neq \D{\modults}$, so there is $u \in \truthset{\modults}{\lnot\varphi}$. Then, \autoref{pro:esmiv-stexec-def-glo:i} implies $u \in \stexec(\strategy) = \bigcap_{\sigma \in \strategy} \stexec(\sigma)$. But $\strategy \in \S_i$, so $\strategy \neq \emptyset$, that is, there is $\sigma \in \strategy$ with $u \in \stexec(\sigma)$; thus, $\R_\sigma(u) \neq \emptyset$, so $\R_\strategy(u) \neq \emptyset$ and hence $\R_\strategy(\truthset{\modults}{\lnot\varphi}) \neq \emptyset$, that is, $\emptyset \subset \R_\strategy(\truthset{\modults}{\lnot\varphi})$. But then, from \autoref{pro:esmiv-stexec-def-glo:ii}, $\emptyset \subset \R_\strategy(\truthset{\modults}{\lnot\varphi}) \subseteq \truthset{\modults}{\bot}$, i.e., $\emptyset \subset \truthset{\modults}{\bot}$, a contradiction. Therefore, $\truthset{\modults}{\varphi} = \D{\modults}$. 
  
  {\proofrl} Suppose $\truthset{\modults}{\varphi} = \D{\modults}$. Then $\truthset{\modults}{\lnot\varphi} = \emptyset$ and hence \ITMKHi in the semantic clause of $\khi(\lnot\varphi, \bot)$ holds for every $\strategy \in 2^{\ACT^*}$. Moreover, $\R_\strategy(\truthset{\modults}{\lnot\varphi}) = \bigcup_{u \in \truthset{\modults}{\lnot\varphi}} \R_\strategy(u) = \bigcup_{u \in \emptyset} \R_\strategy(u) = \emptyset$, so \ITMKHii also holds for any such $\strategy$. Finally, $\S_i \neq \emptyset$ (so there is $\strategy \in \S_i$) and $\AGT \neq \emptyset$ (so there is $i \in \AGT$); therefore, there is $i \in \AGT$ with $\modults, w \models \khi(\neg\varphi,\bot)$.
\end{demostracion}
\end{proof:defining.A.in.ULTS}
\end{proposicion}
 
Hence, one can take $\A \varphi := \bigvee_{i \in \AGT} \khi(\neg\varphi,\bot)$ (recall: $\AGT$ is non-empty and finite) and $\E\varphi:=\neg\A\neg\varphi$. 
 
%\begin{nota} \label{rem:interaction}
Now, clearly different agents have different awareness about their own abilities. At the same time, because of the global nature of the modality of knowing how, it holds that 
\begin{center}
$\model,w\models\kh_i(\psi,\varphi)$ if and only if $\model,w\models\A\kh_i(\psi,\varphi)$,
\end{center}
 or equivalenty,  
\begin{center} 
 $\model,w\models\kh_j(\neg\kh_i(\psi,\varphi),\bot)$, for some agent $j$. 
\end{center}
But this does not imply that agents \emph{know that} ``agent $i$ knows how to achive $\varphi$ given $\psi$.'' It is only the case that $\kh_i(\psi,\varphi)$ becomes an objective true, and hence assuming its negation naturally leads to contradiction. There is no notion of epistemic indistinguishability over states in our models, which could lead to a notion of ``knows that''. 
 
Lastly, one can argue that since models are equipped with a notion of epistemic indistinguishability between plans, an agent should know that a certain plan is (or is not) distinguishable from another, or that an agent is aware of the availability of a certain course of action. However, knowing how modalities cannot talk about the relation itself, only about the existence of a set of indistinguishable plans, and the effects of executing those plans. 

%The rest of the section is devoted to explore the properties of the logic with our new semantics. Moreover, we compare it to the well-known framework from~\cite{Wang15lori,Wang16,Wang2016}.

%%% Local Variables:
%%% mode: latex
%%% TeX-master: "main"
%%% End:

%------------------------------------------------------------------------------------------------
\section{Bisimulations}\label{sec:bisim}
%\label{subsec:bisim}
%!TEX root = main.tex

% \subsection{Bisimulations over \lts}
% \label{subsec:bisim}

%\ssparagraph{Bisimulations over \lts.}
Bisimulation is a crucial tool for understanding the expressive power of a formal language. In ~\cite{FervariVQW17,FervariVQW21}, bisimulation notions for $\KHlogic$ over \ltss have been introduced. This section discusses similar ideas for $\KHilogic$ over \ultss.

\medskip

First, a useful abbreviation.

\begin{definicion}\label{def:notation}
  Let $\model=\tup{\W,\R,\cset{\S_i}_{i\in\AGT},\V}$ be an \ults over \PROP, \ACT and \AGT. Take a set of plans $\strategy \in 2^{(\ACT^*)}$, sets of states $U, T \subseteq \W$ and an agent $i \in \AGT$.
  \begin{itemize}
    \item Write $U \ultsExecStrat{\strategy} T$ $\siffdefs$ $U \subseteq \stexec(\strategy)$ and $\R_{\strategy}(U) \subseteq T$.

    \item Write $U \ultsExecAgi T$ $\siffdefs$ there is $\strategy \in \S_i$ such that $U \ultsExecStrat{\strategy} T$.
  \end{itemize}
  Additionally, $U \subseteq \W$ is $\KHilogic$-definable (respectively, propositionally definable) in $\model$ if and only if there is an $\KHilogic$-formula (propositional formula) $\varphi$ such that $U = \truthset{\model}{\varphi}$.
\end{definicion}

Two quick observations. First, note how the abbreviation simplifies the semantic clause for \emph{knowing how} formulas: $\model, w \models \khi(\psi, \varphi)$ if and only if $\truthset{\model}{\psi} \ultsExecAgi \truthset{\model}{\varphi}$. Second, under the \ults-based semantics, $\KHilogic$-definability implies propositional definability. Its proof, analogous to the \lts-based semantics case in \cite{FervariVQW17,FervariVQW21}, relies on the fact that $\khi$ acts globally.

\begin{proposicion}\label{pro:kh-definable}
  Let $\model$ be an \ults. For all $U \subseteq \D{\model}$, if $U$ is $\KHilogic$-definable, then it is propositionally definable.
  % \begin{demostracion}
  %   The only non-propositional formulas in $\KHilogic$ use the modality $\khi$. But $\khi$'s semantic interpretation is global, so for every \ults $\model$ and every formula $\khi(\psi,\varphi)$, the set $\truthset{\model}{\khi(\psi,\varphi)}$ is either $\D{\model}$ or $\emptyset$. Thus, any $\KHilogic$-formula can be transformed into a semantically equivalent propositional one by replacing its subformulas of the form $\khi(\psi,\varphi)$ for either $\top$ or $\bot$, according to the case.
  % \end{demostracion}
  % 
  % RF - 23/02/2023  - I commented this proof as we said is analogous to others, and taking into account one reviewer's comment about removing trivial proofs.
\end{proposicion}

We now introduce the notion of bisimulation. Note how, although the collection of binary relations of a model is not explicitly mentioned, it is referred to through the abstract relation ``$\ultsExecAgi$'' (\autoref{def:notation}).

\begin{definicion}[$\KHilogic$-bisimulation]\label{def:bisim-khi}
  Let $\model$ and $\model'$ be two $\ultss$, their domains being $\W$ and $\W'$, respectively. Take $Z \subseteq \W \times \W'$.
  \begin{itemize}\itemsep 0cm
    \item For $u \in \W$ and $U \subseteq \W$, define %$Z(u)$, $Z(U) \subseteq \W'$ as
    \begin{nscenter}
      \begin{tabular}{@{}c@{}}
        $Z(u) := \csetsc{u' \in \W'}{uZu'}$, \qquad $Z(U) := \bigcup_{u \in U} Z(u)$.
      \end{tabular}
    \end{nscenter}

    \item For $u' \in \W'$ and $U' \subseteq \W'$, define % $Z^{-1}(u)$, $Z^{-1}(U) \subseteq \W$ as
    \begin{nscenter}
      \begin{tabular}{@{}c@{}}
        $Z^{-1}(u') := \csetsc{u \in \W}{uZu'}$; \qquad $Z^{-1}(U') := \bigcup_{u' \in U'} Z(u')$.
      \end{tabular}
    \end{nscenter}
  \end{itemize}

  A non-empty $Z \subseteq \W \times \W'$ is called an $\KHilogic$-bisimulation between $\model$ and $\model'$ if and only if $wZw'$ implies all of the following.
  \begin{itemize}
    \item \textbf{Atom}: $\V(w)=\V'(w')$.

    \item \textbf{$\khi$-Zig}: for any \emph{propositionally} definable $U \subseteq \W$, if $U \ultsExecAgi T$ for some $T \subseteq \W$, then there is $T' \subseteq \W'$ satisfying both
    \begin{multicols}{2}
      \begin{cond-bisim}
        \item $Z(U) \ultsExecAgi T'$, 
        \item $T' \subseteq Z(T)$.
      \end{cond-bisim}
    \end{multicols}
    
    \item \textbf{$\khi$-Zag}: for any \emph{propositionally} definable $U' \subseteq \W'$, if $U' \ultsExecAgi T'$ for some $T' \subseteq \W'$, then there is $T \subseteq \W$ satisfying both
    \begin{multicols}{2}
      \begin{cond-bisim}
        \item $Z^{-1}(U') \ultsExecAgi T$,
        \item $T \subseteq Z^{-1}(T')$.
      \end{cond-bisim}
    \end{multicols}

    \item \textbf{$\A$-Zig}: for all $u$ in $\W$ there is a $u'$ in $\W'$ such that $uZu'$.

    \item \textbf{$\A$-Zag}: for all $u'$ in $\W'$ there is a $u$ in $\W$ such that $uZu'$.
  \end{itemize}
  We write $\model,w \bisim \model',w'$ when there is an $\KHlogic$-bisimulation $Z$ between $\model$ and $\model'$ such that $wZw'$.
\end{definicion}  

The two requirements in $\khi$-Zig are equivalent to a single one: $Z(U) \ultsExecAgi Z(T)$. They are split to resemble more closely the definition of a standard bisimulation: if $U$ has an `$i$-successor' $T$, then its `bisimulation image' $U'$ also has an `$i$-successor', namely $T'$ (clause $Z(U) \ultsExecAgi T'$), and these successors are a `bisimilar match' (clause $T' \subseteq Z(T)$). The case of $\khi$-Zag is analogous.

\medskip

In order to formalize the crucial properties of a bisimulation, we define the notion of model equivalence with respect to $\KHilogic$.

\begin{definicion}[$\KHilogic$-equivalence]\label{def:equiv-khi}
  Two pointed $\ultss$ $\model, w$ and $\model', w'$ are \emph{$\KHilogic$-equivalent} (written $\model, w \modequiv \model', w'$) if and only if, for every $\varphi \in \KHilogic$,
  \begin{center}
    $\model, w \models \varphi$ \quad iff \quad $\model', w' \models \varphi$.
  \end{center}
\end{definicion}  

Then, we can state the intended correspondence between $\bisim$ and $\modequiv$. %The proofs are similar to the ones presented in~\cite{FervariVQW17,FervariVQW21}.
  
\begin{teorema}[$\KHilogic$-bisimilarity implies $\KHilogic$-equivalence]\label{th:khbisim-to-khequiv}
  Let $\model, w$ and $\model', w'$ be pointed \ultss. Then,
  \begin{center}
    $\model,w \bisim \model', w' \quad\text{implies}\quad \model,w \modequiv \model',w'$.
  \end{center}
  %If $\model$ and $\model'$ are finite then $\model,w \modequiv \model', w' \text{ implies } \model,w \bisim \model', w'$.
  \begin{demostracion}
     Take $\model=\tup{\W,\R,\cset{\S_i}_{i\in\AGT},\V}$ and $\model'=\langle\W',\R',\cset{\S'_i}_{i\in\AGT},$
    $\V'\rangle$. From the given $\model,w \bisim \model',w'$, there is an $\KHilogic$-bisimulation $Z \subseteq (\W \times \W')$ with
    $wZw'$. The proof of $\KHilogic$-equivalence is by structural induction on $\KHilogic$-formulas. The cases for atomic propositions
    and Boolean operators are standard, and only formulas of the form $\khi(\psi,\varphi)$ are left. Note how, for this case, the inductive hypothesis (IH) states that, for $u \in \W$, $u' \in \W'$ and $\chi$ a subformula of $\khi(\psi,\varphi)$, if $uZu'$ then $u \in \truthset{\model}{\chi}$ iff $u' \in \truthset{\model'}{\chi}$.

    \smallskip

    Suppose $w \in \truthset{\model}{\khi(\psi,\varphi)}$. Then, by semantic interpretation, %there is $\strategy \in \S_i$ s.t. $\truthset{\model}{\psi} \ultsExecStrat{\strategy} \truthset{\model}{\varphi}$, that is,
    $\truthset{\model}{\psi} \ultsExecAgi \truthset{\model}{\varphi}$. It is useful to notice that $Z(\truthset{\model}{\chi}) = \truthset{\model'}{\chi}$ holds for $\chi \in \cset{\psi, \varphi}$.
    \begin{itemize}[leftmargin=2.5em,topsep=2pt]
      \item[$\bm{(\subseteq)}$] If $v' \in Z(\truthset{\model}{\chi})$, then there is $v \in \truthset{\model}{\chi}$ such that $vZv'$. Thus, from IH we have $v' \in \truthset{\model'}{\chi}$.
      \item[$\bm{(\supseteq)}$] If $v' \in \truthset{\model'}{\chi}$ then, by $\A$-Zag, there is $v$ with $vZv'$; thus, from IH we have $v \in \truthset{\model}{\chi}$. Hence, $v' \in Z(\truthset{\model}{\chi})$.
    \end{itemize}

    Now, the proof. We have $\truthset{\model}{\psi} \ultsExecAgi \truthset{\model}{\varphi}$. The set $\truthset{\model}{\psi}$ is obviously $\KHilogic$-definable, and hence propositionally definable too (\autoref{pro:kh-definable}). Then, from the $\khi$-Zig clause, there is $T' \subseteq \W'$ such that \begin{inline-cond-bisim} \item $Z(\truthset{\model}{\psi}) \ultsExecAgi T'$ and \item $T' \subseteq Z(\truthset{\model}{\varphi})$\end{inline-cond-bisim}. Therefore, $Z(\truthset{\model}{\psi}) \ultsExecAgi Z(\truthset{\model}{\varphi})$ and hence, by the result above, $\truthset{\model'}{\psi} \ultsExecAgi \truthset{\model'}{\varphi}$. Thus, $w' \in \truthset{\model'}{\khi(\psi,\varphi)}$.

    {\smallskip}

    The direction from $w' \in \truthset{\model'}{\khi(\psi,\varphi)}$ to $w \in \truthset{\model}{\khi(\psi,\varphi)}$ follows a similar   argument, using $\A$-Zig and $\khi$-Zag instead.
  \end{demostracion}
\end{teorema}

It is easy to see that the converse of~\autoref{th:khbisim-to-khequiv} does not hold over arbitrary models: in fact, the counterexample provided in~\cite[Section 2]{FervariVQW21} serves also here to make our point. To satisfy the converse, we usually need to restrict ourselves to some particular classes of models, that are in general known as Hennessy-Milner classes. 
In many modal logics, one typically works only with image-finite models: those in which, at every state, every basic relation has only finitely many successors. For languages in which the global universal modality $\A$ is definable, as $\KHilogic$, this requirement needs to be strengthened, as every state can reach (via the relation underlying $\A$) every other state. For instance, we can take the class of models with a finite domain. Precisely, here we show that the class of finite models (taken as those with a finite domain) forms a Hennessy-Milner class. Note that we do not impose any restriction over the uncertainty relation of the models (or the sets $\S_i$).

\begin{teorema}[$\KHilogic$-equivalence implies $\KHilogic$-bisimilarity]\label{th:khequiv-to-khbisim}
  Let $\model, w$ and $\model', w'$ be \emph{finite} pointed \ultss. Then,
  \begin{center}
    $\model,w \modequiv \model', w' \quad\text{implies}\quad \model,w \bisim \model', w'$.
  \end{center}
\begin{demostracion}
    Take $\model=\tup{\W,\R,\cset{\S_i}_{i\in\AGT},\V}$ and $\model'=\tup{\W',\R',\cset{\S'_i}_{i\in\AGT},\V'}$. The strategy is to show that the relation $\modequiv$ is already a $\KHilogic$--bisimulation. Thus, define
    \begin{nscenter}
      $Z := \csetc{(v,v')}{(\W \times \W')}{\model, v \modequiv \model', v'}$
    \end{nscenter}
    so $wZw'$ implies $w$ and $w'$ satisfy exactly the same $\KHilogic$-formulas. In order to show that $Z$ satisfies the requirements, take any $(w, w') \in Z$.
    \begin{itemize}
      \item \textbf{Atom}. States $w$ and $w'$ agree in all $\KHilogic$-formulas, and thus in all atoms.
      
      \item \textbf{$\A$-Zig}. Take $v \in \W$ and suppose, for the sake of a contradiction, that there is no $v' \in \W'$ such that $vZv'$. Then, from $Z$'s definition, for each $v_i'\in \W' = \cset{v'_1,\ldots,v'_n}$ (recall: $\model'$ is finite) there is an $\KHlogic$-formula $\theta_i$ such that $\model,v \models \theta_i$ but $\model',v'_i \not\models\theta_i$. Now take $\theta := \theta_1 \land \cdots \land \theta_n$. Clearly, $\model,v \models \theta$; however, $\model', v_i' \not\models \theta$ for each $v_i'\in \W'$, as each one of them makes `its' conjunct $\theta_i$ false. Then, $\model,w \models \E \theta$ but $\model',w' \not\models \E \theta$, contradicting the assumption $wZw'$.

      \item \textbf{$\A$-Zag}. Analogous to the $\A$-Zig case.
      
      \item \textbf{$\khi$-Zig}. Take any propositionally definable set $\truthset{\model}{\psi} \subseteq \W$ (thus, $\psi$ is propositional), and suppose $\truthset{\model}{\psi} \ultsExecAgi T$ for some $T \subseteq \W$. We need to find a $T' \subseteq \W'$ satisfying both
      \begin{multicols}{2}
        \begin{cond-bisim}
          \item $Z(\truthset{\model}{\psi}) \ultsExecAgi T'$,
          \item $T' \subseteq Z(T)$.
        \end{cond-bisim}
      \end{multicols}
      Note that $Z(\truthset{\model}{\psi}) = \truthset{\model'}{\psi}$. For $\boldsymbol{(\supseteq)}$, suppose $u' \in \truthset{\model'}{\psi}$. From $\A$-Zag (proved above), there is $u \in \W$ such that $uZu'$; then, from $Z$'s definition, $u \in \truthset{\model}{\psi}$ so $u' \in Z(\truthset{\model}{\psi})$. For $\boldsymbol{(\subseteq)}$, suppose $u' \in Z(\truthset{\model}{\psi})$. Then, there is $u \in \truthset{\model}{\psi}$ such that $uZu'$, and therefore, from $Z$'s definition, $u' \in \truthset{\model'}{\psi}$. Thus, we actually require a $T' \subseteq \W'$ satisfying both
      \begin{multicols}{2}
        \begin{cond-bisim}
          \item $\truthset{\model'}{\psi} \ultsExecAgi T'$,
          \item $T' \subseteq Z(T)$.
        \end{cond-bisim}
      \end{multicols}
      Now, consider two alternatives.
      \begin{enumerate}
        \item Assume $\truthset{\model}{\psi} = \emptyset$. Then, $\truthset{\model'}{\psi} = Z(\truthset{\model}{\psi}) = \emptyset$ and hence $T' = \emptyset$ does the job, as the following hold
        \begin{multicols}{2}
          \begin{cond-bisim}
            \item $\emptyset \ultsExecAgi \emptyset$ (as $\S_i \neq \emptyset$),
            \item $\emptyset \subseteq Z(T)$.
          \end{cond-bisim}
        \end{multicols}

        \item Assume $\truthset{\model}{\psi} \neq \emptyset$. This gives us $T \neq \emptyset$ (from $\truthset{\model}{\psi} \ultsExecAgi T$), which will be useful later.
        %\fer{In the previous version we also had $\truthset{\model'}{\psi} \neq \emptyset$ (which follows from $\A$-Zag and $Z(\truthset{\model}{\psi}) = \truthset{\model'}{\psi}$), but I do not see where it is used.} 
        To show that there is a $T' \subseteq \W'$ satisfying both \ITMBi and \ITMBii, we proceed by contradiction, so suppose there is no $T'$ satisfying both requirements: every $T' \subseteq \W'$ satisfying \ITMBi fails at \ITMBii. In other words, every $T' \subseteq \W'$ satisfying $\truthset{\model'}{\psi} \ultsExecAgi T'$ has a state $v'_{T'} \in T'$ that is not the $Z$-image of some state $v \in T$ (i.e., $vZv'_{T'}$ fails for every $v \in T$). From $Z$'s definition, the latter means that every state in $T$ can be distinguished from this $v'_{T'}$ by an $\KHilogic$-formula. Thus, given any $T' \subseteq \W'$ with $\truthset{\model'}{\psi} \ultsExecAgi T'$, one can find a state $v'_{T'} \in T'$ such that, for each $v \in T$, there is a formula $\theta^v_{v'_{T'}}$ with $\model, v \models \theta^v_{v'_{T'}}$ but $\model', v'_{T'} \not\models\theta^v_{v'_{T'}}$. Then, for each such $v'_{T'}$ in each such $T'$ define
        \begin{nscenter}
          \renewcommand{\arraystretch}{1.8}
          \begin{tabular}{c@{\qquad{and then}\qquad}c}
            $\displaystyle \theta_{T'} := \bigvee_{v \in T} \theta^v_{v'_{T'}}$
            & 
            $\displaystyle \theta := \bigwedge_{\csetsc{T' \subseteq \W'}{\truthset{\model'}{\psi} \ultsExecAgi T'}} \theta_{T'}$,
          \end{tabular}
          \renewcommand{\arraystretch}{1}
        \end{nscenter}
        %so $\theta_{T'}$ is the disjunction of formulas distinguishing $\theta_{v'_{T'}}$ from every state in $T$, and $\theta$ is the conjunction of the latter.
        Observe the following. First, $\theta_{T'}$ is indeed a formula, as $\W$ is finite and thus so is $T$. Equally important, $T \neq \emptyset$, and thus $\theta_{T'}$ does not collapse to $\bot$. Second, $\theta$ is also a formula, as $\W'$ is finite and thus so is $\csetsc{T' \subseteq \W'}{\truthset{\model'}{\psi} \ultsExecAgi T'}$. However, the latter set might be empty. This is what creates the following two cases.
        \begin{itemize}
          \item Suppose $\csetsc{T' \subseteq \W'}{\truthset{\model'}{\psi} \ultsExecAgi T'} = \emptyset$. Then, consider the formula $\khi(\psi,\top)$. Since $\truthset{\model}{\psi} \ultsExecAgi T$ and $T \subseteq \W = \truthset{\model}{\top}$, it follows that $\model,w \models \khi(\psi, \top)$. However, $\model',w' \not\models \khi(\psi, \top)$ as, according to this case, there is no $T' \subseteq \W'$ with $\truthset{\model'}{\psi} \ultsExecAgi T'$. This contradicts the $\KHilogic$-equivalence of $w$ and $w'$.

          \item Suppose $\csetsc{T' \subseteq \W'}{\truthset{\model'}{\psi} \ultsExecAgi T'} \neq \emptyset$. Then, $\theta$ does not collapse to $\top$. Now, note how every $v \in T$ satisfies its `own' disjunct $\theta^v_{v'_{T'}}$ in each conjunct $\theta_{T'}$, and thus it satisfies $\theta$. Thus, $T \subseteq \truthset{\model}{\theta}$ and hence, from $\truthset{\model}{\psi} \ultsExecAgi T$ and the fact that $\khi$-formulas are global, it follows that $\model, w \models \khi(\psi, \theta)$. However, for each $T'$ in $\cset{T' \subseteq \W' \mid \truthset{\model'}{\psi} \ultsExecAgi T'}$, the state $v'_{T'}$ that cannot be matched with any state $v \in T$ makes all disjuncts in $\theta_{T'}$ false, thus falsifying $\theta_{T'}$ and therefore falsifying $\theta$ too. In other words, every $T' \subseteq \W'$ with $\truthset{\model'}{\psi} \ultsExecAgi T'$ contains a state $t'_{T'}$ with $\model', t'_{T'} \not\models \theta$, that is, $\truthset{\model'}{\psi} \ultsExecAgi T'$ implies $T' \not\subseteq \truthset{\model}{\theta}$. Hence, using again the fact that $\khi$-formulas are global, $\model',w' \not\models \khi(\psi, \theta)$, contradicting the $\KHilogic$-equivalence of $w$ and $w'$.
        \end{itemize}
      \end{enumerate}

      \item \textbf{$\khi$-Zag}. Analogous to the $\khi$-Zig case.
    \end{itemize} 
  \end{demostracion}
\end{teorema}

%------------------------------------------------------------------------------------------------
\section{Axiomatization}\label{sec:axiom}
%\msparagraph{Axiomatization.} 

We now present a sound and complete axiom system for $\KHilogic$ under the \ults-based semantics. Recall that $\A \varphi := \bigvee_{i \in \AGT} \khi(\neg\varphi,\bot)$ and $\E\varphi:=\neg\A\neg\varphi$. With this, it turns out that formulas and rules in $\axset$ (the first block of \autoref{tab:khaxiom}) are still sound under \ults (provided $\kh$ is replaced by $\khi$). They will constitute the first part of an axiom system for $\KHilogic$ over \ults (first block in~\autoref{tab:khiaxiom}). Still, this is not enough for a complete axiom system. The axioms on the second block of \autoref{tab:khiaxiom}, \axm{KhE} and \axm{KhA}, are the missing pieces\footnote{\axm{KhE} and \axm{KhA} are also valid under \lts semantics (\autoref{pro:subsystem}). In fact, the next section will show that $\KHiaxiom$ (for \ults) is strictly weaker than $\KHaxiom$ (for \lts).}.

\begin{table}[t]
  $$
  \begin{array}{l@{\quad \quad  }l@{\quad}l}
     \toprule
     \mbox{\underline{Block $\axset$:}}        & \axm{TAUT}   & \vdash \varphi \mbox{ for $\varphi$ a propositional tautology} \\
                             & \axm{DISTA}  & \vdash \A(\varphi\ra\psi) \ra (\A\varphi \ra \A\psi) \\
                              
                              & \axm{TA}     & \vdash \A\varphi \ra \varphi \\
                              & \axm{4KhA}   & \vdash \khi(\psi,\varphi) \ra \A\khi(\psi,\varphi) \\
                              &  \axm{5KhA}   & \vdash \neg\khi(\psi,\varphi) \ra \A\neg\khi(\psi,\varphi) \\
                         &  \axm{MP}     & \mbox{From $\vdash \varphi$ and $\vdash \varphi \limp \psi$ infer $\vdash \psi$ }\\
                         &  \axm{NECA}   & \mbox{From $\vdash \varphi$ infer $\vdash \A\varphi$} \\
     \midrule
     \mbox{\underline{Block $\axset_{\ults}$:}}  & \axm{KhE} & \vdash \left(\E\psi \land \khi(\psi,\varphi)\right) \limp \E\varphi \\
                                          & \axm{KhA} & \vdash \left(\A(\chi \limp \psi) \land \khi(\psi,\varphi) \land \A(\varphi \limp \theta)\right) \limp \khi(\chi, \theta) \\
     \bottomrule
  \end{array}
  $$
  \caption{Axioms $\axset_{\ults}$, for $\KHilogic$ w.r.t. $\ultss$  .}\label{tab:khiaxiom}
\end{table}

\smallskip

We should start by discussing the newly introduced axioms. On the one hand, axiom \axm{KhA} can be subjected to some of the criticisms that apply to \axm{EMP} and \axm{COMPKh} but, in our opinion, to a lesser extent. It implies certain level of idealization, as it entails that the abilities of the agent are, in a sense, closed under global entailment. On the other hand, axiom \axm{KhE} states a simple reasonable requirement: if $\khi(\psi,\varphi)$ is not trivial (given that $\E\psi$ holds), then $\E\varphi$ should be assured.

\medskip

%!TEX root = main.tex

Define the system $\KHiaxiom$ := $\axset$ (first block of \autoref{tab:khiaxiom}) + $\axset_{\ults}$ (second block of \autoref{tab:khiaxiom}). Now we will show that $\KHiaxiom$ is sound and strongly complete for $\KHilogic$ over \ultss. Proving soundness is rather straightforward, so we will focus on strong completeness. Following~\cite[Proposition 4.12]{mlbook}, the strategy is to build, for any \KHiaxiom-consistent set of formulas, an \ults satisfying them. In particular, the notions of theoremhood, (local) consequence, inconsistency, and maximally consistent sets are defined as usual~\cite{mlbook}. We will rely on ideas from~\cite{Wang15lori,Wang2016}; the following theorems will be useful.

\begin{proposicion}
  Formulas $\A\lnot \psi \limp \khi(\psi, \varphi)$ (called \axm{SCOND}) and $\khi(\bot, \varphi)$ (called \axm{COND}) are \KHiaxiom-derivable. That is, \begin{inlineenum} \item $\vdash \A\lnot \psi \limp \khi(\psi, \varphi)$ and \item $\vdash \khi(\bot, \varphi)$\end{inlineenum}.
\begin{demostracion}
  \begin{inlineenum} \item Take $\vdash \A(\psi \limp \psi) \limp \left( \A(\bot \limp \varphi) \limp (\khi(\psi,\bot) \limp \khi(\psi, \varphi)) \right)$, an instance of \axm{KhA}. Using \axm{TAUT} and \axm{NECA} we get $\vdash \A(\psi \limp \psi)$; analogously, we get $\vdash \A(\bot \limp \varphi)$. Then, using \axm{MP} twice yields $\vdash \khi(\psi,\bot) \limp \khi(\psi, \varphi)$, which by $\A$'s definition is $\vdash \A\lnot\psi \limp \khi(\psi, \varphi)$. \item Take $\vdash \A\lnot\bot \limp \khi(\bot, \varphi)$, an instance of the previous item. Using \axm{TAUT} and \axm{NECA} we get $\vdash \A\lnot\bot$ so, by \axm{MP}, $\vdash \khi(\bot, \varphi)$.\end{inlineenum}
\end{demostracion}
\end{proposicion}

Here it is, then, the definition of the required \ults.

\begin{definicion}\label{def:cm-ults-lkhi}
  Let $\smcs$ be the set of all maximally \KHiaxiom-consistent sets (MCS) of formulas in \KHilogic. For any $\Delta \in \smcs$, define 
  \begin{nscenter}
  $
    \begin{array}{r@{\;:=\;}l@{\qquad\qquad}r@{\;:=\;}l}
      \restkhi{\Delta}  & \csetc{\xi}{\Delta}{\xi \;\text{is of the form}\; \khi(\psi,\varphi)}, &
      \restkh{\Delta}   & \bigcup_{i \in \AGT} \restkhi{\Delta}. \\
      \restnkhi{\Delta} & \csetc{\xi}{\Delta}{\xi \;\text{is of the form}\; \lnot \khi(\psi,\varphi)}, &
      \restnkh{\Delta}  & \bigcup_{i \in \AGT} \restnkhi{\Delta}. \\
    \end{array}
    $
  \end{nscenter}
  Let $\Gamma$ be a set in $\smcs$; we will define a structure satisfying its formulas. Define a set of basic actions $\ACT^\Gamma_i := \csetsc{\tup{\psi,\varphi}}{\khi(\psi,\varphi) \in \Gamma}$ associated to each agent $i \in \AGT$, and then their union $\ACT^\Gamma := \bigcup_{i \in \AGT} \ACT^\Gamma_i$. Notice that $\khi(\bot, \varphi) \in \Gamma$ for every $i \in \AGT$ and every $\varphi \in \KHilogic$ (by \axm{COND}); since $\AGT$ is finite and non-empty, this implies that $\ACT^\Gamma$ is denumerable, and thus it is an adequate set of actions for building a model. It is worth noticing that $\ACT^\Gamma$ fixes a new signature. However, since the operators of the language cannot talk explicitly about the names of the actions, we can define a mapping from $\ACT^\Gamma$ to any particular $\ACT$, to preserve the original signature, provided that the cardinalities match.

  \smallskip

  \noindent Then, the structure $\modults^\Gamma=\tup{\W^\Gamma, \R^\Gamma, \cset{\S^\Gamma_i}_{i \in \AGT}, \V^\Gamma}$ over $\ACT^\Gamma$, $\AGT$ and $\PROP$ is defined as follows.
  \begin{itemize}
    \item $\W^\Gamma := \csetc{\Delta}{\smcs}{\restkh{\Delta} = \restkh{\Gamma}}$.

    \item $\R^{\Gamma}_{\tup{\psi,\varphi}} := \bigcup_{i \in \AGT} \R^{\Gamma} _{\tup{\psi,\varphi}^i}$, with
    \begin{nscenter}
      $\R^{\Gamma}_{\tup{\psi,\varphi}^i}
      :=
      \csetc{(\Delta_1, \Delta_2)}{\W^\Gamma \times \W^\Gamma}{\khi(\psi,\varphi) \in \Gamma, \psi \in \Delta_1, \varphi \in \Delta_2}$.
    \end{nscenter}

    \item $\S^\Gamma_i := \left\{ \cset{\tup{\psi,\varphi}} \mid \tup{\psi,\varphi} \in \ACT^\Gamma_i \right\}$.

    \item $\V^\Gamma(\Delta) := \csetc{p}{\PROP}{p \in \Delta}$.
  \end{itemize}
\end{definicion}

Since $\Gamma \in \smcs$, the structure $\modults^\Gamma$ is of the required type, as the following proposition states.

\begin{proposicion}\label{pro:cm-ults-lkhi}
  The structure $\modults^\Gamma = \tup{\W^\Gamma, \R^\Gamma, \cset{\S^\Gamma_i}_{i \in \AGT}, \V^\Gamma}$ is an $\ults$.
  \begin{demostracion}
    It is enough to show that each $\S^\Gamma_i$ defines a partition over a non-empty subset of $2^{(\ACT^*)}$. First, \axm{COND} implies $\khi(\bot, \bot) \in \Gamma$, so $\tup{\bot, \bot} \in \ACT^\Gamma_i$ and hence $\cset{\tup{\bot, \bot}} \in \S^\Gamma_i$; thus, $\bigcup_{\strategy \in \S_i} \strategy \neq \emptyset$. Then, $\S_i$ indeed defines a partition over $\bigcup_{\strategy \in \S_i} \strategy$: its elements are mutually disjoint (they are singletons with different elements), collective exhaustiveness is immediate and, finally, $\emptyset \notin \S^\Gamma_i$.
  \end{demostracion}
\end{proposicion}

Let $\Gamma \in \smcs$; the following properties of $\modults^\Gamma$ will be useful (proofs are similar to the ones in~\cite{Wang2016}).

\begin{proposicion}\label{pro:cm-ults-lkhi-allsameKH}
  For any $\Delta_1, \Delta_2 \in \W^\Gamma$ we have $\restkh{\Delta_1} = \restkh{\Delta_2}$.
  \begin{demostracion}
    Straightforward from the definition of $\W^\Gamma$.
  \end{demostracion}
\end{proposicion}

\begin{proposicion}\label{pro:cm-ults-lkhi-oneall}
  Take $\Delta \in \W^\Gamma$. If $\Delta$ has a $\R^\Gamma_{\tup{\psi,\varphi}}$-successor, then every $\Delta' \in \W^\Gamma$ with $\varphi \in \Delta'$ can be $\R^\Gamma_{\tup{\psi,\varphi}}$-reached from $\Delta$.
  \begin{demostracion}
    If $\Delta$ has a $\R^\Gamma_{\tup{\psi,\varphi}}$-successor, then it has a $\R^\Gamma_{\tup{\psi,\varphi}^i}$-successor for some $i \in \AGT$; thus, $\psi \in \Delta$ and $\khi(\psi,\varphi) \in \Gamma$. Hence, every $\Delta' \in \W^\Gamma$ with $\varphi \in \Delta'$ is such that $(\Delta, \Delta') \in \R^\Gamma_{\tup{\psi,\varphi}^i}$, and thus such that $(\Delta, \Delta') \in \R^\Gamma_{\tup{\psi,\varphi}}$.
  \end{demostracion}
\end{proposicion}

\begin{proposicion}\label{pro:cm-ults-lkhi-allall}
  Let $\varphi$ be an \KHilogic-formula. If $\varphi \in \Delta$ for every $\Delta \in \W^\Gamma$, then $\A\varphi \in \Delta$ for every $\Delta \in \W^\Gamma$.
  \begin{demostracion}
    First, some facts for any $\Delta$ in $\W^\Gamma \subseteq \smcs$. By definition, $\restkh{\Delta} \cup \restnkh{\Delta}$ is a subset of $\Delta$, and therefore it is consistent. Moreover: any maximally consistent extension of $\restkh{\Delta} \cup \restnkh{\Delta}$, say $\Delta'$, should satisfy $\restkh{\Delta} = \restkh{\Delta'}$. For \proofsb, note that $\khi(\psi, \varphi) \in \restkh{\Delta}$ implies $\khi(\psi, \varphi) \in (\restkh{\Delta} \cup \restnkh{\Delta})$, and thus $\khi(\psi, \varphi) \in \Delta'$, i.e., $\khi(\psi, \varphi) \in \restkh{\Delta'}$. For \proofsp, use the contrapositive. If $\khi(\psi, \varphi) \not\in \restkh{\Delta}$ then $\khi(\psi, \varphi) \not\in \Delta$, so $\lnot\khi(\psi, \varphi) \in \Delta$ (as $\Delta$ is an MCS). Thus, $\lnot \khi(\psi, \varphi) \in (\restkh{\Delta} \cup \restnkh{\Delta})$ and hence $\lnot \khi(\psi, \varphi) \in \Delta'$; therefore, $\khi(\psi, \varphi) \not\in \Delta'$ (as $\Delta$ is consistent) and thus $\khi(\psi, \varphi) \not\in \restkh{\Delta'}$.

    \smallskip

    For the proof of the proposition, suppose $\varphi \in \Delta$ for every $\Delta \in \W^\Gamma$. Take any $\Delta \in \W^\Gamma$, and note how $\restkh{\Delta} = \restkh{\Gamma}$. Then, the set $\restkh{\Delta} \cup \restnkh{\Delta} \cup \cset{\lnot \varphi}$ is inconsistent. Otherwise it could be extended into an MCS $\Delta' \in \smcs$. By the result in the previous paragraph, this would imply $\restkh{\Delta'} = \restkh{\Delta}$, so $\restkh{\Delta'} = \restkh{\Gamma}$ and therefore $\Delta' \in \W^\Gamma$. But then, by the assumption, $\varphi \in \Delta'$, and by construction, $\lnot \varphi \in \Delta'$. This would make $\Delta'$ inconsistent, a contradiction.

    Thus, given that $\restkh{\Delta} \cup \restnkh{\Delta} \cup \cset{\lnot \varphi}$ is inconsistent, there should be sets $\cset{\kh_{b_1}(\psi_1,\varphi_1), \ldots, \kh_{b_n}(\psi_n,\varphi_n)} \subseteq \restkh{\Delta}$ and $\cset{\lnot \kh_{b'_1}(\psi'_1,\varphi'_1), \ldots, \lnot \kh_{b'_m}(\psi'_m,\varphi'_m)}  \subseteq \restnkh{\Delta}$ such that
    \begin{csmalltabular}{c}
      $\vdash
          \displaystyle
          \left( \bigwedge_{k=1}^{n} \kh_{b_k}(\psi_k, \varphi_k) \land \bigwedge_{k=1}^{m} \lnot \kh_{b'_k}(\psi'_k, \varphi'_k) \right)
          \limp
          \varphi$.
    \end{csmalltabular}
    Hence, by \axm{NECA},
    \begin{csmalltabular}{@{}c@{}}
      $\vdash
          \displaystyle
          \A
            \left(
              \left( \bigwedge_{k=1}^{n} \kh_{b_k}(\psi_k, \varphi_k) \land \bigwedge_{k=1}^{m} \lnot \kh_{b'_k}(\psi'_k, \varphi'_k) \right)
              \limp
              \varphi
            \right)$
    \end{csmalltabular}
    and then, by \axm{DISTA} and \axm{MP}, 
   \begin{csmalltabular}{@{}c@{}}
        $\vdash
          \displaystyle
          \A\left( \bigwedge_{k=1}^{n} \kh_{b_k}(\psi_k, \varphi_k) \land \bigwedge_{k=1}^{m} \lnot \kh_{b'_k}(\psi'_k, \varphi'_k) \right)
          \limp
          \A\varphi$.
    \end{csmalltabular}

    Now, $\kh_{b_k}(\psi_k, \varphi_k) \in \restkh{\Delta}$ implies (\axm{4KhA} and \axm{MP}) that $\A\kh_{b_k}(\psi_k, \varphi_k) \in \Delta$ (for each $k \in \intint{1}{n}$). Similarly, $\lnot \kh_{b'_k}(\psi'_k, \varphi'_k) \in \restkh{\Delta}$ implies (\axm{5KhA} and \axm{MP}) that $\A \lnot \kh_{b'_k}(\psi'_k, \varphi'_k) \in \Delta$ (for each $k \in \intint{1}{m}$). Thus,
    \begin{csmalltabular}{c}
      $\displaystyle
      \bigwedge_{k=1}^{n} \A\kh_{b_k}(\psi_k, \varphi_k) \in \Delta
      \qquad\text{and}\qquad
      \bigwedge_{k=1}^{m} \A\lnot\kh_{b'_k}(\psi'_k, \varphi'_k) \in \Delta$
    \end{csmalltabular}
    and hence
    \begin{csmalltabular}{@{}c@{}}
      $\displaystyle
      \bigwedge_{k=1}^{n} \A\kh_{b_k}(\psi_k, \varphi_k)
      \land
      \bigwedge_{k=1}^{m} \A\lnot\kh_{b'_k}(\psi'_k, \varphi'_k)
      \in \Delta, 
      \,
      \text{so}
      \,
      \A
        \left(
          \bigwedge_{k=1}^{n} \kh_{b_k}(\psi_k, \varphi_k)
          \land
          \bigwedge_{k=1}^{m} \lnot\kh_{b'_k}(\psi'_k, \varphi'_k)
        \right)
      \in \Delta$
    \end{csmalltabular}
    and therefore $\A\varphi \in \Delta$.
  \end{demostracion}
\end{proposicion}

\begin{proposicion}\label{pro:cm-ults-lkhi-succpre}
  Take $\psi, \psi', \varphi'$ in \KHilogic. Suppose that every $\Delta \in \W^\Gamma$ with $\psi \in \Delta$ has a $\R^{\Gamma}_{\tup{\psi',\varphi'}}$-successor. Then, $\A(\psi \limp \psi') \in \Delta$ for all $\Delta \in \W^\Gamma$.
  \begin{demostracion}
    Take any $\Delta \in \W^\Gamma$. On the one hand, if $\psi \in \Delta$ then, by the supposition, $(\Delta, \Delta') \in \R^{\Gamma}_{\tup{\psi',\varphi'}}$ for some $\Delta'$. Hence, from $\R^{\Gamma}_{\tup{\psi',\varphi'}}$'s definition, $\psi' \in \Delta$ and thus (maximal consistency) $\psi \limp \psi' \in \Delta$. On the other hand, if $\psi \not\in \Delta$ then $\lnot \psi \in \Delta$ (again, maximal consistency) and thus $\psi \limp \psi' \in \Delta$. Thus, $\psi \limp \psi' \in \Delta$ for every $\Delta \in \W^\Gamma$; then, by \autoref{pro:cm-ults-lkhi-allall}, $\A(\psi \limp \psi') \in \Delta$ for every $\Delta \in \W^\Gamma$.
  \end{demostracion}
\end{proposicion}

With these properties at hand, we can prove the truth lemma for $\modults^\Gamma$.

\begin{lema}[Truth lemma for $\modults^\Gamma$]\label{tlm:cm-ults-lkhi}
  Given $\Gamma \!\in\! \smcs$, take $\modults^\Gamma = \tup{\W^\Gamma, \R^\Gamma, \cset{\S^\Gamma_i}_{i \in \AGT}, \V^\Gamma}$. Then, for every $\Theta \in \W^\Gamma$ and every $\varphi \in \KHilogic$,  % $\modults^\Gamma, \Theta \models \varphi$ if and only if $\varphi \in \Theta$. 
  \begin{ctabular}{l@{\qquad\text{if and only if}\qquad}l}
     $\modults^\Gamma, \Theta \models \varphi$ & $\varphi \in \Theta$.
  \end{ctabular}
  \begin{demostracion}
    The proof is by induction on $\varphi$. The atom and Boolean cases as usual, so we focus on the \emph{knowing how} case.
    \begin{itemizenl}
      \item \textbf{Case $\bm{\khi(\psi,\varphi)}$.} {\prooflr} Suppose $\modults^\Gamma, \Theta \models \khi(\psi,\varphi)$, and consider two cases.
      \begin{itemize}
        \item $\bm{\truthset{\modults^\Gamma}{\psi} = \emptyset}$. Then, each $\Delta \in \W^\Gamma$ is such that $\Delta \not\in \truthset{\modults^\Gamma}{\psi}$, which implies $\psi \not\in \Delta$ (by IH) and thus $\lnot\psi \in \Delta$ (by maximal consistency). Hence, by \autoref{pro:cm-ults-lkhi-allall}, $\A\lnot\psi \in \Delta$ for every $\Delta \in \W^\Gamma$. In particular, $\A\lnot\psi \in \Theta$ and thus, by \axm{SCOND} and \axm{MP}, $\khi(\psi, \varphi) \in \Theta$.

        \item $\bm{\truthset{\modults^\Gamma}{\psi} \neq \emptyset}$. From $\modults^\Gamma, \Theta \models \khi(\psi,\varphi)$, there is $\cset{\tup{\psi',\varphi'}} \in \S^\Gamma_i$ such that
        \begin{nscenter}
          \noindent\begin{inline-cond-kh} \item $\truthset{\modults^\Gamma}{\psi} \subseteq \stexec(\cset{\tup{\psi',\varphi'}})$ \;and\; \item $\R^\Gamma_{\cset{\tup{\psi',\varphi'}}}(\truthset{\modults^\Gamma}{\psi}) \subseteq \truthset{\modults^\Gamma}{\varphi}$\end{inline-cond-kh}.
        \end{nscenter}
         In other words, there is $\tup{\psi',\varphi'} \in \ACT^\Gamma_a$ such that
        \begin{cond-kh}
          \item\label{tlm:cm-esmiv-stexec-lkhi-itm:i} for all $\Delta \in \W^\Gamma$, if $\Delta \in \truthset{\modults^\Gamma}{\psi}$ then $\Delta \in \stexec(\cset{\tup{\psi',\varphi'}})$, so $\Delta \in \stexec(\tup{\psi',\varphi'})$ and therefore $\Delta$ has a $\R^\Gamma_{\tup{\psi',\varphi'}}$-successor.

          \item\label{tlm:cm-esmiv-stexec-lkhi-itm:ii} for all $\Delta' \in \W^\Gamma$, if $\Delta' \in \R^\Gamma_{\cset{\tup{\psi',\varphi'}}}(\truthset{\modults^\Gamma}{\psi})$ then $\Delta' \in \truthset{\modults^\Gamma}{\varphi}$.
        \end{cond-kh}
        This case requires three pieces.
        \begin{enumerate}
          \item Take any $\Delta \in \W^\Gamma$ with $\psi \in \Delta$. Then, by IH, $\Delta \in \truthset{\modults^\Gamma}{\psi}$ and thus, by \autoref{tlm:cm-esmiv-stexec-lkhi-itm:i}, $\Delta$ has a $\R^\Gamma_{\tup{\psi',\varphi'}}$-successor. Thus, every $\Delta \in \W^\Gamma$ with $\psi \in \Delta$ has such successor; then (\autoref{pro:cm-ults-lkhi-succpre}), it follows that $\A(\psi \limp \psi') \in \Delta$ for every $\Delta \in \W^\Gamma$. In particular, $\A(\psi \limp \psi') \in \Theta$.

          \item From $\tup{\psi',\varphi'} \in \ACT^\Gamma_i$ it follows that $\khi(\psi',\varphi') \in \Gamma$. But $\Theta \in \W^\Gamma$, so $\restkh{\Theta} = \restkh{\Gamma}$ (by definition of $\W^\Gamma$). Hence, $\khi(\psi',\varphi') \in \Theta$.

          \item Since $\truthset{\modults^\Gamma}{\psi} \neq \emptyset$, there is $\Delta \in \truthset{\modults^\Gamma}{\psi}$. By \autoref{tlm:cm-esmiv-stexec-lkhi-itm:i}, $\Delta$ should have at least one $\R^\Gamma_{\tup{\psi',\varphi'}}$-successor. Then, by \autoref{pro:cm-ults-lkhi-oneall}, every $\Delta' \in \W^\Gamma$ satisfying $\varphi' \in \Delta'$ can be $\R^\Gamma_{\tup{\psi',\varphi'}}$-reached from $\Delta$; in other words, every $\Delta' \in \W^\Gamma$ satisfying $\varphi' \in \Delta'$ is in $\R^\Gamma_{\tup{\psi',\varphi'}}(\Delta)$. But $\Delta \in \truthset{\modults^\Gamma}{\psi}$, so every $\Delta' \in \W^\Gamma$ satisfying $\varphi' \in \Delta'$ is in $\R^\Gamma_{\tup{\psi',\varphi'}}(\truthset{\modults^\Gamma}{\psi})$. Then, by \autoref{tlm:cm-esmiv-stexec-lkhi-itm:ii}, every $\Delta' \in \W^\Gamma$ satisfying $\varphi' \in \Delta'$ is in $\truthset{\modults^\Gamma}{\varphi}$. By IH on the latter part, every $\Delta' \in \W^\Gamma$ satisfying $\varphi' \in \Delta'$ is such that $\varphi \in \Delta'$. Thus, $\varphi' \limp \varphi \in \Delta'$ for every $\Delta' \in \W^\Gamma$, and hence (\autoref{pro:cm-ults-lkhi-allall}) $\A(\varphi' \limp \varphi) \in \Delta'$ for every $\Delta' \in \W^\Gamma$. In particular, $\A(\varphi' \limp \varphi) \in \Theta$.
        \end{enumerate}
        Thus, $\cset{\A(\psi \limp \psi'), \khi(\psi',\varphi'), \A(\varphi' \limp \varphi)} \subset \Theta$. Therefore, by \axm{KhA} and \axm{MP}, $\khi(\psi, \varphi) \in \Theta$.
      \end{itemize}

      {\proofrl} Suppose $\khi(\psi, \varphi) \in \Theta$. Thus (\autoref{pro:cm-ults-lkhi-allsameKH}), $\khi(\psi, \varphi) \in \Gamma$, so $\tup{\psi, \varphi} \in \ACT^\Gamma_i$ and therefore $\cset{\tup{\psi, \varphi}} \in \S^\Gamma_i$. The rest of the proof is split into two cases.
      \begin{itemize}
        \item Suppose there is no $\Delta_\psi \in \W^\Gamma$ with $\psi \in \Delta$. Then, by IH, there is no $\Delta_\psi \in \W^\Gamma$ with $\Delta_\psi \in \truthset{\modults^\Gamma}{\psi}$, that is, $\truthset{\modults^\Gamma}{\lnot\psi} = \D{\W^\Gamma}$. Since $\modults^\Gamma$ is an $\ults$ (\autoref{pro:cm-ults-lkhi}), the latter yields $(\modults^\Gamma, \Delta) \models \khi(\psi, \chi)$ for any $i \in \AGT$, $\chi \in \KHilogic$ and $\Delta \in \W^\Gamma$ (cf. \autoref{pro:ults:universal}); hence, $(\modults^\Gamma, \Theta) \models \khi(\psi, \varphi)$.

        \item Suppose there is $\Delta_\psi \in \W^\Gamma$ with $\psi \in \Delta_\psi$. It will be shown that the set of plans $\cset{\tup{\psi, \varphi}} \in \S^\Gamma_i$ satisfies the requirements.
        \begin{cond-kh}
          \item Take any $\Delta \in \truthset{\modults^\Gamma}{\psi}$. By IH, $\psi \in \Delta$. Moreover, from $\khi(\psi, \varphi) \in \Theta$ and \autoref{pro:cm-ults-lkhi-allsameKH} it follows that $\khi(\psi, \varphi) \in \Delta$. Then, from $\R^\Gamma_{\tup{\psi,\varphi}^i}$'s definition, every $\Delta' \in \W^\Gamma$ with $\varphi \in \Delta'$ is such that $(\Delta, \Delta') \in \R^\Gamma_{\tup{\psi,\varphi}^i}$, and therefore such that $(\Delta, \Delta') \in \R^\Gamma_{\tup{\psi,\varphi}}$. Now note how, since there is $\Delta_\psi \in \W^\Gamma$ with $\psi \in \Delta_\psi$, there should be $\Delta_\varphi \in \W^\Gamma$ with $\varphi \in \Delta_\varphi$. Suppose otherwise, i.e., suppose there is no $\Delta'' \in \W^\Gamma$ with $\varphi \in \Delta''$. Then, $\lnot \varphi \in \Delta''$ for every $\Delta'' \in \W^\Gamma$, and hence (\autoref{pro:cm-ults-lkhi-allall}) $\A\lnot\varphi \in \Delta''$ for every $\Delta'' \in \W^\Gamma$. In particular, $\A\lnot\varphi \in \Delta_\psi$. Moreover, from $\khi(\psi, \varphi) \in \Theta$ and \autoref{pro:cm-ults-lkhi-allsameKH} it follows that $\khi(\psi, \varphi) \in \Delta_\psi$. Then, \axm{KhE} (written as $\khi(\psi, \varphi) \limp (\A\lnot \varphi \limp \A \lnot \psi)$) and \axm{MP} yield $\A\lnot\psi \in \Delta_\psi$, and thus (axiom \axm{TA}) $\lnot\psi \in \Delta_\psi$. Hence, $\cset{\psi, \lnot\psi} \subset \Delta_\psi$, contradicting $\Delta_\psi$'s consistency.

          Let us continue with the proof of the lemma. The existence of $\Delta_\varphi \in \W^\Gamma$ with $\varphi \in \Delta_\varphi$ implies that $(\Delta, \Delta_\varphi) \in \R^\Gamma_{\tup{\psi,\varphi}}$ and thus, since $\tup{\psi,\varphi}$ is a basic action, $\Delta \in \stexec(\tup{\psi,\varphi})$, and so $\Delta \in \stexec(\cset{\tup{\psi,\varphi}})$. Since $\Delta$ is an arbitrary state in $\truthset{\modults^\Gamma}{\psi}$, the required $\truthset{\modults^\Gamma}{\psi} \subseteq \stexec(\cset{\tup{\psi,\varphi}})$ follows.

          \item Take any $\Delta' \in \R^\Gamma_{\cset{\tup{\psi,\varphi}}}(\truthset{\modults^\Gamma}{\psi})$. Then, there is $\Delta \in \truthset{\modults^\Gamma}{\psi}$ such that $(\Delta, \Delta') \in \R^\Gamma_{\tup{\psi,\varphi}}$. By definition of $\R^\Gamma$, it follows that $\varphi \in \Delta'$ so, by IH, $\Delta' \in \truthset{\modults^\Gamma}{\varphi}$. Since $\Delta'$ is an arbitrary state in $\R^\Gamma_{\cset{\tup{\psi,\varphi}}}(\truthset{\modults^\Gamma}{\psi})$, the required $\R^\Gamma_{\cset{\tup{\psi,\varphi}}}(\truthset{\modults^\Gamma}{\psi}) \subseteq \truthset{\modults^\Gamma}{\varphi}$ follows.
        \end{cond-kh}
      \end{itemize}
    \end{itemizenl}
  \end{demostracion}
\end{lema}

Finally, we present the intended result.

\begin{teorema}\label{teo:khi-sound-complete-ults}
  The axiom system $\KHiaxiom$ (\autoref{tab:khiaxiom}) is sound and strongly complete for \KHilogic w.r.t. the class of all $\ultss$.
  \begin{demostracion}
    For soundness, it is enough to show that the system's axioms are valid and that its rules preserve validity (which, as mentioned before, is straightforward). For completeness, take any \KHiaxiom-consistent set of formulas $\Gamma' \subseteq \KHilogic$. Since \KHilogic is enumerable, $\Gamma'$ can be extended into a maximally \KHiaxiom-consistent set $\Gamma \supseteq \Gamma'$ by a standard Lindenbaum's construction (see, e.g., \cite[Lemma 4.17]{mlbook}). By \autoref{tlm:cm-ults-lkhi}, $\Gamma'$ is satisfiable in $\modults^\Gamma$ at $\Gamma$. The fact that $\modults^\Gamma$ is an \ults (\autoref{pro:cm-ults-lkhi}) completes the proof. 
  \end{demostracion}
\end{teorema}

One detail in the construction of the canonical model might be surprising: each set of indistinguishable plans for a given agent is a singleton set.  Hence, the logic is also complete with respect to this particular class of models.  On the other hand, $\ultss$ are a more general and accurate representation from a conceptual point of view.  
We could, for instance, extend the language so that this representation is also reflected by the logic (i.e., the language can explicitly refer to plans), or define public announcements-like modalities to refine the indistinguishability relation of each agent (see, e.g.,~\cite{arec:firt23}).

% \bigskip

% We conclude this section with some comments about the impact of including the indistinguishability relation between plans for each agent. As shown above, in the canonical model each set of indistinguishable plans for a given agent is a singleton set. By comparing the axiom systems from Tables~\ref{tab:khaxiom} and~\ref{tab:khiaxiom}, it is evident that including these  indistinguishability sets in important to characterize the intended logic (see the discussion at the beginning of this section). However, one can argue that including singletons is enough from a purely logical standpoint, and that what  only matters is the awareness of the agent about the existence or availability of a plan. As said, this is true from the perspective of the logic we obtain, but we stick to the decision of having set of indistinguishable plans as we argue that it is a more general and accurate representation from a conceptual point of view. We could instance, extend the language so that this representation is also reflected by the logic (i.e., the language can talk explicitly of the plans), or to define public announcements-like modalities to refine the information of each agent~\cite{arec:firt23}. Thus, as we will discuss in the concluding section, our semantics paves the way for a more general setting.

%------------------------------------------------------------------------------------------------
\section{Expressive power} \label{sec:comparing}
%!TEX root = main.tex

This section compares the original logic from~\cite{Wang15lori,Wang16,Wang2016} with the one introduced in this paper. More precisely, first it will be shown that our logic is weaker than the original one. Then, we will explore two different classes of \ultss under which we can capture the exact semantics of~\cite{Wang15lori,Wang16,Wang2016}. As a consequence, the axiom system in \autoref{tab:khaxiom} is sound and complete for these two classes of models. This shows that our framework generalizes the one based on \ltss.

For the comparison to be meaningful, we will restrict the \ults setting to its single-agent case: a single modality $\kh$ and no subindexes for $\DS{i}$ and $\S_i$.

The provided axiom system can be used to compare the notion of \emph{knowing how} under \ltss with that under \ultss. Here is a first observation.

\begin{proposicion}\label{pro:subsystem}
  \axm{KhE} and \axm{KhA} are theorems of \KHaxiom.
  \begin{demostracion}
    \axm{KhE} can be rewritten as $\left(\kh(\psi,\varphi) \land \A\lnot\varphi\right) \limp \A\lnot\psi$, which is an instance of \axm{COMPKh} in \KHaxiom (just unfold $\A$).
\begin{proof:KhE.KhA.derivables.short}
    For \axm{KhA}, use \axm{EMP} and then \axm{COMPKh} \cite[Proposition 2]{Wang2016}.
\end{proof:KhE.KhA.derivables.short}
\begin{proof:KhE.KhA.derivables.detailed}
    For \axm{KhA}, take $\vdash (\khi(\chi, \psi) \land \khi(\psi,\varphi)) \limp \khi(\chi,\varphi)$ and $\vdash (\khi(\chi, \varphi) \land \khi(\varphi,\theta)) \limp \khi(\chi,\theta)$, two instances of \axm{COMPKh}. By propositional reasoning using the first, the latter becomes $\vdash (\khi(\chi, \psi) \land \khi(\psi,\varphi) \land \khi(\varphi,\theta)) \limp \khi(\chi,\theta)$. On this latter formula, use $\vdash \A(\chi \limp \psi) \limp \khi(\chi,\varphi)$ and $\vdash \A(\varphi \limp \theta) \limp \khi(\varphi,\theta)$ (both instances of \axm{EMP}) to obtain, by propositional reasoning, the required $\vdash (\A(\chi \limp \psi) \land \khi(\psi,\varphi) \land \A(\varphi \limp \theta)) \limp \khi(\chi,\theta)$.
\end{proof:KhE.KhA.derivables.detailed}
  \end{demostracion}
\end{proposicion}

Hence, the \emph{knowing how} operator under \ltss is at least as strong as its \ults-based counterpart: every formula valid under \ultss is also valid under \ltss. The following fact shows that the converse is not the case.

\begin{proposicion}\label{fact:axiom-fail}
  Within $\ults$, axioms \axm{EMP} and \axm{COMPKh} are not valid.
  \begin{demostracion}
    Consider the \ults shown below, with the collection of sets of plans for the agent (i.e., the set $\S$) depicted on the right. Recall that $\kh$ acts globally.
    
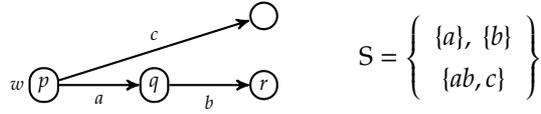
\begin{nscenter}
      \begin{tikzpicture}[->]
        \node [state, label = {[label-state]left:$w$}] (w1) {$p$};
        \node [state, right = of w1] (w2) {$q$};
        \node [state, right = of w2] (w3) {$r$};
        \node [state, above = of w3] (w4) {};

        \path (w1) edge node [label-edge, below] {$a$} (w2)
                   edge node [label-edge, above] {$c$} (w4)
              (w2) edge node [label-edge, below] {$b$} (w3);
      \end{tikzpicture}
      \hspace{1.5cm}
      \begin{picture}(60,0)
        \put(-20,20){
          $\S = \left\{
            \begin{array}{c}
              \{a\},\ \{b\}\\
              \{ab, c\}
            \end{array}
            \right\}$}
        \end{picture}
    \end{nscenter}
    
    With respect to \axm{EMP}, notice that $\A(p\ra p)$ holds; % (recall: $\A$ is interpreted as truth in every state);
    yet, $\kh(p,p)$ fails since there is no $\strategy \in \S$ leading from $p$-states to $p$-states. More generally, \axm{EMP} is valid over \ltss because the empty plan $\epsilon$, strongly executable everywhere, is always available. However, in an $\ults$, the plan $\epsilon$ might not be available to the agent (i.e., $\epsilon \notin \DS{}$), and even if it is, it might be indistinguishable from other plans with different behaviour.
      % \Cref{tab:khaxiom}.  In all pointed LTS $\modlts,w$, if
      % $\modlts,w\models\A(p\ra q)$ then $\modlts,w\models\kh(p,q)$. As shown in
      % \cite{Wang15lori}, the witness SE plan for satisfying $\kh(p,q)$ when
      % $\modlts,w\models\A(p\ra q)$ is the empty plan $\epsilon$.  However, in an $\ults$ $\model=\tup{\W,\R,\S,\V}$,
      % the plan $\epsilon$ could
      % belong to a strategy $\strategy\in\S$ containing other plans leading to $\neg q$-states, even if $\model \models \A(p\ra q)$. In fact,
      % $\epsilon$ might not even be in $\DS{\S}$. Thus, it is easy to
      % find counter examples to {\sf EMP} over {\ults}s.

    With respect to \axm{COMPKh}, notice that $\kh(p,q)$ and $\kh(q,r)$ hold, witness $\cset{a}$ and $\cset{b}$, respectively. However, there is no $\strategy\in\S$ containing only plans that, when started on $p$-states, lead only to $r$-states. Thus, $\kh(p,r)$ fails. More generally, \axm{COMPKh} is valid over \lts because the sequential composition of the plans that make true the conjuncts in the antecedent is a witness that makes true the consequent. However, in an $\ults$, this composition might be unavailable or else indistinguishable from other plans.
%  as we just showed, this is not the case in an \ults.
%  in an \ults, the existence of two sets of plans $\strategy_1$ and $\strategy_2$ does not guarantee the existence of a third one containing the sequential composition of the plans in $\strategy_1$ with the plans in $\strategy_2$. And even if such a third set of plans exists, it might contain additional plans that have a non-adequate behaviour.
  \end{demostracion}
\end{proposicion}

From these two observations it follows that $\kh$ under $\ultss$ is strictly weaker than $\kh$ under \ltss: adding uncertainty about plans changes the logic.

\subsection{A very simple class of \ultss} \label{subsec:simple-models}

Still, the uncertainty-based framework is general enough to capture the \lts semantics. Given the discussion in \autoref{fact:axiom-fail}, there is an obvious class of \ultss in which \axm{EMP} and \axm{COMPKh} are valid: the class of \ultss in which the agent has every plan available and can distinguish between any two of them. Below, we define formally this class.

\begin{definicion}\label{def:class-lts-nu}
Define the class of models: $\cultsnu := \csetsc{\modults}{\modults \mbox{ is an } \ults \mbox{ and } \S = \csetsc{\cset{\sigma}}{\sigma\in\ACT^*}}$.
\end{definicion}

Indeed, for models in $\cultsnu$, the plan $\epsilon$ is available and distinguishable from other plans (witnessing \axm{EMP}) and from $\cset{\sigma_1} \in \S$ and $\cset{\sigma_2} \in \S$ it follows that $\cset{\sigma_1\sigma_2} \in \S$ (witnessing \axm{COMPKh}). Thus, as the following proposition states, an agent in \lts is exactly an agent in \ults that can use every plan and has no uncertainty and full awareness about them. This class is enough to show how the uncertainty-based framework can capture the original one.

\begin{proposicion} \label{prop:simple-lts-ltsu-corresp}
  The following properties hold.
  \begin{enumerate}
    \item\label{prop:simple-lts-ltsu-corresp:itm:1} Given a model $\modults=\tup{\W,\R,\S,\V}$ in $\cultsnu$, the \lts $\modlts_\modults=\tup{\W,\R,\V}$ is such that $\truthset{\modults}{\varphi} = \truthset{\modlts_{\modults}}{\varphi}$ for every $\varphi \in \KHlogic$.

    \item\label{prop:simple-lts-ltsu-corresp:itm:2} Given an \lts $\modlts=\tup{\W,\R,\V}$, the model $\modults_\modlts = \tup{\W, \R, \S, \V}$ with $\S = \csetsc{\cset{\sigma}}{\sigma\in\ACT^*}$, is in $\cultsnu$ and is such that $\truthset{\modlts}{\varphi} = \truthset{\modults_{\modlts}}{\varphi}$ for every $\varphi \in \KHlogic$.
  \end{enumerate}
\begin{proof:simple-lts-ltsu-corresp}
\begin{demostracion}
  In both cases, the proof is by induction on $\varphi \in \KHlogic$.
  \begin{enumerate}
    \item The case for atomic propositions is straightforward ($\modults$ and $\modlts_\modults$ have the same atomic valuation); the Boolean cases follow from their respective inductive hypotheses. For the $\kh$ case, $\proofsb$ $w \in \truthset{\modults}{\kh(\psi,\varphi)}$ implies there is $\strategy \in \S$ with \begin{inline-cond-kh}\item\label{itm:corresp.ii.1} $\truthset{\modults}{\psi} \subseteq \stexec(\strategy)$ and \item\label{itm:corresp.ii.2} $\R_\strategy(\truthset{\modults}{\psi}) \subseteq \truthset{\modults}{\varphi}$\end{inline-cond-kh}. But $\modults$ is in $\cultsnu$, so $\strategy = \cset{\sigma}$ for some $\sigma \in \ACT^*$. Now, note how not only $\truthset{\modlts_\modults}{\psi} = \truthset{\modults}{\psi} \subseteq \stexec(\strategy) = \stexec(\cset{\sigma}) = \stexec(\sigma)$ (by IH, \autoref{itm:corresp.ii.1}, $\strategy = \cset{\sigma}$ and definition of $\stexec$ for sets of actions, respectively), but also $\R_{\sigma}(\truthset{\modlts_\modults}{\psi}) = \R_{\sigma}(\truthset{\modults}{\psi}) = \R_{\cset{\sigma}}(\truthset{\modults}{\psi}) = \R_{\strategy}(\truthset{\modults}{\psi}) \subseteq \truthset{\modults}{\varphi} = \truthset{\modlts_\modults}{\varphi}$ (by IH, definition of $\R$ for sets of actions, $\strategy = \cset{\sigma}$, \autoref{itm:corresp.i.2} and IH, respectively). Hence, \begin{inline-cond-kh}\item $\truthset{\modlts_\modults}{\psi} \subseteq \stexec(\sigma)$ and \item $\R_{\sigma}(\truthset{\modlts_\modults}{\psi}) \subseteq \truthset{\modlts_\modults}{\varphi}$\end{inline-cond-kh}, and thus $w \in \truthset{\modlts_\modults}{\kh(\psi,\varphi)}$. The direction $\proofsp$ is analogous.

    \item The cases for atoms and Boolean operators are as above. For the $\kh$ case, $\proofsb$ $w \in \truthset{\modlts}{\kh(\psi,\varphi)}$ implies there is $\sigma \in \ACT^*$ with \begin{inline-cond-kh}\item\label{itm:corresp.i.1} $\truthset{\modlts}{\psi} \subseteq \stexec(\sigma)$ and \item\label{itm:corresp.i.2} $\R_\sigma(\truthset{\modlts}{\psi}) \subseteq \truthset{\modlts}{\varphi}$\end{inline-cond-kh}. By the construction of $\modults_\modlts$, it follows that $\cset{\sigma} \in \S$. Moreover, note how not only $\truthset{\modults_\modlts}{\psi} = \truthset{\modlts}{\psi} \subseteq \stexec(\sigma) = \stexec(\cset{\sigma})$ (by IH, \autoref{itm:corresp.i.1} and definition of $\stexec$ for sets of actions, respectively), but also $\R_{\cset{\sigma}}(\truthset{\modults_\modlts}{\psi}) = \R_{\cset{\sigma}}(\truthset{\modlts}{\psi}) = \R_\sigma(\truthset{\modlts}{\psi}) \subseteq \truthset{\modlts}{\varphi} = \truthset{\modults_\modlts}{\varphi}$ (by IH, definition of $\R$ for sets of actions, \autoref{itm:corresp.i.2} and IH, respectively). Hence, \begin{inline-cond-kh}\item $\truthset{\modults_\modlts}{\psi} \subseteq \stexec(\cset{\sigma})$ and \item $\R_{\cset{\sigma}}(\truthset{\modults_\modlts}{\psi}) \subseteq \truthset{\modults_\modlts}{\varphi}$\end{inline-cond-kh}, and thus $w \in \truthset{\modults_\modlts}{\kh(\psi,\varphi)}$. The direction $\proofsp$ is analogous.    
  \end{enumerate}
\end{demostracion}
\end{proof:simple-lts-ltsu-corresp}
\end{proposicion}
%Thus, any \lts can be transformed into a pointwise equivalent $\ults$ by taking its strongly executable plans (the ones that define her abilities, which include $\epsilon$) and turn them into basic actions that can be distinguished from any other.

%\begin{textonuevo}
%Thus, any $\ults$ \emph{in $\cults$} can be transformed into a pointwise equivalent \lts by collapsing each set of plans available to the agent into a single basic action.
%\end{textonuevo}

This correspondence, showing that every \lts has a point-wise equivalent model in $\cultsnu$ and vice-versa, gives us a direct completeness result.

\begin{teorema}\label{teo:kh-sound-complete-cults}
  The axiom system \KHaxiom (\autoref{tab:khaxiom}) is sound and strongly complete for \KHlogic w.r.t. the class $\cultsnu$.
\begin{proof:kh-sound-complete-cults}
\begin{demostracion}
  For soundness, we look at both blocks in \autoref{tab:khaxiom}. For the first, \autoref{teo:khi-sound-complete-ults} shows that those axioms and rules are sound for all \ults, and thus in particular sound for those in the class $\cultsnu$. For the second, \autoref{prop:simple-lts-ltsu-corresp:itm:1} of \autoref{prop:simple-lts-ltsu-corresp} shows that every model in $\cultsnu$ is point-wise $\KHlogic$-equivalent to an \lts, thus (\autoref{teo:kh-sound-complete-lts}) making sound such axioms.

  To prove that $\KHaxiom$ is strongly complete over the class $\cultsnu$, we need to show that, given $\Gamma\cup\cset{\varphi}$ a set of formulas in $\KHlogic$, $\Gamma\models\varphi$ implies $\Gamma\vdash\varphi$. 
  Let $\Gamma$ be a consistent set of formulas. As in~\cite[Lemma
  1]{Wang2016}, $\Gamma$ can be extended to an MCS $\Gamma'$, and as a
  consequence, there exists an LTS $\modlts^{\Gamma'}$ such that
  $\modlts^{\Gamma'},\Gamma'\models\Gamma$ (notice that states in the
  canonical model are MCS). Then, by \autoref{prop:simple-lts-ltsu-corresp:itm:2} of \autoref{prop:simple-lts-ltsu-corresp}, we can
  obtain an \ults $\model_{\modlts^{\Gamma'}}$, such
  that $\model_{\modlts^{\Gamma'}},\Gamma'\models\Gamma$. Moreover,
  from \autoref{prop:simple-lts-ltsu-corresp:itm:2} of \autoref{prop:simple-lts-ltsu-corresp} we also know that $\model_{\modlts^{\Gamma'}}$ is in $\cultsnu$.
  % For completeness, take any set of \KHlogic-consistent formulas $\Gamma'$. By the completeness of \KHlogic over \ltss (\autoref{teo:kh-sound-complete-lts}), $\Gamma'$ is satisfiable on a \lts, say $\modlts$. Then, by \autoref{prop:simple-lts-ltsu-corresp:itm:2} of \autoref{prop:simple-lts-ltsu-corresp}, one can build the model $\modults_\modlts$ in $\cultsnu$, which is point-wise equivalent to $\modlts$ and thus satisfies $\Gamma'$. 
\end{demostracion}
\end{proof:kh-sound-complete-cults}
\end{teorema}

%!TEX root = main.tex

\subsection{Active and $\stexec$-compositional \ultss} \label{subsec:active-compositional}

We presented above a very simple class of models that enables us to establish a direct relation between both semantics. However, the result is somewhat trivial: \ultss generalize \ltss by adding uncertainty among plans, and the class $\cultsnu$ contains those \ultss in which the agent does not have uncertainty. %One may wonder whether a more general class can be defined, for which the same correspondence holds.
The rest of this section will discuss a larger and very general class (with very weak constraints) for which the same correspondence holds.

Let us start by introducing some preliminary definitions.

\begin{definicion}\label{def:composition}
  Let $\modults=\tup{\W,\R,\S,\V}$ be an $\ults$.
  \begin{itemize}
    \item The \emph{composition} of $\strategy_1,\strategy_2 \in 2^{\ACT^*}$ is the set of plans $\strategy_1\strategy_2 \in 2^{\ACT^*}$ given by
    \begin{equation*}
      \strategy_1\strategy_2 := \csetc{\sigma_1\sigma_2}{\ACT^*}{\sigma_1 \in \strategy_1 \text{ and } \sigma_2 \in \strategy_2}.
    \end{equation*}

    \item The \emph{$\stexec$-composition} of $\strategy_1,\strategy_2 \in 2^{\ACT^*}$ \emph{in $\modults$} is the set of plans $\strategy_1 \compose \strategy_2 \in 2^{\ACT^*}$ given by
    \[
      \strategy_1 \compose \strategy_2 : =
      \begin{cases*}
         \strategy_1\strategy_2 & if $\stexec(\strategy_1) \neq \emptyset$ and $\R_{\strategy_1}(\stexec(\strategy_1))\subseteq \stexec(\strategy_2)$ \\
        \emptyset               & otherwise.
      \end{cases*}
    \]
  \end{itemize}
\end{definicion}

Thus, the $\stexec$-composition $\strategy_1 \compose \strategy_2$ is the sequential composition of $\strategy_1$ and then $\strategy_2$ (i.e., $\strategy_1\strategy_2$) when $\strategy_1$ is strongly executable somewhere in the model and $\strategy_2$ is strongly executable at all the states that are reachable via $\strategy_1$ from states where $\strategy_1$ is strongly executable. Otherwise, $\strategy_1 \compose \strategy_2 =\emptyset$. This guarantees that $\strategy_1\compose\strategy_2$ contains only suitable plans. For multiple sets of plans $\strategy_1,\dots,\strategy_k \in \S$, the $\stexec$-composition $\strategy_1\compose\cdots\compose\strategy_k$ is the set of plans $\strategy_1\cdots\strategy_k$ if and only if $\stexec(\strategy_1)\neq \emptyset$ and $\R_{\strategy_i}(\stexec(\strategy_i)) \subseteq \stexec(\strategy_{i+1})$ for all $i=1,\dots,k-1$, and $\emptyset$ otherwise.

The following lemma establishes important properties of the just defined $\stexec$-composition. They will be helpful in the rest of the section.

\begin{lema}\label{lemma:composition_properties}
  Let $\model=\tup{\W,\R,\S,\V}$ be an $\ults$ with $\strategy_1,\dots,\strategy_k \in \S$. Then,
  \begin{enumerate}
    \item $\strategy_1\compose\cdots\compose\strategy_k \neq \emptyset$ if and only if $\strategy_i\compose\strategy_{i+1} \neq \emptyset$ for all $i=1,\dots,k-1$.
    \item $\strategy_1\compose\cdots\compose\strategy_k \neq \emptyset$ implies $\stexec(\strategy_1) = \stexec(\strategy_1\compose\dots\compose\strategy_k)$.
  \end{enumerate}
  \begin{demostracion}
    For the first, consider the left-to-right direction. If $\strategy_1\compose\cdots\compose\strategy_k \neq \emptyset$, then $\stexec(\strategy_1) \neq \emptyset$. Moreover, for all $i = 1,\dots,k-1$, if $\stexec(\strategy_i) \neq \emptyset$, then $\stexec(\strategy_{i+1}) \neq \emptyset$ (because $\R_{\strategy_i}(\stexec(\strategy_i)) \subseteq \stexec(\strategy_{i+1})$). Therefore, $\stexec(\strategy_i)\neq \emptyset$ for all $i=1,\dots,k-1$. Using \autoref{def:composition}, for all $i=1,\dots,k-1$ we have $\strategy_i\compose\strategy_{i+1} \neq \emptyset$. The other direction is direct.

    For the second, suppose $\strategy_1\compose\cdots\compose\strategy_k \neq \emptyset$. For $\bm{(\subseteq)}$, proceed by contradiction: assume there is $w \in \stexec(\strategy_1)$ with $w \not\in \stexec(\strategy_1\compose\cdots\compose\strategy_k)$. From the latter, $w \not \in \stexec(\sigma)$ for some $\sigma = \sigma_1 \dots \sigma_k \in \strategy_1\compose\cdots\compose\strategy_k$. Thus, there are $d<k$ and $w=v_1,\dots,v_{d+1}=v$ such that $v_{i+1} \in \R_{\sigma_i}(v_i)$ and $v \not\in \stexec(\sigma_{d+1})$. By hypothesis, $v_1=w \in \stexec(\strategy_1)$ and for all $i=1,\dots,d$, if $v_i \in \stexec(\strategy_i)$, then we have $v_{i+1} \in \stexec(\strategy_{i+1})$ (since $v_{i+1} \in \R_{\strategy_i}(v_i)$ and $\R_{\strategy_i}(\stexec(\strategy_i)) \subseteq \stexec(\strategy_{i+1})$). Hence, $v_{d+1}=v \in \stexec(\strategy_{d+1})$ and therefore $v \in \stexec(\sigma_{d+1})$, a contradiction. Thus, $w \in \stexec(\strategy_1\compose\dots\compose\strategy_k)$, and therefore $\stexec(\strategy_1) \subseteq \stexec(\strategy_1\compose\dots\compose\strategy_k)$. The direction $\bm{(\supseteq)}$ is rather immediate.
  \end{demostracion}
\end{lema}
  
Now, here are the crucial properties we will require of \ults, to establish the intended correspondence with \ltss. 

\begin{definicion}\label{def:esm-req}
  We say that an $\ults$ $\model=\tup{\W,\R,\S,\V}$ is:
  \begin{itemize}
  \item \emph{active} if and only if there exists $\strategy\in\S$ such that
    $\stexec(\strategy)=\W$ and, for all $u,v\in\W$,
    $v\in\R_{\strategy}(u)$ implies $\model,u\bisim\model,v$.

  \item \emph{$\stexec$-compositional} if and only if for all $\strategy_1,\strategy_2\in\S$ with $\strategy_1\compose\strategy_2\neq\emptyset$ there exists $\strategy\in\S$ such that:
    \begin{enumerate}
      \item $\R_{\strategy_1\compose\strategy_2}\subseteq \R_{\strategy}$,
      \item $\stexec(\strategy_1\compose\strategy_2) \subseteq \stexec(\strategy)$, and 
      \item for all $(w,v) \in \R_\strategy$ there exists $(w',v')\in\R_{\strategy_1\compose\strategy_2}$ such that $\model,w\bisim\model,w'$ and $\model,v\bisim\model,v'$.
    \end{enumerate}
  \end{itemize}
  We define the class $\cultsac := \csetsc{\model}{\model \mbox{ is active and $\stexec$-compositional}}$.
\end{definicion}

While activeness ensures that there is a set of plans doing what the empty plan $\epsilon$ does in an \lts, $\stexec$-compositionality ensures that $\S$ is closed under a suitable notion of composition of sets of plans. The use of bisimilarity gives us a slightly more general class of models.

The next lemma establishes that the requirements for $\stexec$-compositionality generalize to an arbitrary number of sets of plans.

\begin{lema}\label{lemma:compositional_properties}
  Let $\model=\tup{\W,\R,\S,\V}$ be an $\stexec$-compositional $\ults$, and take $\strategy_1,\dots,\strategy_k \in \S$ (with $k\geq2$) such that $\strategy_1\compose\cdots\compose\strategy_k \neq \emptyset$. Then, there is $\strategy \in \S$ such that:
  \begin{enumerate}
    \item\label{itm:comp:1} $\R_{\strategy_1\compose\dots\compose\strategy_k}\subseteq \R_{\strategy}$,
    \item\label{itm:comp:2} $\stexec(\strategy_1\compose\dots\compose\strategy_k) \subseteq \stexec(\strategy)$, and 
    \item\label{itm:comp:3} for all $(w,v) \in \R_\strategy$, there exists $(w',v')\in\R_{\strategy_1\compose\dots\compose\strategy_k}$ such that $\model,w\bisim\model,w'$ and $\model,v\bisim\model,v'$.
  \end{enumerate}
  \begin{demostracion}
    We prove the existence of $\strategy$ by induction on $k\geq 2$; then we will show that this witness does the work. The base case $k=2$ follows from the definition, so take sets of plans in $\S$ such that $\strategy_1\compose\cdots\compose\strategy_k\compose\strategy_{k+1} \neq \emptyset$. By \autoref{lemma:composition_properties}, $\strategy_2\compose\cdots\compose\strategy_k\compose\strategy_{k+1} \neq \emptyset$ and thus, by inductive hypothesis, there is a $\strategy' \in \S$ such that \begin{inlineenum} \item $\R_{\strategy_2\compose\dots\compose\strategy_{k+1}}\subseteq \R_{\strategy'}$, \item $\stexec(\strategy_2\compose\dots\compose\strategy_{k+1}) \subseteq \stexec(\strategy')$ , and \item for all $(w,v) \in \R_{\strategy'}$ there exists $(w',v')\in\R_{\strategy_2\compose\dots\compose\strategy_{k+1}}$ such that $\model,w\bisim\model,w'$ and $\model,v\bisim\model,v'$.\end{inlineenum} Note also how $\stexec(\strategy_1) \neq \emptyset$ and $\R_{\strategy_1}(\stexec(\strategy_1)) \subseteq \stexec(\strategy_2) = \stexec(\strategy_2\compose\dots\compose\strategy_{k+1}) \subseteq \stexec(\strategy')$ (by definition of $\stexec$-composition, \autoref{lemma:composition_properties} and the second property of $\strategy'$). Thus, $\strategy_1\compose\strategy' \neq \emptyset$ and hence there is a $\strategy \in \S$ such that \begin{inlineenum} \item $\R_{\strategy_1\compose\strategy'}\subseteq \R_{\strategy}$, \item $\stexec(\strategy_1\compose\strategy') \subseteq \stexec(\strategy)$, and \item for all $(w,v) \in \R_{\strategy}$ there exists $(w',v')\in\R_{\strategy_1\compose\strategy'}$ such that $\model,w\bisim\model,w'$ and $\model,v\bisim\model,v'$.\end{inlineenum} We will prove that $\strategy$ is the witness we are looking for.

    For \autoref{itm:comp:1}, take $(w,v) \in \R_{\strategy_1\compose\dots\compose\strategy_{k+1}}$. Then, there exists $u \in \W$ such that $(w,u) \in \R_{\strategy_1}$ and $(u,v) \in \R_{\strategy_2\compose\dots\compose\strategy_{k+1}}$. Hence, we have $(w,u) \in \R_{\strategy_1}$ and $(u,v) \in \R_{\strategy'}$, so $(w,v) \in \R_{\strategy_1\compose\strategy'}$ and therefore $(w,v) \in \R_{\strategy}$.

    For \autoref{itm:comp:2}, take $w \in \stexec(\strategy_1\compose\dots\compose\strategy_{k+1})$. By \autoref{lemma:composition_properties}, $w \in \stexec(\strategy_1)$ and, by the same lemma, $w \in \stexec(\strategy_1\compose\strategy')$. Hence, $w \in \stexec(\strategy)$.

    For \autoref{itm:comp:3}, take $(w,v) \in \R_{\strategy}$. Then, there exists $(w',v')\in\R_{\strategy_1\compose\strategy'}$ such that $\model,w\bisim\model,w'$ and $\model,v\bisim\model,v'$. Thus, $(w',u')\in\R_{\strategy_1}$ and $(u',v')\in\R_{\strategy'}$ for some $u' \in \W$. Again, since $(u',v') \in \R_{\strategy'}$, there exists $(u'',v'')\in\R_{\strategy_2\compose\dots\compose\strategy_{k+1}}$ such that $\model,u'\bisim\model,u''$ and $\model,v'\bisim\model,v''$. Using that $\bisim$ is transitive, there exists $(w',v'') \in \R_{\strategy_1\compose\dots\compose\strategy_{k+1}}$ such that $\model,w\bisim\model,w'$ and $\model,v\bisim\model,v''$.
  \end{demostracion}
\end{lema}

With these tools at hand, we will show that for every \lts there is an $\KHlogic$-equivalent $\ults$ in $\cultsac$, and vice-versa. First, we present the mapping from \ultss to \ltss.

\begin{proposicion}\label{prop:esmtolts}
  Let $\model=\tup{\W,\R,\S,\V}$ be an \ults in $\cultsac$, over $\ACT$. Take $\ACT' := \csetsc{a_{\strategy}}{\strategy \in \S }$, and then define the \lts $\modlts_{\model} = \tup{\W, \R', \V}$ over $\ACT'$ by taking $\R'_{a_{\strategy}} := \csetsc{(w,v) \in \R_{\strategy}}{w \in \stexec(\strategy)}$ (so basic actions in $\modlts_\model$ correspond to sets of SE plans in $\model$). Then, $\truthset{\model}{\varphi} = \truthset{\modlts_\model}{\varphi}$ for every $\varphi \in \KHlogic$.
\begin{demostracion}
    It is clear that $\modlts_\model$ is an \lts. To obtain a proper signature, we can extend $\ACT'$  (in case it is finite) into an arbitrary $\ACT$. 

    The rest of the proof of is by structural induction on the formulas in $\KHlogic$. The cases for the Boolean fragment are straightforward. We will discuss the case for formulas of the shape $\kh(\psi,\varphi)$. In doing so, the following property will be useful: for every $\strategy \in S$, we have $\stexec(\strategy) = \stexec(a_{\strategy})$. Indeed, $\bm{(\subseteq)}$ if $u \in \stexec(\strategy)$ then there is $v \in \W$ such that $(u,v) \in \R_\strategy$, so $(u,v) \in \R'_{a_\strategy}$ and therefore, being $a_\strategy$ a basic action, $u \in \stexec(a_\strategy)$. Moreover, $\bm{(\supseteq)}$, if $u \in \stexec(a_\strategy)$ then there is $v \in \W$ such that $(u,v) \in \R'_{a_\strategy}$, so $u \in \stexec(\strategy)$.

    \ssparagraph{$\bm{(\subseteq)}$} Suppose $w \in \truthset{\model}{\kh(\psi,\varphi)}$; then there is $\strategy \in \S$ satisfying both
    \begin{multicols}{2}
      \begin{cond-kh}
        \item\label{pro1:lr:itm:i} $\truthset{\model}{\psi} \subseteq \stexec(\strategy)$ and 
        \item\label{pro1:lr:itm:ii} $\R_\strategy(\truthset{\model}{\psi}) \subseteq \truthset{\model}{\varphi}$.
      \end{cond-kh}
    \end{multicols}
    \noindent We will prove $w \in \truthset{\modlts_\model}{\kh(\psi,\varphi)}$ using $a_\strategy \in \ACT'$ as our witness. First, for showing that $a_\strategy$ has the right properties, suppose $v \in \truthset{\modlts_\model}{\psi}$. Then $v \in \truthset{\model}{\psi}$ (by IH), so $v \in \stexec(\strategy)$ (by \autoref{pro1:lr:itm:i}) and hence $v \in \stexec(a_{\strategy})$ (property discussed above). Therefore, $\truthset{\modlts_\model}{\psi} \subseteq \stexec(a_\strategy)$. Second, for showing that $a_{\strategy}$ does the required work, suppose $u \in \R'_{a_{\strategy}}(\truthset{\modlts_\model}{\psi})$. Then, $u \in \R'_{a_{\strategy}}(\truthset{\model}{\psi})$ (by IH), hence $u \in \R_{\strategy}(\truthset{\model}{\psi})$ (by definition of $\R'$) so $u \in \truthset{\model}{\varphi}$ (by \autoref{pro1:lr:itm:ii}), and then $u \in \truthset{\modlts_\model}{\varphi}$ (by IH). Thus, $\R'_{a_{\strategy}}(\truthset{\modlts_\model}{\psi}) \subseteq \truthset{\modlts_\model}{\varphi}$. From the two pieces, it follows that  $w \in \truthset{\modlts_\model}{\kh(\psi,\varphi)}$.
      
    \ssparagraph{$\bm{(\supseteq)}$} Suppose $w \in \truthset{\modlts_\model}{\kh(\psi,\varphi)}$; then there is $\sigma \in (\ACT')^*$ satisfying both
    \begin{multicols}{2}
      \begin{cond-kh}
        \item\label{pro1:rl:itm:i} $\truthset{\modlts_\model}{\psi} \subseteq \stexec(\sigma)$ and
        \item\label{pro1:rl:itm:ii} $\R'_\sigma(\truthset{\modlts_\model}{\psi}) \subseteq \truthset{\modlts_\model}{\varphi}$.
      \end{cond-kh}
    \end{multicols}  
    \noindent There are two main cases. First, assume $\sigma=\epsilon$. Since $\model$ is active, there is $\strategy\in\S$ s.t. $\stexec(\strategy)=\W$ and for all $u,v\in\W$, $v\in\R_{\strategy}(u)$ implies $\model,u\bisim\model,v$. It is not hard to show that this is the witness we need.
    %
    %% Keep the text below for a long version of the paper
    %
    % \begin{enumerate}[\bfseries \itshape (i')]
    %   \item Take any $u \in \W$; note how $\R_\epsilon (u) \neq \emptyset$. Moreover, from $\epsilon \in \strategy$ it follows that $\R_\epsilon (u) \subseteq \R_\strategy (u)$. Hence, $\R_\strategy (u) \neq \emptyset$ (i.e., $u \in \stexec(\strategy)$) for all $u \in \W$, and thus in particular for every $u \in \truthset{\model}{\psi}$. Therefore, $\truthset{\model}{\psi} \subseteq \stexec(\strategy)$. 
    %  
    %   \item Since $\sigma = \epsilon$, we have $\R'_\sigma(\truthset{\modlts_\model}{\psi}) = \truthset{\modlts_\model}{\psi}$; hence, \autoref{pro1:rl:itm:ii} above actually says $\truthset{\modlts_\model}{\psi} \subseteq \truthset{\modlts_\model}{\varphi}$. Moreover, from $\strategy$'s constrains, $(u,v) \in \R_{\strategy}$ implies $u,v$ are bisimilar, so (\autoref{th:khbisim-to-khequiv}) both satisfy exactly the same formulas in $\model$. Thus, $u \in \R_\strategy(\truthset{\model}{\psi})$ implies $u \in \truthset{\model}{\psi}$ (condition on $\strategy$), which implies $u \in \truthset{\modlts_\model}{\psi}$ (IH), which implies $u \in \truthset{\modlts_\model}{\varphi}$ (see above), which implies $u \in \truthset{\model}{\varphi}$ (IH). Hence, $\R_\strategy(\truthset{\model}{\psi}) \subseteq \truthset{\model}{\varphi}$.
    % \end{enumerate}
    % Hence, $w \in \truthset{\model}{\kh(\psi,\varphi)}$.
    %        
    % {\smallskip}
    
    Second, assume $\sigma\neq\epsilon$, i.e., $\sigma=a_{\strategy_1}{\cdots}a_{\strategy_k}$ with $a_{\strategy_i} \in \ACT'$ (so $\strategy_i \in \S$). Then, there are two possibilities.
    \begin{itemize}
      \item If $\strategy_1\compose\cdots\compose\strategy_k = \emptyset$, by \autoref{lemma:composition_properties} there is $i \in \cset{1, \ldots, k-1}$ s.t. $\strategy_i\compose\strategy_{i+1}=\emptyset$. Then (by \autoref{def:composition}), either $\stexec(\strategy_i)=\emptyset$ (hence $\stexec(a_{\strategy_i})=\emptyset$) or $\R_{\strategy_i}(\stexec(\strategy_i))\not\subseteq\stexec(\strategy_{i+1})$ (so there is $v \in \R_{\strategy_i}(\stexec(\strategy_i))$ with $v \notin \stexec(\strategy_{i+1})$, i.e., there are $u,v \in W$ such that $u \in \stexec(\strategy_i)$, $(u,v) \in \R_{\strategy_i}$ and $v \notin \stexec(\strategy_{i+1})$, and hence $u \in \stexec(a_{\strategy_i})$ [from the first], $(u,v) \in \R'_{a_{\strategy_i}}$ [from the first and the second] and $v \notin \stexec(a_{\strategy_{i+1}})$ [from the third]; thus, $\R'_{a_{\strategy_i}}(\stexec(a_{\strategy_i}))\not\subseteq\stexec(a_{\strategy_{i+1}})$). In both cases we get $\stexec(\sigma)=\stexec(a_{\strategy_1}{\cdots}a_{\strategy_k})=\emptyset$, and hence $\truthset{\modlts_\model}{\psi} = \emptyset$ (by \autoref{pro1:rl:itm:i}). By IH, this implies $\truthset{\model}{\psi} = \emptyset$, so to get $w \in \truthset{\model}{\kh(\psi,\varphi)}$ we only need $\S \neq \emptyset$, which we have as $\model$ is an $\ults$.
      
      \item If $\strategy_1 \compose \cdots \compose \strategy_k \neq \emptyset$, we contemplate two scenarios. If $\truthset{\modlts_\model}{\psi}=\emptyset$ then, by IH, $\truthset{\model}{\psi}=\emptyset$; thus, as before, any $\strategy\in\S$ works as a witness. Otherwise, $\truthset{\modlts_\model}{\psi} \neq \emptyset$ and then, since $\model$ is $\stexec$-compositional, by \autoref{lemma:compositional_properties} there exists $\strategy\in\S$ such that $\R_{\strategy_1\compose\cdots\compose\strategy_k}\subseteq\R_\strategy$,  $\stexec(\strategy_1\compose \cdots\compose \strategy_k) \subseteq \stexec(\strategy)$, and for all $(v,u) \in \R_{\strategy}$, we have $(v',u') \in \R_{\strategy_1\compose\cdots\compose\strategy_k}$ for some $v', u' \in \W$ satisfying $\model,v\bisim\model,v'$ and $\model,u\bisim\model,u'$. Let's show that this $\strategy$ does the work.
        
      For the first $\kh$-clause, take $w \in \truthset{\model}{\psi}$. Then, by IH, $w \in \truthset{\modlts_\model}{\psi}$, and by \autoref{pro1:rl:itm:i}, $w \in \stexec(a_{\strategy_1}{\cdots}a_{\strategy_k})$. For a contradiction, suppose $w \not\in \stexec(\strategy_1\compose\cdots\compose\strategy_k)$; then $w \not\in \stexec(\strategy_1)$ (by \autoref{lemma:composition_properties}), 
      %  by reasoning inductively it must be the case that either for some $0<i\leq k$,  $w \not\in \stexec(\strategy_i)$, or  for some $0<i < k$, $\R_{\strategy_i}(\stexec(\strategy_i))\not\subseteq\stexec(\strategy_{i+1})$. In both cases, 
      so $w \not\in \stexec(a_{\strategy_1})$ and hence $\R'_{a_{\strategy_1}}(w) = \emptyset$. Hence, $w \not \in \stexec(a_{\strategy_1}{\cdots}a_{\strategy_k})$, which is a contradiction. Therefore, $w \in \stexec(\strategy_1\compose\cdots\compose\strategy_k)$, so $w \in \stexec(\strategy)$ (since $\model$ is $\stexec$-compositional). Thus, $\truthset{\model}{\psi} \subseteq \stexec(\strategy)$.

      For the second $\kh$-clause, take $u, v \in \W$ such that $v \in \truthset{\model}{\psi}$ and $(v,u) \in \R_{\strategy}$. By \autoref{def:esm-req}, there are $(v',u')\in \R_{\strategy_1\compose \cdots\compose \strategy_k}$ such that $\model,v\bisim\model,v'$ and $\model,u\bisim\model,u'$. By \autoref{th:khbisim-to-khequiv}, $v' \in \truthset{\model}{\psi}$ so, by inductive hypothesis, $v' \in \truthset{\modlts_\model}{\psi}$. By \autoref{def:composition}, $u' \in (\R_{\strategy_1} \circ \cdots \circ \R_{\strategy_k})(v')$. Now, let $v'=w_1,\dots,w_{k+1}=u'$ be such that $w_{i+1} \in \R_{\strategy_i}(w_i)$ for all $i=1,\dots,k$. Since $w_1 \in \truthset{\model}{\psi}$, from $\truthset{\model}{\psi} \subseteq \stexec(\strategy_1\compose\cdots\compose\strategy_k)$ (proved in the paragraph above) and \autoref{lemma:composition_properties}, we have $w_1 \in \stexec(\strategy_1\compose\cdots\compose\strategy_k) = \stexec(\strategy_1)$. Moreover, for all $i=1,\dots,k-1$, if $w_i \in \stexec(\strategy_i)$, since $\R_{\strategy_i}(\stexec(\strategy_i)) \subseteq \stexec(\strategy_{i+1})$ and $w_{i+1} \in \R_{\strategy_i}(w_i)$, then $w_{i+1} \in \stexec(\strategy_{i+1})$. Therefore, $w_i \in \stexec(\strategy_i)$ for all $i=1,\dots,k$. Since we assume $(w_i,w_{i+1}) \in \R_{\strategy_i}$ for all $i=1,\dots,k$, we have that $(w_i,w_{i+1}) \in \R_{a_{\strategy_i}}$. Hence, $u' \in (\R'_{a_{\strategy_1}} \circ \cdots \circ \R'_{a_{\strategy_k}})(v')$. In other words, since $v' \in \truthset{\modlts_\model}{\psi}$, $u' \in \R'_{a_{\strategy_1}{\cdots}a_{\strategy_k}}(\truthset{\modlts_\model}{\psi})$. Thus, by \autoref{pro1:rl:itm:ii} we get $u' \in \truthset{\modlts_\model}{\varphi}$. By IH, $u' \in \truthset{\model}{\varphi}$, which implies that $u \in \truthset{\model}{\varphi}$ (by $\model,u\bisim\model,u'$, and \autoref{th:khbisim-to-khequiv}). Therefore $\R_{\strategy}(\truthset{\model}{\psi}) \subseteq \truthset{\model}{\varphi}$. From the two pieces, $w \in \truthset{\model}{\kh(\psi,\varphi)}$.  
    \end{itemize}  

This finishes the proof. 
  \end{demostracion}
\end{proposicion}

Now we will prove the other direction. From an \lts we can obtain an active, $\stexec$-compositional and point-wise equivalent $\ults$. 

\begin{proposicion}\label{prop:ltstoesm}
  Let $\modlts=\tup{\W,\R,\V}$ be an \lts over $\ACT$. Take $\ACT' := \csetsc{a_{\sigma}}{\sigma \in \ACT^* \;\text{and}\; \stexec(\sigma) \neq \emptyset }$, and then define the \ults $\model_{\modlts} = \tup{\W,\R', \S', \V}$ over $\ACT'$ by taking $\R'_{a_{\sigma}} := \csetsc{(w,v) \in \R_{\sigma}}{w \in \stexec(\sigma)}$ (so basic actions in $\model_{\modlts}$ are strongly executable plans in $\modlts$) and $\S':=\csetsc{\cset{a_\sigma}}{a_\sigma \in \ACT'}$. Then,
  \begin{enumerate}
    \item\label{itm:ltstoesm:1} $\model_{\modlts}$ is an active and $\stexec$-compositional $\ults$ (i.e., is in $\cultsac$);
    \item\label{itm:ltstoesm:2} for every $\varphi \in \KHlogic$, 
    $\truthset{\modlts}{\varphi} = \truthset{\model_{\modlts}}{\varphi}$.
  \end{enumerate}
\begin{demostracion}
  First, \autoref{itm:ltstoesm:1}. For showing that $\model_{\modlts}$ is an $\ults$, note how $\DS{}' = \bigcup_{\strategy \in \S'} \strategy$ is non-empty ($\epsilon \in \ACT^*$ and $\stexec(\epsilon) = \W$, so $\epsilon \in \ACT'$ and hence $\cset{a_{\epsilon}} \in \S'$) and, moreover, $\S'$ does not contain the empty set and its elements are pairwise disjoint (the latter two by definition). Moreover, we can map elements from $\ACT'$ into $\ACT$ to preserve the same signature (as their cardinalities match). Activeness is straightforward, as $\cset{a_\epsilon}$ is in $\S'$ and behaves exactly as $\epsilon$. 
  
  For $\stexec$-compositionality, take $\cset{a_{\sigma_1}}, \cset{a_{\sigma_2}}\in\S'$ s.t. $\cset{a_{\sigma_1}}\compose \cset{a_{\sigma_2}}\neq\emptyset$.
  Then, $\cset{a_{\sigma_1}}\compose \cset{a_{\sigma_2}}=\cset{a_{\sigma_1}a_{\sigma_2}}$ and, moreover, $\stexec(\cset{a_{\sigma_1}}) \neq \emptyset$ and $\R'_{\cset{a_{\sigma_1}}}(\stexec(\cset{a_{\sigma_1}})) \subseteq \stexec(\cset{a_{\sigma_2}})$ ($\bm{\otimes_1}$). We need to provide a $\strategy \in \S'$ satisfying the $\stexec$-compositionality conditions; it will be shown that $\strategy=\cset{a_{\sigma_1\sigma_2}}$ does the work. In doing so, it is useful to notice that $\R'_{a_{\sigma_1}a_{\sigma_2}} = \R'_{a_{\sigma_1\sigma_2}}$ (the proof is straightforward).

  First, we need to show that $\strategy=\cset{a_{\sigma_1\sigma_2}}$ is in $\S$, which boils down to showing that $a_{\sigma_1\sigma_2} \in \ACT'$, that is, $\stexec(\sigma_1\sigma_2)\neq\emptyset$. For this, recall that $\stexec(\cset{a_{\sigma_1}}) \neq \emptyset$, so we know that $\cset{a_{\sigma_1}}$ is strongly executable at some $u \in \W$. Moreover, from $\R'_{\cset{a_{\sigma_1}}}(\stexec(\cset{a_{\sigma_1}})) \subseteq \stexec(\cset{a_{\sigma_2}})$ it follows that any such execution ends in states where $a_{\sigma_2}$ is strongly executable. Then, $a_{\sigma_1}a_{\sigma_2}$ is strongly executable at $u$, which implies $\R'_{a_{\sigma_1}a_{\sigma_2}}(u) \neq \emptyset$. But $\R'_{a_{\sigma_1}a_{\sigma_2}} = \R'_{a_{\sigma_1\sigma_2}}$, so $\R'_{a_{\sigma_1\sigma_2}}(u) \neq \emptyset$, which by definition of $\R'$ implies $u \in \stexec(\sigma_1\sigma_2)$, that is, $\stexec(\sigma_1\sigma_2) \neq \emptyset$, as required.

  Then, the $\stexec$-compositionality conditions. The first and the third, $\R'_{a_{\sigma_1}a_{\sigma_2}}\subseteq\R'_{a_{\sigma_1\sigma_2}}$ and the bisimilarity one, follow from $\R'_{a_{\sigma_1}a_{\sigma_2}} = \R'_{a_{\sigma_1\sigma_2}}$. For the second, $\stexec(a_{\sigma_1}a_{\sigma_2}) \subseteq \stexec(a_{\sigma_1\sigma_2})$, take $u \in \stexec(a_{\sigma_1}a_{\sigma_2})$; we need to show that $u \in \stexec(a_{\sigma_1\sigma_2})$. For this, it is enough to show that $\R'_{a_{\sigma_1\sigma_2}}(u) \neq \emptyset$ (as $a_{\sigma_1\sigma_2}$ is a basic action), i.e., that $\R_{\sigma_1\sigma_2}(u) \neq \emptyset$ (which implies $a_{\sigma_1\sigma_2}$ exists) and $u \in \stexec(\sigma_1\sigma_2)$. Now, the assumption $u \in \stexec(a_{\sigma_1}a_{\sigma_2})$ implies $u \in \stexec(a_{\sigma_1})$ and $\R'_{a_{\sigma_1}}(u) \subseteq \stexec(a_{\sigma_2})$. The first implies not only $\R'_{a_{\sigma_1}}(u) \neq \emptyset$ (so $u \in \stexec(\sigma_1)$) but also $\R'_{a_{\sigma_1}}(u) = \R_{\sigma_1}(u)$. From the second and the latter, $\R_{\sigma_1}(u) \subseteq \stexec(a_{\sigma_2})$. But note: $v \in \stexec(a_{\sigma_2})$ implies there is $v' \in \R'_{a_{\sigma_2}}(v)$, so $v \in \stexec(\sigma_2)$. Thus, $\stexec(a_{\sigma_2}) \subseteq \stexec(\sigma_2)$ and hence $\R_{\sigma_1}(u) \subseteq \stexec(\sigma_2)$. Then, the now latter and $u \in \stexec(\sigma_1)$ imply $u \in \stexec(\sigma_1\sigma_2)$ (the second goal) and thus, by definition of $\stexec$, it follows that $\R_{\sigma_1\sigma_2}(u) \neq \emptyset$ (the first goal).
  
  For \autoref{itm:ltstoesm:2}, the proof is by structural induction; again, only the case of $\kh(\psi,\varphi)$ is discussed.
    
  \ssparagraph{$\bm{(\subseteq)}$} Suppose $w \in \truthset{\modlts}{\kh(\psi,\varphi)}$; then there is $\sigma \in \ACT^*$ satisfying both
  \begin{multicols}{2}
    \begin{cond-kh}
       \item\label{pro2:lr:itm:i} $\truthset{\modlts}{\psi} \subseteq \stexec(\sigma)$ and 
       \item\label{pro2:lr:itm:ii} $\R_\sigma(\truthset{\modlts}{\psi}) \subseteq \truthset{\modlts}{\varphi}$.
    \end{cond-kh}
  \end{multicols}
  There are two cases. First, assume $\stexec(\sigma)=\emptyset$. From this, \autoref{pro2:lr:itm:i} implies $\truthset{\modlts}{\psi} = \emptyset$, so $\truthset{\model_\modlts}{\psi} = \emptyset$ (by IH) and $\R'_{\strategy}(\truthset{\model_\modlts}{\psi}) = \emptyset$ for any $\strategy \in 2^{(\ACT')^*}$. Hence, to obtain $w \in \truthset{\model_{\modlts}}{\kh(\psi,\varphi)}$ it is enough to have $\S' \neq \emptyset$, which we do as $\cset{a_{\epsilon}} \in \S'$.
  
  Second, assume $\stexec(\sigma) \neq \emptyset$. Then, $a_\sigma \in \ACT'$ and $\cset{a_\sigma} \in \S'$; this will be shown to be our witness. For the first $\kh$-clause, if $u \in \truthset{\model_{\modlts}}{\psi}$ then $u \in \truthset{\modlts}{\psi}$ (IH), so $u \in \stexec(\sigma)$ (\autoref{pro2:lr:itm:i}), which implies $\R_{\sigma}(u) \neq \emptyset$. The last two together imply $\R'_{a_{\sigma}}(u) \neq \emptyset$ (definition of $\R'$), so $u \in \stexec(a_{\sigma}) = \stexec(\cset{a_{\sigma}})$. Hence, $\truthset{\model_{\modlts}}{\psi} \subseteq \stexec(\cset{a_{\sigma}})$. For the second $\kh$-clause, suppose  $u \in \R'_{\cset{a_{\sigma}}}(\truthset{\model_{\modlts}}{\psi})$. Then, $u \in \R'_{a_\sigma}(\truthset{\model_\modlts}{\psi})$ so $u \in \R'_{a_\sigma}(\truthset{\modlts}{\psi})$ (IH) and then $u \in \R_{\sigma}(\truthset{\modlts}{\psi})$ (definition of $\R'$) so, by \autoref{pro2:lr:itm:ii}, $u \in \truthset{\modlts}{\varphi}$ and thus $u \in \truthset{\model_{\modlts}}{\varphi}$ (IH). Consequently, $\R'_{\strategy}(\truthset{\model_{\modlts}}{\psi}) \subseteq \truthset{\model_{\modlts}}{\varphi}$. From the two clauses, $w \in \truthset{\model_{\modlts}}{\kh(\psi,\varphi)}$.

  \ssparagraph{$\bm{(\supseteq)}$} Suppose $w \in \truthset{\model_{\modlts}}{\kh(\psi,\varphi)}$. Then there is an element of $\S'$ fulfilling the $\kh$-clauses, which by definition of $\S'$ implies there is $\cset{a_{\sigma}} \in \S'$ (with $\sigma \in \ACT^*$) satisfying both
  \begin{multicols}{2}
    \begin{cond-kh}
      \item\label{pro2:rl:itm:i} $\truthset{\model_{\modlts}}{\psi} \subseteq \stexec(\cset{a_{\sigma}})$ and
      
      \item\label{pro2:rl:itm:ii} $\R'_{\cset{a_{\sigma}}}(\truthset{\model_{\modlts}}{\psi}) \subseteq \truthset{\model_{\modlts}}{\varphi}$.
    \end{cond-kh}
  \end{multicols}
  {\noindent}It will be shown that $\sigma$ is our witness. For the first $\kh$-clause, $u \in \truthset{\modlts}{\psi}$ implies $u \in \truthset{\model_{\modlts}}{\psi}$ (IH), hence $u \in \stexec(\cset{a_\sigma})$ (\autoref{pro2:rl:itm:i}) and then $\R'_{a_{\sigma}}(u) \neq \emptyset$, which implies $u \in \stexec(\sigma)$ (definition of $\R'$). Thus, $\truthset{\modlts}{\psi} \subseteq \stexec(\sigma)$. For the second $\kh$-clause, take $u \in \R_{\sigma}(\truthset{\modlts}{\psi})$, so $u \in \R_{\sigma}(\truthset{\modults_\modlts}{\psi})$ (IH). Then $u \in \R_{\sigma}(v)$ for some $v \in \truthset{\modults_\modlts}{\psi}$ By \autoref{pro2:rl:itm:i}, $v \in \stexec(\cset{a_{\sigma}})$, i.e., $v \in \stexec(a_{\sigma})$ so $\R'_{a_\sigma}(v) \neq \emptyset$ and then $v \in \stexec(\sigma)$ (definition of $\R'$). This, together with $u \in \R_{\sigma}(v)$ imply $u \in \R'_{a_\sigma}(v)$, i.e., $u \in \R'_{\cset{a_\sigma}}(v)$. Hence, $u \in \R'_{a_\sigma}(\truthset{\modults_\modlts}{\psi})$, so $u \in \truthset{\model_{\modlts}}{\varphi}$ (\autoref{pro2:rl:itm:ii}) and then $u \in \truthset{\modlts}{\varphi}$ (IH). Thus, $\R_{\sigma}(\truthset{\modlts}{\psi}) \subseteq \truthset{\modlts}{\varphi}$.
  From the two clauses, $w \in \truthset{\modlts}{\kh(\psi,\varphi)}$.
\end{demostracion}

\end{proposicion}
 
From these results, the axiom system for $\KHlogic$ over \lts (\autoref{tab:khaxiom}) is also sound and complete for $\KHlogic$ over \emph{active} and \emph{$\stexec$-compositional} \ultss.

\begin{teorema}
\label{th:kh-complete}
The axiom system $\KHaxiom$ (\autoref{tab:khaxiom}) is sound and strongly complete w.r.t. the class $\cultsac$. % of active and $\stexec$-compositional \ultss.
\begin{demostracion}
  The arguments are exactly as in~\autoref{teo:kh-sound-complete-cults}, by using this time~\autoref{prop:esmtolts} and~\autoref{prop:ltstoesm}.
% For soundness, let $\model,w$ be an active and $\stexec$-compositional pointed \ults. 
% By \autoref{prop:esmtolts}, there exists an \lts $\modlts_{\model}$
% s.t. $\model,w\models\varphi$ iff $\modlts_\model,w\models\varphi$,
% for all $\varphi$.  Since $\modlts_\model$ is an \lts, and every axiom
% $\psi$ in $\KHaxiom$ holds in every \lts,
% $\modlts_\model,w\models\psi$. Hence $\model,w\models\psi$, for all
% $\psi$ in $\KHaxiom$, i.e., axioms and rules are sound over \ultss.

% To prove that $\KHaxiom$ is strongly complete over the class of active and $\stexec$-compositional \ults we need to show that, given $\Gamma\cup\cset{\varphi}$ a set of formulas in $\KHlogic$, $\Gamma\models\varphi$ implies $\Gamma\vdash\varphi$. 
%   Let $\Gamma$ be a consistent set of formulas. As in~\cite[Lemma
%   1]{Wang2016}, $\Gamma$ can be extended to an MCS $\Gamma'$, and as a
%   consequence, there exists an LTS $\modlts^{\Gamma'}$ such that
%   $\modlts^{\Gamma'},\Gamma'\models\Gamma$ (notice that states in the
%   canonical model are MCS). Then, by \autoref{prop:ltstoesm}, we can
%   obtain an \ults $\model_{\modlts^{\Gamma'}}$, such
%   that $\model_{\modlts^{\Gamma'}},\Gamma'\models\Gamma$. Moreover,
%   from \autoref{prop:ltstoesm} we know that $\model_{\modlts^{\Gamma'}}$ is
%   active and $\stexec$-compositional.
\end{demostracion}
\end{teorema}

%%% Local Variables:
%%% mode: latex
%%% TeX-master: "main"
%%% End:

%\input{modelequiv.tex}

%------------------------------------------------------------------------------------------------
\section{Finite model property and complexity}\label{sec:complexity}

%!TEX root = main.tex

This section is devoted to the study of the computational complexity of the logic \KHilogic over \ultss. To do so, we will use two standard tools from modal logic: filtration and selection (see, e.g.,~\cite{mlbook} for details). First, we define a notion of filtration that, given an arbitrary model and a formula, allows us to obtain a finite model that satisfies the formula if and only if the original model satisfies it. This proves that the satisfiability problem for \KHilogic is decidable.  Then, we define a (more specialized) selection function which, from a canonical model, enables us to extract a polynomial-size model. Thus, we show that  the satisfiability problem for \KHilogic is \NP-complete (given that we provide a model checking algorithm running in \Poly).

\subsection{Finite model property via filtrations}\label{subsec:filtrations}
%!TEX root = main.tex

We start by introducing two relations that will be crucial to define a proper notion of filtration, given a set of formulas $\Sigma$ and a model $\model$. 
% Notice that, in what follws, we call a set of plans $\strategy$ a `witness' for a formula $\khi(\psi,\varphi)$ if $\strategy$~makes the satisfiability conditions of~\autoref{def:sem-esm} true for $\khi(\psi,\varphi)$.

\begin{definicion}[$\Sigma$-equivalence]
Let $\model=\tup{\W,\R, \{\S_i\}_{i\in \AGT}, \V}$ be an \ults and let $\Sigma$ be a set of $\KHilogic$-formulas closed under subformulas. Define the relations ${\modequiv_\Sigma} \subseteq \W\times\W$ and ${\planequiv_\Sigma} \subseteq \S_\AGT \times \S_\AGT$ (with $\S_\AGT := \bigcup_{i\in\AGT}\S_i$) as:
\begin{nscenter}
\begin{tabular}{lcl}
$w \modequiv_\Sigma v$ &  \iffdef &
for all $\psi\in\Sigma$, $\model,w\models \psi$ iff $\model,v\models \psi$,\\
$\strategy \planequiv_\Sigma \strategy'$ & \iffdef &
for all $i\in\AGT$ and $\khi(\psi,\varphi)\in\Sigma$,  $\strategy$ is a witness \\ 
& & for $\khi(\psi,\varphi)$ in $\model$ iff $\strategy'$ is a witness for $\khi(\psi,\varphi)$ in $\model$.
\end{tabular}
\end{nscenter}

Notice that $\modequiv_\Sigma$ (a generalisation of $\modequiv$ in \autoref{def:equiv-khi} to a given set of formulas) and $\planequiv_\Sigma$ are equivalence relations over $\W$ and $\S_\AGT$, respectively. For $w \in \W$ (resp., $\strategy\in2^{(\ACT^*)}$), we use $[w]_\Sigma$ (resp., $[\strategy]_\Sigma$) to denote $w$'s (resp., $\strategy$'s) $\Sigma$-equivalence class;  i.e., 
\begin{nscenter}
$[w]_\Sigma := \csetc{v}{\W}{w \modequiv_\Sigma v}$; \qquad  $[\strategy]_\Sigma := \csetc{\strategy'}{2^{(\ACT^*)}}{\strategy \planequiv_\Sigma \strategy'}$.
\end{nscenter}
% Finally, denote $\W_{/\modequiv_\Sigma}$ and $(\S_i)_{/\planequiv_\Sigma}$ are the quotients of $\W$ via $\modequiv_\Sigma$, and of $\S_i$ via $\planequiv_\Sigma$, respectively.
\end{definicion}

Although the notation $[\_]_\Sigma$ is overloaded, its argument will always disambiguate its use.

\begin{definicion} %[Filtration preliminaries]
Let $\model = \tup{\W,\R, \{\S_i\}_{i\in \AGT}, \V}$ be an \ults and let $\Sigma$ be a set of $\KHilogic$-formulas that is closed under subformulas. For $i\in\AGT$ define $\ACT^{\Sigma}_i := \cset{a_{[\strategy]_\Sigma} \mid \strategy\in\S_i \text{ is a witness of some } \khi(\psi,\varphi)\in\Sigma \text{ in } \model}$;
and  $\ACT^\Sigma := \bigcup_{i\in\AGT} \ACT^{\Sigma}_i$.
%\begin{itemize}
%\item $\W_\Sigma := \csetsc{[w]_\Sigma}{w \in \W}$,
%\item $\AGT := \csetc{i}{\AGT}{\khi(\psi,\varphi)\in\Sigma}$,
%\item $\PROP' := \csetc{p}{\PROP}{p \in \Sigma}$,
%\item for each $i\in\AGT$ and $\strategy\in\S_i$,

%\quad$[\strategy,i]$ := \cset{$\strategy' \mid$ for each $\khi(\psi,\varphi)\in\Sigma$, $\strategy$ is a witness of $\khi(\psi,\varphi)$ iff $\strategy'$ is a witness of $\khi(\psi,\varphi)$},

%\item $\ACT^{\Sigma}= \cset{a^i_{[\strategy]_\Sigma} \mid \strategy \text{ is a witness of some } \khi(\psi,\varphi)\in\Sigma \text{ and }, i\in\AGT}$.
%\end{itemize}
%If there are no witness for a certain $\khi(\psi,\varphi)$, it is not added.
\end{definicion}

The idea behind the definition of $\ACT^{\Sigma}$ is that, for each $\khi(\psi,\varphi)\in\Sigma$ that is true at $\model$, we consider an action mimicking the behaviour of those sets of plans $\strategy$ that witness the satisfiability of $\khi(\psi,\varphi)$ in $\model$. 

Now, we are in position of defining the notion of filtration. 

\begin{definicion}[Filtration of $\model$ through $\Sigma$] \label{def:filtrations}
Let $\model = \tup{\W,\R, \{\S_i\}_{i\in \AGT}, \V}$ be an \ults;  let $\Sigma$ be a set of $\KHilogic$-formulas that is closed under subformulas. An \ults $\model^f = \tup{\W^f,\R^f, \{\S^f_i\}_{i\in \AGT}, \V^f}$ is \emph{a filtration of $\model$ through $\Sigma$} if and only if it satisfies the following conditions:

\begin{enumerate}
\item $\W^f := \csetsc{[w]_\Sigma}{w \in \W}$;
\item $\V^f([w]_\Sigma) := \csetc{p}{\Sigma}{\model,w\models p}$;
\item for all $i\in\AGT$,  $a_{[\strategy]_{\Sigma}}\in\ACT^{\Sigma}_i$ implies $\cset{a_{[\strategy]_{\Sigma}}}\in\S^f_i$;
\item  $\S_i^f$ is finite and well-defined (as per~\autoref{rem:corresp}), and each $\strategy\in\S^f_i$ is finite;
\item \label{item:5} for all $\strategy\in\S^f_i$, if $\strategy$ is a witness of some $\khi(\psi,\varphi)\in\Sigma$ in $\model^f$, then there is $\strategy'\in\S_i$ such that $\strategy'$ is a witness of $\khi(\psi,\varphi)$ in $\model$;
\item \label{item:6} if $(w,v)\in\R_\strategy$ and $a_{[\strategy]_\Sigma}\in\ACT^\Sigma_i$, then $([w]_\Sigma,[v]_\Sigma) \in \R^f_{a_{[\strategy]_\Sigma}}$;
\item \label{item:7} if $([w]_\Sigma,[v]_\Sigma) \in \R^f_{a_{[\strategy]_\Sigma}}$, and $\strategy$ is a witness of some $\khi(\psi,\varphi)\in\Sigma$ in $\model$, then $w\in\truthset{\model}{\psi}$ implies $v\in\truthset{\model}{\varphi}$.
% \item $\R^f_{a_{(\psi,\varphi),[\strategy,i]}}([w]_\Sigma,[v]_\Sigma)$ iff
% \begin{itemize}
% \item for each $\strategy'\in[\strategy,i]$ and $w'\in[w]_\Sigma$, $w\in\stexec(\strategy')$, and
% \item there are $\strategy'\in[\strategy,i]$, $w'\in[w]_\Sigma$ and $v'\in[v]_\Sigma$ such that $\R_{\strategy'}(w',v')$
% \end{itemize}
\end{enumerate}
\end{definicion}

Note that $\V^f$ is well-defined: given $p \in \Sigma$, if $[w]_\Sigma = [v]_\Sigma$ and $\model,w\models p$, then $\model,v\models p$. Also, as $\S^f_i$ is well-defined (by definition), we have that $\model^f$ is an \ults over $\ACT^\Sigma$, $\PROP$ and $\AGT$.   Also, note that if \ref{item:5}, \ref{item:6} and \ref{item:7} above are turned into if and only if conditions, they always define an \ults which is a filtration.

\autoref{def:filtrations} deserves further comments. Notice that, for the \lts part, the filtration is defined similarly as for the basic modal logic (see, e.g.,~\cite{mlbook}). The most significant difference is the change in the labelling of the relations, since we now use $\ACT^\Sigma$ as the set of action names, instead of $\ACT$. But this has a consequence on the definition of $\S^f_i$. The relation $\planequiv_\Sigma$ enables us to obtain a finite set of witnesses for the formulas $\khi(\psi,\varphi)\in\Sigma$, from which we also get that $\ACT^\Sigma$ and (together with $\modequiv_\Sigma$) $\W^f$ are finite. However, the new set $\S^f_i$ is defined in terms of a new set of action names, so there are potentially infinite new available plans to consider. Thus, we need to state that $\S^f_i$ is any finite set, satisfying the minimum and maximum conditions, whose members are also finite, and that is well-defined. Finally, $\ACT^\Sigma$ may be finite, unlike the original set of actions $\ACT$ which is infinite by definition. However, this poses no problem in the construction, as $\ACT^\Sigma$ can be extended to an infinite set, without breaking the finiteness of the filtration (recall that the accessibility relation is defined over a, potentially finite, subset of the set of actions).

% \begin{proposicion}
% Let $\Sigma$ be a set of $\KHilogic$-formulas that is closed under subformulas, let $\model$ be an \ults with $\model^f$ its filtration through $\Sigma$. 
% %Then, $card(\W^f)\leq 2^{card(\Sigma)}$, $\ACT^{\Sigma}$ is polynomial in the number of $\khi(\psi,\varphi)$ in $\Sigma$ and $\S^f_i$ is finite. Thus, 
% Then, $\model^f$ is a finite model.
% \begin{demostracion}
% First, note that $card(\W^f)\leq 2^{card(\Sigma)}$. By definition, for all $i\in\AGT$, $\ACT^{\Sigma}_i$ is at most the number of $\khi(\psi,\varphi)\in\Sigma$, since if there are two groups of witnesses $[\strategy]_\Sigma$ and $[\strategy']_\Sigma$ for some $\khi(\psi,\varphi)$, $[\strategy]_\Sigma=[\strategy']_\Sigma$. Hence, $\ACT^{\Sigma}$ is polynomial in the number of $\khi(\psi,\varphi)$ in $\Sigma$. By definition, $\S^f_i$ is finite. Thus, $\model^f$ is finite.
% \end{demostracion}
% \end{proposicion}

\begin{teorema}
Let $\model = \tup{\W,\R, \{\S_i\}_{i\in \AGT}, \V}$ be an \ults and
let $\Sigma$ be a set of $\KHilogic$-formulas that is closed under subformulas. Then, for all $\psi\in\Sigma$ and $w\in\W$, $\model,w\models \psi$ iff $\model^f,[w]_\Sigma\models \psi$. Moreover, if $\Sigma$ is finite then $\model^f$ is a finite model.
\begin{demostracion}
Boolean cases work as expected. So, we will only show that  $\model,w\models \khi(\psi,\varphi)$ iff $\model^f,[w]_\Sigma\models \khi(\psi,\varphi)$.

{\prooflr} Suppose that $\model \models \khi(\psi,\varphi)$: let $\strategy \in \S_i$ be such that $\truthset{\model}{\psi} \subseteq \stexec(\strategy)$ and $\R_{\strategy}(\truthset{\model}{\psi}) \subseteq \truthset{\model}{\varphi}$.
By definition, $a_{[\strategy]_\Sigma} \in \ACT^{\Sigma}_i$ and therefore, $\cset{a_{[\strategy]_\Sigma}} \in \S^f_i$. If $\truthset{\model^f}{\psi}=\emptyset$, the result trivially follows. Otherwise, let $[w]_\Sigma \in \truthset{\model^f}{\psi}$.
By IH, $w \in \truthset{\model}{\psi}$, and since $\strategy$ is SE at $w$, $\R_{\strategy}(w) \neq \emptyset$.
Since $\model^f$ is a filtration, we have that $\R_{a_{[\strategy]_\Sigma}}(w) \neq \emptyset$ and $\cset{a_{[\strategy]_\Sigma}}$ is SE at $[w]_\Sigma$. Thus, $\truthset{\model^f}{\psi} \subseteq \stexec(\cset{a_{[\strategy]_\Sigma}})$.

Let $([w]_\Sigma,[v]_\Sigma) \in \R^f_{a_{[\strategy]_\Sigma}}$ be such that $[w]_\Sigma \in \truthset{\model^f}{\psi}$. By IH, $w \in \truthset{\model}{\psi}$. Since $\strategy$ is a witness of $\khi(\psi,\varphi) \in \Sigma$ in $\model$ (by assumption), by the definition of $\model^f$ we get $v \in \truthset{\model}{\varphi}$. Again, by IH, $[v]_\Sigma \in \truthset{\model}{\varphi}$.
Thus, $\R^f_{\cset{a_{[\strategy]_\Sigma}}}(\truthset{\model}{\psi}) \subseteq \truthset{\model}{\varphi}$. Therefore, $\model^f \models \khi(\psi,\varphi)$.

{\proofrl} Suppose that $\model^f \models \khi(\psi,\varphi)$: let $\strategy \in \S^f_i$  be such that $\truthset{\model^f}{\psi} \subseteq \stexec(\strategy)$ and $\R_{\strategy}(\truthset{\model^f}{\psi}) \subseteq \truthset{\model^f}{\varphi}$.
By definition of $\model^f$, since $\strategy$ is a witness of $\khi(\psi,\varphi) \in \Sigma$ in $\model^f$, we have that there is $\strategy'\in\S_i$ such that $\strategy'$ is a witness of $\khi(\psi,\varphi)$ in $\model$.
Thus, $\model \models \khi(\psi,\varphi)$.

\smallskip

It remains to show that $\model^f$ is finite. First, note that the number of elements in $\W^f$ is $2^m$, with $m$ being the number of formulas in $\Sigma$. By definition, for all $i\in\AGT$, $\ACT^{\Sigma}_i$ is at most the number of $\khi(\psi,\varphi)\in\Sigma$, since if there are two groups of witnesses $[\strategy]_\Sigma$ and $[\strategy']_\Sigma$ for some $\khi(\psi,\varphi)$, $[\strategy]_\Sigma=[\strategy']_\Sigma$. Hence, $\ACT^{\Sigma}$ is polynomial in the number of $\khi(\psi,\varphi)$ in $\Sigma$. Finally, by definition, $\S^f_i$ is finite. Thus, $\model^f$ is finite.
\end{demostracion}
\end{teorema}

The last theorem states that every satisfiable formula of \KHilogic, is satisfiable in a finite model. As a consequence, the satisfiability problem for \KHilogic is decidable. In the next section we will refine this result and provide exact complexity bounds.

\subsection{Complexity via selection}\label{subsec:complexity}
%!TEX root = main.tex

Here we investigate the computational complexity of the satisfiability
problem of $\KHilogic$ under the $\ults$-based semantics.  We will
establish membership in \NP by showing a polynomial-size model property.

Given a formula, we will show that it is possible to select just a
piece of the canonical model which is relevant for its evaluation. The
selected model will preserve satisfiability, and moreover, its size
will be polynomial w.r.t.\ the size of the input formula. 

\begin{definicion}[Selection function]\label{def:selection-function}
  Let
  $\cmodel=\tup{\W^\Gamma,\R^\Gamma,\cset{\S_i^\Gamma}_{i\in\AGT},\V^\Gamma}$
  be a canonical model for an MCS $\Gamma$ (see
  \autoref{def:cm-ults-lkhi}); take $w\in\W^\Gamma$ and a formula
  $\varphi\in\KHilogic$. Define
  $\ACT_\varphi := \cset{\tup{\theta_1,\theta_2} \in\ACT^\Gamma \mid
    \kh_i(\theta_1,\theta_2) \text{ is a subformula of } \varphi}$.  A
  \emph{canonical selection function} $\sel^{\varphi}_{w}$ is a
  function that takes $\cmodel$, $w$ and $\varphi$ as input, returns a
  set $\W'\subseteq \W^\Gamma$, and is such that:
    
  \begin{enumerate} 
  \item \label{def:s-f1} $\sel^{\varphi}_{w}(p)=\{w\}$; % \quad \quad
    %$\sel^{\varphi}_{w}(\neg\varphi_1)=\sel^\varphi_w(\varphi_1)$;
    %\quad \quad
    %$\sel^{\varphi}_{w}(\varphi_1\vee\varphi_2)=
    %\sel^{\varphi}_{w}(\varphi_1)\cup\sel^{\varphi}_{w}(\varphi_2)$;
    \item \label{def:s-f2} $\sel^{\varphi}_{w}(\neg\varphi_1)=\sel^\varphi_w(\varphi_1)$
    \item \label{def:s-f3} $\sel^{\varphi}_{w}(\varphi_1\vee\varphi_2)=
         \sel^{\varphi}_{w}(\varphi_1)\cup\sel^{\varphi}_{w}(\varphi_2)$;
    
  \item \label{def:s-f4} If
    $\truthset{\cmodel}{\kh_i(\varphi_1,\varphi_2)}\neq\emptyset
    \text{ and } \truthset{\cmodel}{\varphi_1}=\emptyset$:
    $\sel^{\varphi}_{w}(\kh_i(\varphi_1,\varphi_2)) = \{w\}$;
    
  \item \label{def:s-f5} If
    $\truthset{\cmodel}{\kh_i(\varphi_1,\varphi_2)}\neq\emptyset
    \text{ and } \truthset{\cmodel}{\varphi_1}\neq\emptyset$:
        
    $\sel^{\varphi}_{w}(\kh_i(\varphi_1,\varphi_2)) = \{w_1,w_2\} \cup
    \sel^{\varphi}_{w_1}(\varphi_1) \cup
    \sel^{\varphi}_{w_2}(\varphi_2)$,
    \noindent where $w_1$, $w_2$ are
    s.t. $(w_1,w_2)\in\R^\Gamma_{\tup{\varphi_1,\varphi_2}}$;
    
  \item \label{def:s-f6} If
    $\truthset{\cmodel}{\kh_i(\varphi_1,\varphi_2)}=\emptyset$ (note
    that $\truthset{\cmodel}{\varphi_1}\neq\emptyset$):
        
    For all set of plans $\strategy$, either
    $\truthset{\cmodel}{\varphi_1} \not\subseteq \stexec(\strategy)$
    or
    $\R_\strategy^\Gamma(\truthset{\cmodel}{\varphi_1}) \not\subseteq
    \truthset{\cmodel}{\varphi_2}$. For each $a \in \ACT_\varphi$:

    \begin{enumerate}
    \item if
      $\truthset{\cmodel}{\varphi_1} \not\subseteq \stexec(\cset{a})$:
      we add $\{w_1\} \cup \sel^{\varphi}_{w_1}(\varphi_1)$ to
      $\sel^{\varphi}_{w}(\kh_i(\varphi_1,\varphi_2))$, where
      $w_1 \in\truthset{\cmodel}{\varphi_1}$ and
      $w_1 \notin \stexec(\cset{a})$;
        
    \item if
      $\R_\strategy^\Gamma(\truthset{\cmodel}{\varphi_1})
      \not\subseteq \truthset{\cmodel}{\varphi_2}$ we add
      $\{w_1,w_2\} \cup \sel^{\varphi}_{w_1}(\varphi_1) \cup
      \sel^{\varphi}_{w_2}(\varphi_2)$ to
      $\sel^{\varphi}_{w}(\kh_i(\varphi_1,\varphi_2))$, where
      $w_1\in\truthset{\cmodel}{\varphi_1}$,
      $w_2 \in \R^\Gamma_a(w_1)$ and
      $w_2 \notin \truthset{\cmodel}{\varphi_2}$.
    \end{enumerate}
  \end{enumerate}
\end{definicion}
    
We can now select a small model which preserves the
satisfiability of a given formula.

\begin{definicion}[Selected model]
  Let $\cmodel$ be the canonical model for an MCS $\Gamma$, $w$ a
  state in $\cmodel$, and $\varphi$ an $\KHilogic$-formula. Let
  $\sel^\varphi_w$ be a selection function, we define the
  \emph{model selected by} $\sel^\varphi_w$ as
  $\model^\varphi_w=\tup{\W^\varphi_w,\R^\varphi_w,\cset{(\S^\varphi_w)_i}_{i\in\AGT},\V^\varphi_w}$,
  where

    \begin{compactitemize}
      \item $\W^\varphi_w := \sel^\varphi_w(\varphi)$; 
      \item $(\R^\varphi_w)_{\tup{\theta_1,\theta_2}} := \R^\Gamma_{\tup{\theta_1,\theta_2}}\cap(\W^\varphi_w)^2$ for each $\tup{\theta_1,\theta_2} \in \ACT_\varphi$;
      \item $(\S^\varphi_w)_i := \cset{\cset{a} \ |\ a \in \ACT_\varphi} \cup \cset{\cset{\tup{\bot,\top}}}$, for $i\in\AGT$
        (and $(\R^\varphi_w)_{\tup{\bot,\top}} := \emptyset$);
        \item $\V^\varphi_w$ is the restriction of $\V^\Gamma$ to $\W^\varphi_w$.
     % \item $\V^\varphi_w$ is the restriction of $\V^\Gamma$ to $\W^\varphi_w$.
    \end{compactitemize}
    \end{definicion}
    
    Note that, although $\ACT_\varphi$ can be an empty set, each collection of sets of plans $(\S^\varphi_w)_i$ is not. Moreover, $\ACT_\varphi$ can be extended to an infinite set of actions, to be defined over a proper signature. Therefore, $\model^\varphi_w$ is an $\ults$.
    
    \begin{proposicion}\label{prop:selection-preserves-sat}
    Let $\cmodel$ be a canonical model, $w$ a state in $\cmodel$ and $\varphi$ an $\KHilogic$-formula. Let $\model^\varphi_w$ be the selected model by a selection function $\sel^\varphi_w$.
    Then, $\cmodel,w\models\varphi$ implies that 
    for all $\psi$ subformula of $\varphi$, and for all $v\in\W^\varphi_w$ we have that 
    $\cmodel,v\models\psi$ iff $\model^\varphi_w,v\models\psi$.
    % Let $\sel^\varphi_w$ be a selection function for $\cmodel$, $w$ and $\varphi$. If $\cmodel,w\models \varphi$, iff the generated model by $\sel^{\varphi}_{w}(\varphi)$ satisfies $\varphi$. 
    Moreover, $\model^\varphi_w$ is polynomial on the size of $\varphi$.
    \begin{demostracion}

The proof proceeds by induction in the size of the formula:
\medskip

\noindent
Case $\psi=p$: if $\cmodel,v\models p$, then $p\in\V^\Gamma(v)$. Given that $v\in\W^\varphi_w$, we have $p\in\V^\varphi_w(v)$ and therefore $\model^\varphi_w,v\models p$. The other direction is similar.
\medskip

\noindent
Case $\psi=\neg\psi_1$: if $\cmodel,w\models \neg\psi_1$, then $\cmodel,w\not\models \psi_1$. By IH, $\model^\varphi_w,w\not\models \psi_1$ and therefore $\model^\varphi_w,w\models \neg\psi_1$. The other direction is similar.
\medskip

\noindent
Case $\psi=\psi_1\vee\psi_2$: if $\cmodel,v\models \psi_1\vee\psi_2$, then $\cmodel,v\models \psi_1$ or $\cmodel,v\models \psi_2$. By IH, $\model^\varphi_w,v\models \psi_1$ or $\model^\varphi_w,v\models \psi_2$ and therefore $\model^\varphi_w,v\models \psi_1\vee\psi_2$. The other direction is similar.
\medskip

\noindent
Case $\psi=\kh_i(\psi_1,\psi_2)$: 
Suppose that $\cmodel,v\models \kh_i(\psi_1,\psi_2)$. We consider two possibilities:

%
% \begin{compactitemize}
% \item \textbf{Case $\bm{\psi=\kh_i(\psi_1,\psi_2)}$:}
% \subitem $\cmodel,v\models \kh_i(\psi_1,\psi_2)$ iff
% there is a $\strategy \in \S^\Gamma_i$ s.t.
% \subitem \quad $\truthset{\cmodel}{\psi_1} \subseteq \stexec^{\cmodel}(\strategy)$ and
% $\R^\Gamma_\strategy(\truthset{\cmodel}{\psi_1}) \subseteq \truthset{\cmodel}{\psi_2}$
%
% \subitem $\model^\varphi_w,v\models \kh_i(\psi_1,\psi_2)$ iff
% there is a $\strategy' \in (\S^\varphi_w)_i$ s.t.
% \subitem \quad $\truthset{\model^\varphi_w}{\psi_1} \subseteq \stexec^{\model^\varphi_w}(\strategy')$ and
% $(\R^\varphi_w)_{\strategy'}(\truthset{\model^\varphi_w}{\psi_1}) \subseteq \truthset{\model^\varphi_w}{\psi_2}$
%% \end{compactitemize}
%

%       \item \textbf{Case $\boldsymbol{\khi(\psi,\varphi)}$.} {\prooflr} Consider two cases.
%       \begin{itemize}
%         \item $\bm{\truthset{\modults^\Gamma}{\psi} = \emptyset}$

\begin{itemize}
\item $\bm{\truthset{\cmodel}{\psi_1}=\emptyset}$. Since $\cmodel,v\models \kh_i(\psi_1,\psi_2)$ there is a $\strategy \in \S^\Gamma_i$ s.t. $\emptyset = \truthset{\cmodel}{\psi_1} \subseteq \stexec^{\cmodel}(\strategy)$ and $\emptyset = \R^\Gamma_\strategy(\truthset{\cmodel}{\psi_1}) \subseteq \truthset{\cmodel}{\psi_2}$. 
By IH $\truthset{\model^\varphi_w}{\psi_1} \subseteq \truthset{\cmodel}{\psi_1}$. Notice that, since $\truthset{\cmodel}{\psi_1} =\emptyset$, we also have $\truthset{\model^\varphi_w}{\psi_1}=\emptyset$. Let $\strategy'=\cset{\tup{\bot,\top}}$, we know that $\strategy'\in(\S^\varphi_w)_i$, and $(\R^\varphi_w)_{\strategy'}(\truthset{\model^\varphi_w}{\psi_1})=\emptyset$. So, there is a $\strategy' \in (\S^\varphi_w)_i$ s.t. $\truthset{\model^\varphi_w}{\psi_1} \subseteq \stexec^{\model^\varphi_w}(\strategy')$ and $(\R^\varphi_w)_{\strategy'}(\truthset{\model^\varphi_w}{\psi_1}) \subseteq \truthset{\model}{\psi_2}$. Therefore, $\model^\varphi_w,v\models \kh_i(\psi_1,\psi_2)$.

\item $\bm{\truthset{\cmodel}{\psi_1}\neq\emptyset}$: since 
$\cmodel,v\models \kh_i(\psi_1,\psi_2)$ 
there exists a $\strategy \in \S^\Gamma_i$ s.t.
$\truthset{\cmodel}{\psi_1} \subseteq \stexec^{\cmodel}(\strategy)$ and
$\R^\Gamma_\strategy(\truthset{\cmodel}{\psi_1}) \subseteq \truthset{\cmodel}{\psi_2}$. 
By Truth Lemma, $\kh_i(\psi_1,\psi_2) \in v$, then $\kh_i(\psi_1,\psi_2) \in \Gamma$ and $\tup{\psi_1,\psi_2} \in\ACT_\Gamma$.
By the definition of $\R^\Gamma_{\tup{\psi_1,\psi_2}}$,
we have that for all $w \in \truthset{\cmodel}{\psi_1}$, it holds that
$\R^\Gamma_{\tup{\psi_1,\psi_2}}(w)\neq\emptyset$
and $\R^\Gamma_{\tup{\psi_1,\psi_2}}(w) \subseteq \truthset{\cmodel}{\psi_2}$.
Thus, 
$\truthset{\cmodel}{\psi_1} \subseteq \stexec^{\cmodel}(\cset{\tup{\psi_1,\psi_2}})$ and $\R^\Gamma_{\tup{\psi_1,\psi_2}}(\truthset{\cmodel}{\psi_1}) \subseteq \truthset{\cmodel}{\psi_2}$. 
Since $\truthset{\cmodel}{\psi_1}\neq\emptyset$, 
there exist $w_1,w_2\in\W^\Gamma$ s.t.
$(w_1,w_2) \in\R^\Gamma_{\tup{\psi_1,\psi_2}}$.

Notice that by definition of $\model^\varphi_w$, we have that
$\cset{\tup{\psi_1,\psi_2}} \in (\S^\varphi_w)_i$ and  that $(\R^\varphi_w)_{\tup{\psi_1,\psi_2}}$ is defined. Also, by the definition of $\sel^{\varphi}_{w}$, \autoref{def:s-f5}, there exist $w'_1,w'_2\in\W^\varphi_w$ s.t. $(w'_1,w'_2)\in(\R^\varphi_w)_{\tup{\psi_1,\psi_2}}$.
Let $v_1\in \truthset{\model^\varphi_w}{\psi_1}$
$\subseteq \truthset{\model^\Gamma}{\psi_1}$ (the inclusion holds by IH). Then, we have $v_1 \in \stexec^{\cmodel}(\cset{\tup{\psi_1,\psi_2}})$ and $\R^\Gamma_{\tup{\psi_1,\psi_2}}(v_1) \subseteq \truthset{\cmodel}{\psi_2}$.
Since for all $v_2 \in \R^\Gamma_{\tup{\psi_1,\psi_2}}(v_1)$, we have $v_2\in\truthset{\cmodel}{\psi_2}$, (in particular $v_2=w'_2$), then $w'_2 \in(\R^\varphi_w)_{\tup{\psi_1,\psi_2}}(v_1)$.
Thus, $v_1 \in \stexec^{\model^\varphi_w}(\cset{\tup{\psi_1,\psi_2}})$. 

Aiming for a contradiction, suppose now that $(\R^\varphi_w)_{\tup{\psi_1,\psi_2}}(v_1) = \R^\Gamma_{\tup{\psi_1,\psi_2}}(v_1)\cap \W^\varphi_w \not\subseteq \truthset{\model^\varphi_w}{\psi_2}$; and 
let $v_2\in(\R^\varphi_w)_{\tup{\psi_1,\psi_2}}(v_1)$ s.t. $v_2 \not\in \truthset{\model^\varphi_w}{\psi_2}$. Then we have that $(\R^\varphi_w)_{\tup{\psi_1,\psi_2}}(v_1) \subseteq \R^\Gamma_{\tup{\psi_1,\psi_2}}(v_1)$, but also by IH $v_2 \not\in \truthset{\cmodel}{\psi_2}$. Thus, $\cmodel,v\not\models \kh_i(\psi_1,\psi_2)$, which is a contradiction. 
Then, it must be the case that $(\R^\varphi_w)_{\cset{\tup{\psi_1,\psi_2}}}(v_1) \subseteq \truthset{\model^\varphi_w}{\psi_2}$. 
Since we showed that $\truthset{\model^\varphi_w}{\psi_1} \subseteq \stexec^{\model^\varphi_w}(\cset{\tup{\psi_1,\psi_2}})$ and
$(\R^\varphi_w)_{\cset{\tup{\psi_1,\psi_2}}}(\truthset{\model^\varphi_w}{\psi_1}) \subseteq \truthset{\model^\varphi_w}{\psi_2}$, we conclude $\model^\varphi_w,v\models \kh_i(\psi_1,\psi_2)$.
\end{itemize}

Assume now $\model^\varphi_w,v\models \kh_i(\psi_1,\psi_2)$. Again, we consider two possibilities:

\begin{itemize}
\item $\bm{\truthset{\model^\varphi_w}{\psi_1}=\emptyset}$: 
since $\model^\varphi_w,v\models \kh_i(\psi_1,\psi_2)$, then 
 $\emptyset = \truthset{\model^\varphi_w}{\psi_1} \subseteq \stexec^{\model^\varphi_w}(\strategy')$ and $\emptyset = (\R^\varphi_w)_{\strategy'}(\truthset{\model^\varphi_w}{\psi_1}) \subseteq \truthset{\model^\varphi_w}{\psi_2}$ for some $\strategy' \in (\S^\varphi_w)_i$.
We claim that $\truthset{\cmodel}{\psi_1}=\emptyset$. Because otherwise
%\begin{itemize}
%\item If $\truthset{\cmodel}{\psi_1} \neq\emptyset$ and $\cmodel,v\models %\kh_i(\psi_1,\psi_2)$: by $\sel^{\varphi}_{w}$, \autoref{def:s-f5}, $\emptyset %\neq(\R^\varphi_w)_{\tup{\psi_1,\psi_2}}$ is defined and %$\truthset{\model^\varphi_w}{\psi_1} \neq\emptyset$.
%
if $\cmodel,v\models \kh_i(\psi_1,\psi_2)$, by $\sel^{\varphi}_{w}$, \autoref{def:s-f5}, $\emptyset \neq(\R^\varphi_w)_{\tup{\psi_1,\psi_2}}$ is defined and $\truthset{\model^\varphi_w}{\psi_1} \neq\emptyset$, contradicting hypothesis.  And if $\cmodel,v\not\models \kh_i(\psi_1,\psi_2)$, by $\sel^{\varphi}_{w}$, item \autoref{def:s-f6}, and IH,
$\truthset{\model^\varphi_w}{\psi_1} \neq\emptyset$, again a contradiction.

Let $\strategy$ be any set of plans in $\S^\Gamma_i$; since $\R^\Gamma_\strategy(\truthset{\cmodel}{\psi_1})=\emptyset$, $\truthset{\cmodel}{\psi_1} \subseteq \stexec(\strategy)$ and $\R^\Gamma_\strategy(\truthset{\cmodel}{\psi_1}) \subseteq \truthset{\cmodel}{\psi_2}$. Then, $\cmodel,v\models \kh_i(\psi_1,\psi_2)$.

\item $\bm{\truthset{\model^\varphi_w}{\psi_1}\neq\emptyset}$:  first, notice that
by IH,  $\truthset{\cmodel}{\psi_1}\neq\emptyset$. Also, by
 $\model^\varphi_w,v\models \kh_i(\psi_1,\psi_2)$, we get 
 $\truthset{\model^\varphi_w}{\psi_1} \subseteq \stexec^{\model^\varphi_w}(\strategy')$
and $(\R^\varphi_w)_{\strategy'}(\truthset{\model^\varphi_w}{\psi_1}) \subseteq \truthset{\model^\varphi_w}{\psi_2}$, for some $\strategy' \in (\S^\varphi_w)_i$. 
Aiming for a contradiction, suppose $\cmodel,v\not\models \kh_i(\psi_1,\psi_2)$. 
This implies that for all $\strategy \in \S^\Gamma_i$, 
$\truthset{\cmodel}{\psi_1} \not\subseteq \stexec^{\cmodel}(\strategy)$ or
$\R^\Gamma_\strategy(\truthset{\cmodel}{\psi_1}) \not\subseteq \truthset{\cmodel}{\psi_2}$. 
Also, by definition of $\ACT_\varphi$ we have that for all $\strategy=\cset{a} \in (\S^\varphi_w)_i$, with $a \in \ACT_\varphi$, 
$\truthset{\cmodel}{\psi_1} \not\subseteq \stexec^{\cmodel}(\strategy)$ or
$\R^\Gamma_\strategy(\truthset{\cmodel}{\psi_1}) \not\subseteq \truthset{\cmodel}{\psi_2}$;
i.e., for all $a \in \ACT_\varphi$
$\truthset{\cmodel}{\psi_1} \not\subseteq \stexec^{\cmodel}(\cset{a})$ or $\R^\Gamma_{\cset{a}}(\truthset{\cmodel}{\psi_1}) \not\subseteq \truthset{\cmodel}{\psi_2}$.
Thus, there exists $w_1 \in\truthset{\cmodel}{\psi_1}$ s.t.
$w_1 \not\in \stexec^{\cmodel}(a)$ or there exists $w_2 \in\R^\Gamma_a(w_1)$ s.t. $w_2 \not\in \truthset{\cmodel}{\psi_2}$.
By definition of $\sel^{\varphi}_{w}$, \autoref{def:s-f6}, we add witnesses for each $a \in \ACT_\varphi$. So, let $\strategy' \in (\S^\varphi_w)_i$. If $\strategy'=\cset{\tup{\bot,\top}}$,
trivially we obtain 
$\emptyset \neq\truthset{\model^\varphi_w}{\psi_1}\not\subseteq \stexec^{\model^\varphi_w}(\strategy')=\emptyset$.
Then, take another $\strategy'=\cset{a}$ s.t. $a\in\ACT_\varphi$,
and $w'_1 \in \truthset{\model^\varphi_w}{\psi_1} \subseteq \truthset{\cmodel}{\psi_1}$. 
If $w'_1 \not\in \stexec^{\cmodel}(\cset{a})$, $\R^\Gamma_a(w'_1)=\emptyset$ and thus $(\R^\varphi_w)_a(w'_1)=\emptyset$ and therefore $w'_1 \not\in \stexec^{\model^\varphi_w}(\cset{a})$.
On the other hand, if there exists $w_2 \in\R^\Gamma_a(w'_1)$ s.t. $w_2 \not\in \truthset{\cmodel}{\psi_2}$, then by $\sel^{\varphi}_{w}$ and IH,
there exists $w'_2 \in\W^\varphi_w$ s.t.
$w'_2 \in\R^\Gamma_a(w'_1)$ and $w'_2\not\in\truthset{\model^\varphi_w}{\psi_2}$, and 
consequently, there exists $w'_2 \in(\R^{\varphi}_{w})_a(w'_1)$ s.t. $w'_2 \not\in \truthset{\model^{\varphi}_{w}}{\psi_2}$. In any case, it leads to $\model^{\varphi}_{w},v\not\models \kh_i(\psi_1,\psi_2)$, a contradiction. Therefore, $\cmodel,v\models \kh_i(\psi_1,\psi_2)$.

\end{itemize}

\medskip
Notice now that the selection function adds states from $\cmodel$, only for each $\khi$-formula that 
  appears as a subformula of $\varphi$;
  and the number of states added at each time is polynomial 
  in $|\varphi|$. Hence, the size of $\W^\varphi_w$ is polynomial. Since $(\S^\varphi_w)_i$ is also polynomial, the size of $\model^\varphi_w$ is polynomial in $|\varphi|$.
  Finally, the definition of an \ults requires the set of actions to be infinite. It is easy to see that we can extend $\ACT_\varphi$ to an infinite set. Since~\autoref{def:abmap} defines relations over some subset of the actions, this extension does not alter the size of $\model^\varphi_w$.
\end{demostracion}

    \end{proposicion}

  In order to prove that the satisfiability problem of $\KHilogic$
  is in \NP, it remains to show that the model checking problem is in \Poly. 

\begin{proposicion}\label{prop:modcheck}
  The model checking problem for $\KHilogic$ is in \Poly.
\begin{demostracion}
  Given a pointed $\ults$ $\model,w$ and a formula $\varphi$, we define
  a bottom-up labeling algorithm running in polynomial time which checks whether 
  $\model,w\models\varphi$. We follow
  the same ideas as for the basic modal logic {\sf K} (see
  e.g.,~\cite{blackburn06}). Below we introduce the case for formulas
  of the shape $\kh_i(\psi,\varphi)$, over an $\ults$
  $\model=\tup{\W,\R,\S,\V}$:

  \begin{algorithm}
  \begin{algorithmic}
  \small
    \State{{\bf Procedure} ModelChecking($(\model,w)$, $\kh_i(\psi,\varphi))$} 
    \State $lab(\kh_i(\psi,\varphi))\gets \emptyset$;
    \ForAll{$\strategy\in\S_i$} 
      \State{$kh \gets True$}; 
      \ForAll{$\sigma\in\strategy$}
        \ForAll{$v\in lab(\psi)$}
          \State{$kh\gets (kh \ \& \ v\in\stexec(\sigma) \ \& \ \R_\sigma(v)\subseteq lab(\varphi))$}; 
        \EndFor
      \EndFor

      \If{$kh$}
        \State $lab(\kh_i(\psi,\varphi))\gets \W$;
      \EndIf
    \EndFor 
  \end{algorithmic}
\end{algorithm}

  As $\S_i$ and each $\strategy\in\S_i$ are not empty, the first two
  {\bf for} loops are necessarily executed.  If
  $lab(\psi) =\emptyset$, then the formula $\kh_i(\psi,\varphi)$ is
  trivially true. Otherwise, $kh$ will remain true only if the
  appropriate conditions for the satisfiability of
  $\kh_i(\psi,\varphi))$ hold. If no $\strategy$ succeeds, then the
  initialization of $lab(\kh_i(\psi,\varphi))$ as $\emptyset$ will not
  be overwritten, as it should be.  Both $v\in\stexec(\sigma)$ and
  $\R_\sigma$ can be verified in polynomial time. Hence, the model
  checking problem is in \Poly.
\end{demostracion}
\end{proposicion}

The intended result for satisfiability now follows.

\begin{teorema}\label{th:khcomplexity}
The satisfiability problem for $\KHilogic$ over $\ultss$ is \NP-complete.
\begin{demostracion}  
      Hardness follows from \NP-completeness of propositional logic (a fragment of $\KHilogic$).
      By \autoref{prop:selection-preserves-sat}, each satisfiable formula 
      $\varphi$ has a model of polynomial-size on $\varphi$. Thus, we can guess 
      a polynomial model $\model,w$, and verify $\model,w\models\varphi$ (which can 
      be done in polyonomial time, due to~\autoref{prop:modcheck}). %similarly as for modal logic {\sf K}~\cite{blackburn06}) (see~\autoref{sec:appendix} for details). 
      Thus, the satisfiability problem is in the class \NP.
    \end{demostracion}
\end{teorema}

%%% Local Variables:
%%% mode: latex
%%% TeX-master: "tark21"
%%% End:

%------------------------------------------------------------------------------------------------
\section{Final remarks}\label{sec:final}
In this article, we introduce a new semantics for the \emph{knowing
  how} modality from~\cite{Wang15lori,Wang16,Wang2016}, for multiple
agents. It is defined in terms of \emph{uncertainty-based labeled
  transition systems ($\ults$)}. The novelty in our proposal is that
$\ultss$ are equipped with an indistinguishability relation among
plans.  In this way, the epistemic notion of uncertainty of an agent
--which in turn defines her epistemic state-- is reintroduced, bringing the notion of \emph{knowing how} closer to the notion of \emph{knowing that} from classical epistemic
logics.  We believe that the semantics based on $\ults$ can represent properly the
situation of a shared, objective description of the affordances of a
given situation, together with the different, subjective and personal
abilities of a group of agents; this seems difficult to achieve using
a semantics based on LTSs alone.

We show that the logic of~\cite{Wang15lori,Wang16,Wang2016} can be obtained
by imposing particular conditions over $\ults$; thus, the new semantics is more general.
In particular, it provides counter-examples to \axm{EMP} and \axm{COMP}, which  directly link the knowing how modality $\kh$ to
properties of the universal modality. 
%\footnote{The rest of the axioms and rules in \KHaxiom  (those shown in block $\axset$) merely state properties of the universal modality and the fact that $\kh$ is global.}
Indeed, consider \axm{EMP}: even though $\A(\psi\ra\varphi)$ objectively holds in the underlying
LTS of an $\ults$, it could be argued that an agent might not be aware of actions or plans to turn those facts into knowledge, resulting in $\kh(\psi,\varphi)$ failing in the
model.

 To characterize validities in this language over \ultss, we introduce a sound and strongly complete axiom system. 

We also define a suitable notion of bisimulation  over $\ultss$, following 
 ideas introduced in~\cite{FervariVQW17,FervariVQW21}. We show that bisimilarity implies formula equivalence, and 
that finite models form a Hennessy-Milner class (i.e., that formula equivalence implies
bisimilarity over finite models). 

Finally, we prove that the satisfiability problem for our multi-agent knowing
how logic over the $\ults$-based semantics is \NP-complete. The proof relies on a selection argument on the canonical model, and on the fact that the model checking problem is polynomial. We also provide a filtration technique that, given an arbitrary model satisfying  $\varphi$, returns a finite model that satisfies $\varphi$.

% Finally, we exemplify how the new semantics enables the definition of
% dynamic operations that update the knowledge of an agent. 
% It is natural to define operators 
% updating the knowledge of an agent by reducing her \emph{uncertainty}. In the 
% LTS-based semantics, one can model these updates by removing  
% actions that violate the strong executability of a plan; but it is difficult to control such removal, and ensure that it does not alter the strong executability statu	s of other, unrelated, plans. Similarly, one could add new strongly executable plans, but again this can lead to side effects.
% To exemplify the dynamic possibilities of the \esm based semantics, 
% we defined
% a modality which relies on updates in the
% indistinguishability relation between plans.  
% We also consider some variants of the operator, ranging
% from an (in a multi-agent setup, public) announcement to
% distinguish two specific plans, arbitrary announcements, and a goal-directed learning operator.
% To the best of our knowledge, this article investigates, for the first time, 
% dynamic modalities in the context of a \emph{knowing how} logic. 

\ssparagraph{Future work.}
There are several interesting lines of research to explore in the
future.  First, our framework easily accommodates other notions of
executability. For instance, one could require only some of the plans in
a set $\strategy$ to be strongly executable, or weaken the condition
of \emph{strong} executability, etc. We can also explore the 
effects of imposing different restrictions
on the construction of the indistinguishability relation between plans. 
  It would be interesting to investigate which logics we
obtain in these cases, and their relations with the \lts
semantics. 

Second, to our knowledge, the exact complexity of
the satisfiability problem for knowing how over \lts{s} is open.
It would be interesting to solve this problem, for instance, by following and adapting ideas from~\cite{MeierTVM09}.
% RF: sabemos que esto no funciona.
% see whether an adaptation of our selection procedure works over \lts{s}. 

Third, the $\ults$ semantics, in the multi-agent
setting, leads to natural definitions of concepts of \emph{collective} knowing how, in the spirit of~\cite{RAK}.
For instance, one can easily define a notion of \emph{general knowing how} as
$\EKh_G(\psi,\varphi) := \bigwedge_{i\in G} \khi(\psi,\varphi)$, whose reading is \emph{``everyone in the group $G$ knows how to achieve $\varphi$ given $\psi$''}; and 
\emph{``somebody in the group $G$ knows how to achieve $\varphi$ given $\psi$''}, 
as $\SKh_G(\psi,\varphi) := \bigvee_{i\in G} \khi(\psi,\varphi)$  (see, e.g.,~\cite{AgotnesW21} for a similar approach in standard epistemic logic).
Other, more complex notions such as
%\emph{global}, 
\emph{distributed} and \emph{common knowing how}, 
deserve further exploration.
% Second, in this article, we barely scratch
% the surface of the dynamic possibilities offered by the new semantics.
% We would like to investigate properties like axiomatization,
% expressivity and complexity of the dynamic operators presented.
%% RAUL: commented for long version
% Regarding axiomatizations, one of the main obstacles towards getting a
% complete axiomatization for these dynamic modalities is that the rule
% of uniform substitution fails. This behaviour is not surprising: when
% dealing with modalities for model-changing operations, we usually lose
% uniform substitution (see, e.g.,~\cite{HollidayHI13,ArecesFH15}).
% Regarding expressivity, we would like to know the exact relationship
% between the modalities we proposed, as this issue turned to be
% particularly challenging.  We plan to approach this issue by defining
% Ehrenfeucht-Fra\"iss\'e games to capture their respective
% expressivity.

Finally, dynamic modalities capturing epistemic updates can be defined
via operations that modify the indistinguishability relation among
plans (as is done with other dynamic epistemic
operators, see, e.g.,~\cite{DELbook}).  This would allow to express
different forms of communication, such as \emph{public},
\emph{private} and \emph{semi-private} announcements concerning (sets of) plans.
Some preliminary results have been presented in~\cite{arec:firt23}.

%%% Local Variables:
%%% mode: latex
%%% TeX-master: "tark21"
%%% End:

%\paragraph*{\bf Acknowledgments.} This work is partially supported by projects ANPCyT-PICT-2017-1130, Stic-AmSud 20-STIC-03 `DyLo-MPC', Secyt-UNC, GRFT Mincyt Cba, and by the Laboratoire International Associ\'e SINFIN.

\bibliographystyle{plain}
\bibliography{references}

\end{document}
% end of file template.tex